\documentclass[11pt]{article}
 \usepackage{a4,amssymb,verbatim,amsmath}

 \usepackage{cite}

 \usepackage{hyperref}

 \numberwithin{equation}{section}

\newtheorem{theorem}{Theorem}[section]

\newtheorem{remark}[theorem]{Remark}

\newtheorem{ex}{Example}[section]

\newtheorem{ass}{Assumption}[section]

\numberwithin{equation}{section}

\newcommand{\xbasis}{\kappa}

\begin{document}

\newcommand{\ddt}{\partial \over \partial t}
\newcommand{\dds}{\partial \over \partial s}

\newcommand{\ddx}{\partial \over \partial x}
\newcommand{\ddy}{\partial \over \partial y}
\newcommand{\ddz}{\partial \over \partial z}

\newcommand{\ddu}{\partial \over \partial u}
\newcommand{\ddv}{\partial \over \partial v}
\newcommand{\ddw}{\partial \over \partial w}

\newcommand{\ddrho}{\partial \over \partial \rho }
\newcommand{\ddp}{\partial \over \partial p}

\newcommand{\ddEy}{\partial \over \partial E^y }
\newcommand{\ddEz}{\partial \over \partial E^z }
\newcommand{\ddHy}{\partial \over \partial H^y }
\newcommand{\ddHz}{\partial \over \partial H^z }

\newcommand{\ddphi}{\partial \over \partial  \varphi}

\newcommand{\ddtbar}{\partial \over \partial \bar{t}}
\newcommand{\ddxbar}{\partial \over \partial \bar{x}}

% \textit{\today}  15.10.2021

\bigskip
\par
\bigskip

\begin{center}
{\Large {\bf   Plane  one-dimensional  MHD  flows:   \\
symmetries and conservation laws}}
\end{center}

\begin{center}

 %  {12.10.2021}

\end{center}

\begin{center}

{\large Vladimir A.  Dorodnitsyn}$^{1,a}$, {\large  Evgeniy  I. Kaptsov}$^{2,b}$, \\
\smallskip
{\large Roman V. Kozlov}$^{3,c}$ , {\large Sergey  V. Meleshko}$^{2,d}$,
Potcharapol Mukdasanit$^{2,e}$
\end{center}

\bigskip

\begin{tabbing}
$^1$ \= Keldysh Institute of Applied Mathematics, Russian Academy of Science, \\
\> Miusskaya Pl.~4, Moscow, 125047, Russia;\\
$^2$ \= School of Mathematics, Institute of Science, \\
\> Suranaree University of Technology, Nakhon Ratchasima, 30000, Thailand;\\
$^3$ \= Department of Business and Management Science, Norwegian School of Economics, \\
\> Helleveien 30, 5045, Bergen, Norway;
\\
\\
$^a$ \= Dorodnitsyn@Keldysh.ru \\
$^b$ \= evgkaptsov@math.sut.ac.th  \\
$^c$ \= Roman.Kozlov@nhh.no  \\
$^d$ \= sergey@math.sut.ac.th \\
$^e$ \= sherlock\_nono@hotmail.com
\end{tabbing}

\begin{center}
{\bf Abstract}
\end{center}
\begin {quotation}
{The paper  considers  the plane one-dimensional flows for
magneto{hydro}dynamics in the mass Lagrangian coordinates. The inviscid,
thermally non-conducting medium is modeled by  a polytropic gas. The
equations are examined for symmetries and conservation laws.
For the case of the finite electric conductivity
we establish  Lie group classification, i.e.
we describe all cases of the conductivity $ \sigma ( \rho , p)$
for which there are symmetry extensions.
The conservation laws are derived  by the direct computation.
For the case of the infinite electrical conductivity
 the equations can be brought into a variational form
in the Lagrangian coordinates.
Lie group classification is performed for the entropy function
as an arbitrary element.
Using the variational structure, we employ the Noether theorem for obtaining conservation laws.
The conservation laws are also given in the physical variables.}

% {to add}  invariant solutions

% {to add}  conservative difference schemes

\end{quotation}

\bigskip

Key words:

\smallskip

Lie point symmetries,

conservation laws,

Noether theorem,

Euler-Lagrange equations

\eject

\section{Introduction}

The equations of magneto{hydro}dynamics (MHD)  describe
motion of electrically  conducting fluids  under the action of the internal forces,
which consist of the pressure and electromagnetic forces.
 These equations describe phenomena related to plasma flows,
for example in plasma confinement, as well as physical problems in
astrophysics and fluid metals flows.

In the present paper  we consider plane one-dimensional MHD flows.
The equations which describe such flows will be examined
for Lie point symmetries and conservation laws.
We assume that medium is inviscid and thermally non-conducting.
It is modeled by a polytropic gas.
Both cases of finite and infinite  electric conductivity are analyzed.
The  particular case of the infinite  electric conductivity
corresponds to "freezing"  of the magnetic force  lines
in the  trajectories  of motion.

Lie point symmetries represent an efficient tool to analyze nonlinear differential equations
\cite{bk:Ovsyannikov[1962], bk:Ibragimov1985, bk:Olver, bk:Bluman1989}.
They are related to fundamental physical principles of the considered models
and correspond to important properties of the   differential equations:

\begin{itemize}

\item

Transformations  generated by symmetries transfer solutions into another solutions.
It allows to find new solutions from the known ones.

\item

Symmetries of  PDEs allow to find particular solutions  of a special form,
the so called invariant solutions.

\item

Invariance of variational PDEs   is a necessary condition for application of Noether's theorem,
which can be used to find conservation laws.

\end{itemize}

%  {to list what symmetries can do }

Lie group symmetries of various versions of MHD equations
were considered in many publications.
For example, the case of the finite  conductivity  was investigated in
\cite{bk:Gridnev1968, bk:DorMHDpreprint1976}.
The case of the infinite conductivity  was examined in
\cite{art:Rogers1969, bk:HandbookLie_v2, art:MeleshkoMoyoWebb2021, bk:PaliathanasisMHD2021}.
Invariant solutions were considered
in    \cite{bk:Oliveri_F_2005, bk:Picard_P_Y_2008,  bk:Golovin_2009, bk:Golovin_2011, bk:Golovin_2019}.
It should be noted that most of the papers devoted to applications of
Lie group symmetries to MHD consider the case of the infinite conductivity.

Variational methods have many applications in mathematics and physics.
If differential equations  have a form of Euler-Lagrange equations,
there is a possibility to employ    the Noether theorem  \cite{bk:Noether1918}.
The theorem allows one to use symmetries of the differential equations
which are either variational or divergence symmetries of the  Lagrangian function
to derive conservation laws.
Several other approaches to find conservation laws and other conserved quantities
for MHD were recently reviewed in \cite{bk:Webb2018}.

% The focus is on the conservation laws.

This paper is organized as follows.
The next section
% reviews a method which will be used
% to {facilitate derivation} of Lie group classification and
provides a short description of the Noether theorem,  specified for this paper.
Section \ref{MHD}   describes the equations of MHD and restricts them for plane one-dimensional flows.
In this section we also introduce  mass Lagrangian coordinates.
Symmetries and conservation laws of the MHD equations with the finite conductivity
are obtained in Sections \ref{Symmetries_finite} and \ref{CL_finite}, respectively.
Symmetries and conservation laws for the infinite conductivity are treated
in Sections  \ref{Symmetries_infinite} and \ref{CL_infinite}.
Finally, Section \ref{Concluding} provides concluding remarks.
Some technical details are  extracted into the Appendices.

\section{Background theory}

\label{Background}

In the next section we will describe the
magnetohydrodynamics equations and
specify them for plane one-dimensional flows,
which will be analyzed for admitted Lie point symmetries and
conservation laws.
The conservation laws will be obtained
by direct  computations
% the method which will use used to reduce
% the  amount of work needed to derive
% Lie group classification of the plane one-dimensional MHD flows
and using the Noether theorem,
which can be employed to find conservation laws
if the equations have a variational formulation.

% for the case of infinite conductivity.

\subsection{Lie group classification problem}

The Lie group classification problem consists of finding all Lie
groups admitted by a system of partial differential equations
\cite{bk:Ovsyannikov[1962], bk:Ibragimov1985, bk:Olver}. Admitted
groups can depend  on arbitrary elements (constants and functions of
the independent and dependent variables) included in the equations.
Practically, the groups are presented by their generators.
The generators admitted  for all arbitrary elements are called
the kernel of the admitted Lie algebras.
Lie group classification presents
all non-equivalent extensions of the kernel and the corresponding specific form of the arbitrary elements.
It is performed with respect to equivalence
transformations,  which preserve the structure of the equations but may
change the arbitrary elements.

\subsection{Noether theorem}

The Noether theorem
\cite{bk:Noether1918}
(see also    \cite{bk:Ovsyannikov[1962], bk:Ibragimov1985, bk:Olver, bk:Bluman1989})
can be used to find conservation laws of variational equations  with symmetries.
Here we present a simplified version of this theorem
restricted to second-order PDEs
with two independent variables $(t,s)$, which represent time and one spacial coordinate.
In this case we need to consider
first-order Lagrangian functions
\begin{equation}     \label{Lagrangian}
  {L}={L}( t, s, \varphi   ,   \varphi   _t, \varphi   _s )  ,
\qquad
\varphi  =  ( \varphi ^1 , \ldots  ,      \varphi ^m  )  .
\end{equation}
The Lagrangian provides the second-order  Euler-Lagrange equations
%for Lagrangian~(\ref{Lagrangian_function})
% is computed as
\begin{equation}     \label{EL_equations}
\frac{\delta{L}}{\delta \varphi   ^i  }
= { \partial  L \over \partial \varphi   ^i   }
-  D_t   \left( { \partial  L \over \partial  \varphi   ^i  _t }     \right)
-  D_s   \left( { \partial  L \over \partial  \varphi   ^i  _s }   \right)
 =    0  ,
\qquad
i = 1, \ldots , m ,
\end{equation}
where $ D_t$ and $ D_ s$ are total differentiation operators with respect to $t$ and $s$.
Operators $\displaystyle\frac{\delta}{\delta   \varphi  ^i   } $ are called the variational operators.

Lie point symmetries of these differential equations are given by the operators of the form
\begin{equation}      \label{Symmetry}
X =\xi^{t} (t,s,  \varphi   )  \frac{\partial}{\partial t}
+\xi^{s} (t,s, \varphi   ) \frac{\partial}{\partial s}
+\eta^{i} (t,s,\varphi  ) \frac{\partial}{\partial  \varphi   ^i }   .
\end{equation}
It is assumed that the operator is prolonged to the second-order derivatives,
present in the Euler-Lagrange equations,
according to the standard prolongation formulas
\cite{bk:Ovsyannikov[1962], bk:Ibragimov1985, bk:Olver, bk:Bluman1989}.

The Noether theorem is based on the following identities.
The first identity  \cite{bk:Ibragimov1985}
relates the invariance of the elementary action,
which is also called invariance of the Lagrangian,
to the conservation laws:
% {(Is there an article to refer)}
\begin{equation}      \label{Noether_identity}
X  {L}+{L}(D_{t} \xi^{t}+D_{s}\xi^{s})
= ( \eta^{i}  - \xi ^t  \varphi   ^i _t  -  \xi ^s  \varphi   ^i _s  )
\frac{\delta{L}}{\delta \varphi   ^i  }
+ D_t ( N^t L ) +  D_s ( N^s L )    ,
\end{equation}
where
\begin{equation}     \label{Noether_operators}
N^{t}
=\xi  ^{t}
+  (  \eta^{i} - \xi^{t}    \varphi ^i  _{t} - \xi ^{s}   \varphi ^i  _{s}  )
\frac{\partial  }{\partial     \varphi ^i   _{t} }   ,
\qquad
N^{s}
=\xi  ^{s}
+  (  \eta^{i} - \xi^{t}    \varphi ^i  _{t} - \xi ^{s}   \varphi ^i  _{s}  )
\frac{\partial  }{\partial     \varphi ^i   _{s} }   ,
\end{equation}
are  the Noether operators.

If
$$
X  {L}+{L}(D_{t} \xi^{t}+D_{s}\xi^{s}) = 0 ,
$$
% $B _1 \equiv 0 $ and $  B _2 \equiv  0 $,
the symmetry $X $ is called a {\it variational} symmetry of the Lagrangian.
For
$$
X  {L}+{L}(D_{t} \xi^{t}+D_{s}\xi^{s}) = D_{t}  B _{1}   +  D_{s}  B _{2}
$$
with {non}trivial  $B _1  (t,s, \varphi   )$
and $B _2 (t,s, \varphi   ) $ we say that $X$ is a {\it divergence} symmetry.

The other set of identities \cite{bk:DorodnitsynKozlov[2011]}
(see also \cite{bk:Olver})
relates the invariance of the  Lagrangian
to the invariance of  the Euler-Lagrange equations:
\begin{multline}
\frac{\delta}{\delta \varphi ^j }
\left(
 X  {L}
+ {L}(D_{t} \xi^{t}+D_{s}\xi^{s}) \right)
=
X \left( \frac{\delta L}{\delta  \varphi ^{j}} \right)
\\
+
\left(\frac{\partial\eta^{k}}{\partial \varphi  ^{j}}
-\frac{\partial\xi^{t}}{\partial \varphi  ^{j}}  \varphi  _{t}^{k}
-\frac{\partial\xi^{s}}{\partial \varphi  ^{j}}  \varphi  _{s}^{k}
+     D_{t} \xi^{t}+D_{s}\xi^{s}  \right)
\frac{\delta L}{\delta \varphi  ^{k}}   ,
\quad
j=1,2,  \ldots ,m .
% \label{eq:second}
\end{multline}

The Noether theorem is formulated as follows:

\begin{theorem}    \cite{bk:Noether1918}
Let the Lagrangian function (\ref{Lagrangian}) satisfy the equation
\begin{equation}
   X  {L}+{L}(D_{t} \xi^{t}+D_{s}\xi^{s})
=D_{t}  B _{1}+D_{s}  B _{2}  ,
\label{eq:Noether}
\end{equation}
where $X$ is a generator   (\ref{Symmetry})
and  $ B _{i}(t,s, \varphi   ) $, $ i=1,2$.
Then  the operator $X$ is a symmetry of the Euler-Lagrange equations (\ref{EL_equations}),
and the    Euler-Lagrange equations  possess a conservation law
\begin{equation}
D_{t}  (   N^{t} L - B _{1}  )
+
D_{s}   (   N^{s} L - B _{2} )
= 0  .
\end{equation}
\end{theorem}

% If $B _1 \equiv 0 $ and $  B _2 \equiv  0 $,  the symmetry $X $ is called a {\it variational} symmetry of the Lagrangian. For {non}trivial  $B _1$ and $B _2$ we say that $X$ is a {\it divergence} symmetry.

% The  densities of the conservation laws
% $(T^{t}  ,T^{s})$
% \begin{equation}       \label{conservation_laws_densities}
% T^{t}
% =\xi  ^{t} {L}
% +  (  \eta^{i} - \xi^{t}   u ^i  _{t} - \xi ^{s}  u ^i  _{s}  )  \frac{\partial  L }{\partial    u ^i   _{t} }  - B_1   ,
% \qquad
% T^{s}
% =\xi  ^{s}{L}
% +  (  \eta^{i} - \xi^{t}  u ^i   _{t} - \xi ^{s}  u ^i   _{s}  )  \frac{\partial  L }{\partial    u ^i   _{s}}   -  B_2  .
% \end{equation}

% {Explain:} Noether  = variational  + divergence

\section{Magneto{hydro}dynamics equations  and plane one-dimensional   flows}
%  of an ideal polytropic  gas}

\label{MHD}

\subsection{Three-dimensional MHD}

The magneto{hydro}dynamics equations
in Eulerian coordinates
can be written in different ways \cite{bk:KulikovskiiLubimov1965, bk:SamarskyPopov_book[1992], bk:Landau_Electrodynamics}
(see also  \cite{Davidson2001,  Galtier2016,  Brushlinskii, bk:Webb2018}).

For simplicity we take the  dimensionless (scaled) form of MHD equations with the finite conductivity
\begin{subequations}       \label{3D_system}
\begin{gather}
 \rho _t  +    \mbox{div} (  \rho   \textbf{u} ) =0 ,
\\
\textbf{u}  _t +    (  \textbf{u}  \cdot    \nabla  ) \textbf{u}  =
-  { 1 \over \rho }     \nabla   p
+   {    [  \textbf{i}  \times  \textbf{H} ]     \over  \rho  } ,
\\
% \varepsilon   _ t    +     (  \textbf{u}  \cdot    \nabla  ) \varepsilon   =
% -  { p \over \rho }   \   \mbox{div}   \  \textbf{u}
% +     { \sigma  \over \rho }    \   \textbf{E} ^2     ,
% \qquad
% or
% \qquad
 \varepsilon   _ t    +     (  \textbf{u}  \cdot    \nabla  ) \varepsilon   =
 -  { p \over \rho }   \   \mbox{div}   \  \textbf{u}
 +     { 1 \over \rho }    \   (  \textbf{i}      \textbf{E}  )   ,
 \label{3D_system_energy}
\\
\textbf{H} _t
=    \mbox{rot}  \  [   \textbf{u}   \times   \textbf{H} ]
-     \mbox{rot} \   \textbf{E} ,
\qquad
\mbox{div} \ \textbf{H} = 0 ,
\\
\textbf{i}  =  \sigma  \textbf{E}  =   \mbox{rot}  \ \textbf{H}   .
 \label{relation}
\end{gather}
\end{subequations}
Here $ \rho $  is the density, $p$ is the  pressure
and   $ \varepsilon $ is the internal energy per unit volume.
In the three-dimensional  space
we denote the coordinates and the velocity components
as $\mathbf{x} = (x,y,z)$ and   $\mathbf{u} = (u,v,w)$;
\begin{equation*}
 \nabla = \left(   {\ddx} ,    {\ddy} ,  {\ddz}  \right)
\end{equation*}
is the gradient operator.
The  equation  (\ref{relation})    gives
relations for {the electric   current density}   $\mathbf{i} = (i^x ,i^y,i^z)$,
the electric field  $\mathbf{E} = (E^x ,E^y,E^z)$
and the magnetic field  $\mathbf{H} = (H^x,H^y,H^z)$.
% \begin{equation}     \label{relation}
% \textbf{i}  =  \sigma  \textbf{E}  =   \mbox{rot}  \ \textbf{H}  .
% \end{equation}
It contains the electric conductivity  $\sigma = \sigma    ( \rho , p) {\not \equiv}  0 $.
Note that these relations    allow to write down different forms of the MHD system.
In particular,   $ \textbf{i}  $  and $ \textbf{E}  $ can be eliminated from the system.

The MHD system   (\ref{3D_system})
should be supplemented by the equation of {\it state}
which has  the form
\begin{equation*}
  \varepsilon = \varepsilon  (\rho , p) .
\end{equation*}
We consider a  medium described by an ideal gas
\cite{bk:Chernyi_gas, bk:Ovsyannikov[2003], Landau, Chorin, Toro}
\begin{equation}   \label{ideal}
p = \rho R T,
\end{equation}
where $ T $ is the temperature and  $R$ is the specific gas constant.
The ideal gas is called  polytropic
if the internal energy function $ \varepsilon $ is linear in the temperature
\begin{equation} \label{EpsCvT}
\varepsilon (T) = c_{v} T  = { R T \over \gamma -1} ,
\end{equation}
where $c_v$ is the specific heat capacity measured at constant volume
and
\begin{equation*} \label{gammaRCv}
\gamma = 1 + \frac{R}{c_{v}}  > 1
\end{equation*}
is  the polytropic constant.
Eliminating the temperature from  (\ref{ideal}) and    (\ref{EpsCvT}),
we obtain  the equation of state
\begin{equation} \label{InternalEnergyRel}
\varepsilon = \frac{1}{\gamma - 1} \frac{p}{\rho}.
\end{equation}

The pressure, the density and the entropy $\tilde{S}$ are related by the equation
\cite{bk:Chernyi_gas, bk:Ovsyannikov[2003]}
\begin{equation}   \label{function_S}
p =  S  \rho^{\gamma} ,
\qquad
S = e^{(\tilde{S}-\tilde{S}_{0})/c_{v}} ,
\end{equation}
where   $\tilde{S}_{0}$  is constant.

\begin{remark}
The equation of energy conservation
can be written in different forms.
Instead of the  equation   (\ref{3D_system_energy})
it is possible to consider the equation for  the pressure
\begin{equation}      \label{equation_p_Euler}
p  _ t    +     (  \textbf{u}  \cdot    \nabla  )  p
+ \gamma    p     \   \mbox{div}  \   \textbf{u}
=    ( \gamma  -1 )      (  \textbf{i}      \textbf{E}  )
\end{equation}
or the equation for the function $ S $, which corresponds to the entropy as given in  (\ref{function_S}),
\begin{equation}   \label{equation_S_Euler}
 S _ t    +     (  \textbf{u}  \cdot    \nabla  ) S    =
      {    \gamma  -1    \over \rho   ^{\gamma} }    \   (  \textbf{i}      \textbf{E}  )   .
\end{equation}
\end{remark}

\begin{remark}
For  material derivative
\begin{equation*}
{ d  \over d t }
=
{ \partial  \over \partial t }
+
(  \textbf{u}  \cdot    \nabla  ) ,
% \qquad
%  \nabla = \left(   {\ddx} ,    {\ddy} ,  {\ddz}  \right) ,
\end{equation*}
i.e. the time derivation along the    trajectories,
we can rewrite    the system    (\ref{3D_system}) as
\begin{subequations}
\begin{gather}
   { d  \over d t }  \rho    +     \rho \  \mbox{div}   \  \textbf{u}   =  0 ,
\\
 { d  \over d t }  \textbf{u}
 =
-  { 1 \over \rho }     \nabla   p
+   {    [  \textbf{i}  \times  \textbf{H} ]     \over  \rho  } ,
\\
% \varepsilon   _ t    +     (  \textbf{u}  \cdot    \nabla  ) \varepsilon   =
%  -  { p \over \rho }   \   \mbox{div}   \  \textbf{u}
% +     { \sigma  \over \rho }    \   \textbf{E} ^2     ,
% \qquad
% or
% \qquad
{ d  \over d t }   \varepsilon
 =
 -  { p \over \rho }   \   \mbox{div}   \  \textbf{u}
 +     { 1 \over \rho }    \   (  \textbf{i}      \textbf{E}  )   ,
\\
% { d  \over d t }
% \textbf{H}
% =
%   (  \textbf{u}  \cdot    \nabla  )    \textbf{H}
% +
%   \mbox{rot}  \  [   \textbf{u}   \times   \textbf{H} ]
% -     \mbox{rot} \   \textbf{E} ,
%  \qquad
% \mbox{div} \ \textbf{H} = 0 ,
% \\
{ d  \over d t }
\textbf{H}
=
  (  \textbf{H}  \cdot    \nabla  )    \textbf{u}
-  \textbf{H} \     \mbox{div}  \    \textbf{u}
-     \mbox{rot} \   \textbf{E} ,
\qquad
\mbox{div} \ \textbf{H} = 0 ,
\\
\textbf{i}  =  \sigma  \textbf{E}  =   \mbox{rot}  \ \textbf{H}  .
\end{gather}
\end{subequations}
In this case equations (\ref{equation_p_Euler}) and (\ref{equation_S_Euler})
get rewritten as
\begin{equation}
{ d  \over d t }    p
+ \gamma    p     \   \mbox{div}  \   \textbf{u}
=    ( \gamma  -1 )      (  \textbf{i}      \textbf{E}  )
\end{equation}
and
\begin{equation}
 { d  \over d t }   S     =
      {    \gamma  -1    \over \rho   ^{\gamma} }    \   (  \textbf{i}      \textbf{E}  )   .
\end{equation}
\end{remark}

\subsection{Plane one-dimensional flows}

The plane one-dimensional MHD flows
represent  one-dimensional flows of the system  (\ref{3D_system})
with all dependent variables
being functions of only two independent variables:  $t$ and $x$.
In this case the equations
\begin{subequations}   \label{plane_simplification}
\begin{gather}
H^x   _ t  = 0   ,
\\
\mbox{div} \ \textbf{H}
=   {  \partial H^x  \over \partial x}
% +   {  \partial H^y  \over \partial y}
% +   {  \partial H^z  \over \partial z}
=  0
\end{gather}
\end{subequations}
give
\begin{equation}     \label{condition_Hx}
H ^x = H ^0  = \mbox{const} .
\end{equation}
% The relations   (\ref{relation})
% \begin{equation*}
% \textbf{i}  =  \sigma  \textbf{E}  =   \mbox{rot}  \  \textbf{H}
% \end{equation*}
%  takes the form
% \begin{equation}          \label{relation_Euler}
% i^x = E^x = 0 ,
% \qquad
% t^y = \sigma   E^y    =   - H^z _x  ,
% \qquad
% i^z =  \sigma E^z  =     H^y _x  .
% ( 0 , i^y , i^z )
%   \sigma  ( 0,  E^y , E^z  )
% =  ( 0, - H^z _x ,  H^y _x )
% \end{equation}

The system of equations  (\ref{3D_system})    gets reduced to
\begin{subequations}    \label{Euler_1D_Eqs}
\begin{gather}      \label{Euler_1D_rho}
\rho_{t}+u\rho_{x}+\rho u_{x}=0,
\\
  \label{Euler_1D_u}
\rho(u_{t}+u u_{x}) +p_{x} +H ^y  H  ^y  _{x} + H ^z   H ^z _{x} =0,
% \qquad \mbox{maybe with $E$}
\\
  \label{Euler_1D_v}
\rho(v_{t}+u v_{x})=H ^0  H ^y _{x},
\\
  \label{Euler_1D_w}
\rho(w_{t}+u w_{x})=H ^0 H ^z _{x},
\\
  \label{Euler_1D_p}
p_{t}+up_{x}+\gamma p u_{x}= (\gamma -1) \sigma ( ( E^y )^2 + ( E^z )^2 )  ,
\\
  \label{Euler_1D_Hy}
H ^y _{t}+u H ^y _{x}+H ^y  u_{x}=  H ^0  v_{x}  + E^z _x ,
\\
  \label{Euler_1D_Hz}
H ^z _{t}+u H ^z _{x}+H^z u_{x}= H ^0  w_{x}   - E^y _x ,
\\
%  i^x = E^x = 0 ,
% \qquad
% t^y =
\sigma   E^y    =   - H^z _x  ,
\qquad
% i^z =
\sigma E^z  =     H^y _x  .         \label{relation_Euler}
\end{gather}
\end{subequations}
Here we use the equation for the pressure     (\ref{equation_p_Euler})
instead of the equation for the internal energy   (\ref{3D_system_energy})
and eliminate the electric  current density  $ \textbf{i}  $.
As we noted earlier,  the components  $E^y $  and  $E^z $  can be eliminated.
From now
the equations  (\ref{plane_simplification})
will be discarded because of  (\ref{condition_Hx}).

\subsection{Plane one-dimensional flows in Lagrangian coordinates}

The introduction of the mass Lagrangian coordinates
$ ( s, \eta, \zeta ) $
is described in Appendix \ref{Appendix_Lagrangian_variables}.
In the Lagrangian coordinates  the Eulerian spatial coordinates are given by
\begin{equation}     \label{introduction_xyz}
x= \varphi ( t, s  )  ,
\qquad
y = \eta + \psi ( t, s  ) ,
\qquad
z =  \zeta  + \chi ( t, s  )   ,
\end{equation}
where
the functions   $ \varphi $,  $  \psi  $ and $  \chi  $  satisfy the equations
\begin{subequations}   \label{introductions}
\begin{gather}
\label{introduction_s}
\varphi_{t}(t, s)  = u(t, \varphi(t, s) )  ,
\qquad
\varphi_{s}(t, s)  =  \frac{1}{\rho(t, \varphi(t, s) )} ,
\\
\label{introduction_psi}
\psi _{t} (t, s)  = v (t, \varphi(t, s))  ,
\\
\label{introduction_chi}
\chi  _{t} (t, s)  = w (t, \varphi(t, s))  .
\end{gather}
\end{subequations}

% \begin{remark}
% In the Eulerian coordinates $ (t,x) $
% the mass Lagrangian coordinate can be seen as a potential for the
% conservation of mass   (rewritten equation  (\ref{Euler_1D_rho}))
% \begin{equation*}
% \rho _t  +  ( \rho  u )_x = 0
% \end{equation*}
% introduced by the system
% \begin{equation*}
% s _x = \rho  ,
% \qquad
% s _t  =  - \rho u  ,
% \end{equation*}
% or, equivalently, by the one-form
% \begin{equation*}
% d s = \rho \ d x - \rho u \ d t .
% \end{equation*}
% \end{remark}

In the mass Lagrangian coordinates $ (t,s)$
the equations  (\ref{Euler_1D_Eqs}),
describing the plane one-dimensional   MHD  flows,
take the form
\begin{subequations}
\label{Lagrangian_Eqs}
\begin{gather}
\rho_t = -\rho^2 u_{{s}},
  \label{Lagrangian_Eq_1}
\\
u_t = - p_{{s}} -        H^y    H^y   _s     -    H^z    H^z    _s ,
\qquad
x_t = u ,
%  \qquad \mbox{maybe with $E$}
  \label{Lagrangian_Eq_2}
\\
v_t =     {  H^0   }     H^y   _s ,
\qquad
y_t = v ,
  \label{Lagrangian_Eq_3}
\\
w_t =     {  H^0    }     H^z  _s ,
\qquad
z_t = w ,
  \label{Lagrangian_Eq_4}
\\
p_t = -\gamma  \rho  p  u_{{s}}
+ (\gamma -1) \sigma ( ( E^y )^2 + ( E^z )^2 )  ,
   \label{Lagrangian_Eq_7}
\\
 H^y   _t  =   \rho ( H^0   v _s    -   H^y  u _s   +  E^z_s  )  ,
  \label{Lagrangian_Eq_5}
\\
 H^z   _t  =\rho ( H^0   w _s     -   H^z  u _s  - E^y _s )  ,
  \label{Lagrangian_Eq_6}
\\
\sigma E^y = -\rho H^z_{{s}},
\qquad
\sigma E^z = \rho H^y_{{s}} .    \label{relation_Lagrange}
\end{gather}
\end{subequations}
% The relations    (\ref{relation_Euler}) get rewritten as
%  \begin{equation}      \label{relation_Lagrange}
% \sigma E^y = -\rho H^z_{{s}},
% \qquad
% \sigma E^z = \rho H^y_{{s}}.
% \end{equation}
Note that
the time  differentiation in the system  (\ref{Lagrangian_Eqs})   is the Lagrangian one,
i.e. it is taken along the   trajectories.
For this reason we add the components of the equation $   \textbf{x} _t = \textbf{u} $.

We remark that in the mass Lagrangian coordinates $ (t,s) $
the Eulerian spatial coordinate  $x$ is nonlocal.
It is given by the system
\begin{equation}   \label{nonlocal_x}
x_ t = u ,
\qquad
x_s = {  1 \over  \rho  } .
\end{equation}
We also have
\begin{subequations}     \label{nonlocals_y_z}
\begin{gather}
\label{nonlocal_y}
y_ t = v ,
\\
\label{nonlocal_z}
z_ t = w .
\end{gather}
\end{subequations}

\begin{remark}    \label{remark_magnetic_1}
Using equation   (\ref{Lagrangian_Eq_1}),
it is possible to rewrite equations (\ref{Lagrangian_Eq_5}) and
(\ref{Lagrangian_Eq_6})  as the conservation laws
\begin{equation}
\label{correction_ Hy}
\left(   { H^y  \over \rho } \right) _ t = ( H^0   v   +  E^z  ) _s ,
\end{equation}
\begin{equation}
\label{correction_ Hz}
\left(   { H^z  \over \rho } \right) _ t = ( H^0   w   - E^y  ) _s  .
\end{equation}
\end{remark}

\section{Lie group classification of  equations  (\ref{Lagrangian_Eqs}) with finite conductivity}

\label{Symmetries_finite}

In this section  we perform  a group classification
of  the  system    (\ref{Lagrangian_Eqs})
with  an arbitrary  function   $\sigma ( \rho , p) $ and an arbitrary constant $H^0$.
Infinitesimal generators  of the Lie symmetry group are considered in the form
\begin{multline} \label{X0}
X = \xi^t \frac{\partial}{\partial{t}}
    + \xi^s \frac{\partial}{\partial{s}}
    + \eta^x \frac{\partial}{\partial{x}}
    + \eta^y \frac{\partial}{\partial{y}}
    + \eta^z \frac{\partial}{\partial{z}}
    + \eta^u \frac{\partial}{\partial{u}}
    + \eta^v \frac{\partial}{\partial{v}}
    + \eta^w \frac{\partial}{\partial{w}}
\\
+ \eta^\rho \frac{\partial}{\partial{\rho}}
+ \eta^p \frac{\partial}{\partial{p}}
    + \eta^{E^y} \frac{\partial}{\partial{E^y}}
    + \eta^{E^z} \frac{\partial}{\partial{E^z}}
    + \eta^{H^y} \frac{\partial}{\partial{H^y}}
    + \eta^{H^z} \frac{\partial}{\partial{H^z}} .
\end{multline}
% defined in the space of the coordinates $(t, s, \mathbf{x},  \mathbf{u}, \rho, p,  \mathbf{E}, \mathbf{H})$,
% is the infinitesimal generator of a one-parameter Lie-symmetry group for the system.
The coefficients~$\xi^t$, $\xi^s$, $\eta^x$, ..., $\eta^{H^z}$ of the generator are functions
of the independent and dependent variables
$t$, $s$, $\mathbf{x}$,  $\mathbf{u}$,  $\rho$,  $p$,  $ E ^y $, $ E^z $, $ H^y $  and $ H ^z $.

% {QUESTION: How the symmetries can be computed without  equations for   $ y_s $ and $ z_s $}

The infinitesimal criterion of invariance~\cite{bk:Ovsyannikov[1962], bk:Ibragimov1985,  bk:Olver}
 requires
\begin{equation} \label{classifDetsysF}
\left.  X  ( \mathcal{F} )  \right|_{ \mathcal{F} = 0 }= 0,
\end{equation}
where $\mathcal{F} = 0 $ denotes system~(\ref{Lagrangian_Eqs}).
% and $[\mathcal{F}]$ stands for  system~(\ref{Lagrangian_Eqs}) along with its differential consequences.
Here the generator $X$  is prolonged to  all derivatives
involved in the system   $\mathcal{F} =  0$ according to the standard prolongation formulas
\cite{bk:Ovsyannikov[1962], bk:Ibragimov1985, bk:Olver}.

Splitting equation~(\ref{classifDetsysF}) with respect to the first-order derivatives
and performing standard simplifications,
we derive  the classifying equations
\begin{equation} \label{LagrFlatClassSys}
\def\arraystretch{1.5}
{
\begin{array}{c}
2 (  a_6  - a_7   +  a_8) \rho \sigma_\rho
+  2 a_8 p \sigma_p
 =   (  a_6  -   2 a_7 ) \sigma  ,
\\
H^0 a_8 = 0,
\\
H^0 \eta^y_v = 0, \quad
H^0 \eta^y_w = 0, \quad
H^0 \eta^z_v = 0, \quad
H^0 \eta^z_w = 0 ,
\\
H^0 \eta^v_s = 0, \quad
H^0 \eta^v_y = 0, \quad
H^0 \eta^v_z = 0,
\\
H^0 (\eta^v_v   + a_6  - a_7 ) = 0, \quad
H^0 (\eta^v_w + a_5) = 0,
\\
H^0 \eta^w_s = 0, \quad
H^0 \eta^w_y = 0, \quad
H^0 \eta^w_z = 0,
\\
H^0 (\eta^w_v - a_5) = 0, \quad
H^0 (\eta^w_w    + a_6  - a_7 ) = 0,
\end{array}
} % boxed
\end{equation}
where the coefficients  of the generator are
\begin{equation} \label{LagrFlatClassSyms}
\def\arraystretch{1.75}
\begin{array}{c}
\xi^t = a_6 t + a_1,
\quad
\xi^s = (2 a_6  - a_7   + 2 a_8 )   s + a_2,
\\
\eta^x = a_4 t + a_7 x + a_3,
\quad
\eta^y =  f_3 t   + a_7 y    - a_5 z     + f_1,
\quad
\eta^z =  f_4 t  + a_5  y   + a_7 z    + f_2,
\\
\eta^u =   ( - a_6  +  a_7  ) u     + a_4,
\quad
\eta^v =   (  - a_6  + a_7   ) v    - a_5 w   +     f_3  ,
\quad
\eta^w =  a_5 v  + ( - a_6  +  a_7   ) w   + f_4 ,
\\
\eta^\rho = 2  (a_6 - a_7  + a_8 ) \rho,
\quad
\eta^p = 2 a_8 p,
\\
\eta^{E^y} =  (- a_6   + a_7 + a_8  ) E^y   - a_5 E^z  ,
\quad
\eta^{E^z} = a_5  E^y + (- a_6   + a_7 + a_8 ) E^z,
\\
\eta^{H^y} =  a_8 H^y  - a_5 H^z ,
\quad
\eta^{H^z} = a_5 H^y + a_8 H^z .
\end{array}
\end{equation}
Here $ a_i $, $ i = 1, \ldots, 8 $ are constants and
\begin{equation} \label{arbf}
f_i = f_i(s, v, w, y - t v, z - t w), \qquad i=1, \ldots, 4,
\end{equation}
are arbitrary functions of their arguments.

% {Coefficients were checked 12.09.2021}

According to~(\ref{LagrFlatClassSyms}), a Lie algebra admitted by system~(\ref{Lagrangian_Eqs}) belongs to
the extended Lie algebra with the {basis determined by the generators}
%\begin{equation}
%X = \sum_{j=1}^{8} a_j X_j + \sum_{k=9}^{12} f_{k-8} X_k,
%\end{equation}
%where
\begin{multline} \label{LagrFlatClassOps}
% \def\arraystretch{1.75}
% \begin{array}{c}
Y_1 = {\ddt},
\quad
Y_2 =  {\dds},
\quad
Y_3 = {\ddx},
\quad
Y_4 = t  {\ddx} +  {\ddu},
\\
Y_5
=  z {\ddy} - y  {\ddz}
+ w   {\ddv} - v  {\ddw}
+ E^z {\ddEy} - E^y  {\ddEz}
+ H^z  {\ddHy} - H^y  {\ddHz},
% \qquad
% \mbox{and rotations of  (y,z) and (v,w)}
 \\
Y_6 = t   {\ddt} + 2 s   {\dds}
    - u  {\ddu} - v  {\ddv} - w  {\ddw}
    + 2\rho  {\ddrho}
    - E^y  {\ddEy} - E^z  {\ddEz},
\\
Y_7
= - s   {\dds}
+ x  {\ddx}  + y  {\ddy}  + z  {\ddz}
+ u  {\ddu} + v  {\ddv}  + w  {\ddw}
    - 2\rho  {\ddrho}
    + E^y  {\ddEy} + E^z  {\ddEz},
% \qquad
% \mbox{and scaling of  (y,z) and (v,w)}
\\
Y_8 = 2 s   {\dds}
    + 2\rho  {\ddrho}
    + 2 p   {\ddp}
    + E^y  {\ddEy} +  E^z   {\ddEz}
    + H^y  {\ddHy} +  H^z  {\ddHz},
\\
Y_{9} = f_1  {\ddy},
\quad
Y_{10} = f_2  {\ddz},
\quad
Y_{11} = f_3 \left( t    {\ddy} + {\ddv}  \right),
\quad
Y_{12} = f_4 \left( t  {\ddz} +  {\ddw} \right),
% \end{array}
\end{multline}
where arbitrary functions $f_1$, ..., $f_4$ have the form~(\ref{arbf}).

For the further  analysis  of the symmetries
it is necessary  to consider two cases of   (\ref{condition_Hx}),
namely  $ H^0 \neq  0  $  and $ H^0 = 0  $, separately.

%%%%%%%%%%%%%%%%%%%%%%%%%%%%%%%%
\subsection{Case $H^0 \neq 0$}
%%%%%%%%%%%%%%%%%%%%%%%%%%%%%%%%

In this case the group classification of equations is obtained
with respect to arbitrary  {elements }   % (for system~(\ref{Lagrangian_Eqs}) they are
$\sigma ( \rho , p ) $ and~$H^0$.

Equivalence transformations allow changing  the arbitrary elements while
preserving the structure of the  equations~\cite{bk:Ovsyannikov[1962]}.
The generators of the equivalence transformations for system~(\ref{Lagrangian_Eqs})
are  given in      (\ref{equivalence_2}), Appendix B.
These transformations  can be used to scale the function  $ \sigma $ and the constant $ H^0$.

\subsubsection{Arbitrary $\sigma ( \rho , p) $}

In the most general case of  $\sigma ( \rho , p) $ and $ H^0$
we find the {kernel} of the Lie algebras admitted by system~(\ref{Lagrangian_Eqs}).
It consists of the generators admitted by the system
for an arbitrary function  $\sigma ( \rho , p) $ and an arbitrary constant $H^0$.
In order to obtain the kernel, we
%considers~$\sigma$ and $H^0$ as arbitrary elements and
split (\ref{LagrFlatClassSys}) with respect to  $\sigma$, $\sigma_\rho$, $\sigma_p$  and  $H^0$.
From the resulting equations it immediately follows that
\begin{equation}
a_6 = a_7 = a_8 = 0 .
\end{equation}
We also obtain conditions on the functions  $\eta^y$, $\eta^z$, $\eta^v$ and $\eta^w$.
From these conditions it follows  that the functions  $f_i$
must be less general than given in (\ref{arbf}),
namely  they are   functions of $s$.
Finally,
we obtain  the following {kernel of the admitted Lie algebras}
\begin{multline} \label{kern01}
% \def\arraystretch{1.75}
% \begin{array}{c}
 {X_1} = {\ddt},
\quad
 {X_2} = {\dds},
\quad
 {X_3} = {\ddx},
\quad
 {X_4} = t {\ddx} + {\ddu},
\\
 {X_5} =  z {\ddy} - y {\ddz}
+ w  {\ddv} - v  {\ddw}
    + E^z {\ddEy} - E^y {\ddEz} + H^z {\ddHy} - H^y {\ddHz},
\\
 {X}_{6}= h_1(s) {\ddy},
\quad
 {X}_{7}= h_2(s) {\ddz},
\quad
 {X}_{8} = t {\ddy} + {\ddv},
\quad
 {X}_{9} = t {\ddz} + {\ddw},
% \end{array}
\end{multline}
where $ h_1 (s) $ and $ h_2 (s) $  are arbitrary functions.

\subsubsection{Special cases of $ \sigma  ( \rho , p) $}

Consider system~(\ref{LagrFlatClassSys}) with $H^0 \neq 0$.
Taking into account that functions $ f_i $ have the form   (\ref{arbf}),
%  \[
% f_i = f_i(s, v, w, y - t v, z - t w),
% \qquad
% i = 1, \dots, 4,
% \]
we get
\begin{equation}
\def\arraystretch{1.75}
\begin{array}{c}
a_8 = 0,
\\
\eta^y = b_{1} t +   a_7 y   - a_5 z  + h_1(s),
\quad
\eta^z = b_{2} t +  a_5 y + a_7 z + h_2(s),
\\
\eta^v = ( - a_6  + a_7 ) v - a_5 w + b_1,
\quad
\eta^w = a_5 v +  (  - a_6  + a_7 ) w  + b_{2},
\end{array}
\end{equation}
where
$  h_1(s) $  and  $  h_2(s) $   are arbitrary functions and
$b_1$ and $b_{2}$  are constants.

In the case $a_6 = a_7$ there are no extensions of the kernel  (\ref{kern01}).
Hence, the classifying equation can be written as
\begin{equation} \label{LagrFlatClassSysH0neq0}
\displaystyle
{\rho \sigma_\rho }  = \alpha  {\sigma},
\qquad
\alpha = \frac{a_6 - 2 a_7}{2(a_6 - a_7)}.
\end{equation}
% In order to find relations between~$a_6$ and~$a_7$ through~$\alpha$,
The latter equation can be rewritten in the form
\begin{equation}
(2\alpha - 1)a_6  =  2 (\alpha - 1)a_7 .
\end{equation}
%There are only two possibilities:
%\begin{equation}
%\def\arraystretch{1.75}
%\begin{array}{lrl}
%1)&\;
%a_6 = \beta a_7 \neq 0, \quad \beta \neq 1,
%\quad
%\text{and}
%\quad
%&\sigma = \rho^{\frac{1}{2}\frac{\beta - 2}{\beta - 1}} F(p).
%\\
%2)\;
%&a_6 = 1, \quad a_7 = 0,
%\quad
%\text{and}
%\quad
%&\sigma = \sqrt{\rho} F(p).
%\end{array}
%\end{equation}
The resulting classification,  based on this  equation,  is given in Table~1.
The first column of the table gives the dimension $\mbox{dim}   \  L $
of the admitted  Lie algebra.
The extension of the kernel of the admitted Lie algebras (\ref{kern01})  is given in
the second column of the table.
The third column gives the corresponding forms of the function  $\sigma$.
Here and in the next  table  $F$ denotes an arbitrary differentiable function of its argument.

\begin{table}[ht]
\def\arraystretch{1.75}
\centering
\begin{tabular}{|c|c|c|c|}
\hline
$\mbox{dim} \  L $   & Extension of kernel~(\ref{kern01}) & $\sigma(p, \rho)$ \\
%& $\tilde{\sigma}(T)$ \\
\hline
$6$ &
$  X_{10} =   2 (\alpha - 1)   Y_6 +     (2\alpha - 1)   Y_7 $
% &   $\sqrt{\rho} F(p)$
% & $T^{-\frac{1}{2}}$    for   $    F(p) = {p} ^{-1/2} $  \\
%  & $ \frac{2(\alpha - 1)}{2\alpha - 1}X_6 + X_7$
&   $ \rho^{\alpha} F(p)$  \\
% & $T^{-\alpha}$   for   $    F(p) = {p} ^{- \alpha}  $ \\
\hline
\end{tabular}
\label{tab:class_lagr_flat_H_neq_0}
\caption{Lie group extensions for $H^0 \neq 0$.}
\end{table}

\begin{remark}
Using equation
% Eliminating $\varepsilon$ and $c_V$ form~(\ref{EpsCvT}), (\ref{gammaRCv}) and
(\ref{ideal}),
% one gets that~$T =\frac{1}{R}\frac{p}{\rho}$.
it is possible to express some particular  cases $\sigma(p, \rho)$
in the form~$\tilde{\sigma}(T)$.
The presentation of the electric conductivity
as a functions of the temperature
is of interest for  physical applications.
For  the particular case  $  F(p) =  C {p} ^{- \alpha}$
we obtain   $  \tilde{\sigma}(T)   = \tilde {C}T  ^{-\alpha}$.

% The resulting forms of the function~$\tilde{\sigma}(T)$ are given (up to constant multipliers)
% in the fourth column of Table~1.
\end{remark}

%%%%%%%%%%%%%%%%%%%%%%%%%%%%%%%%
\subsection{Case $H^0 = 0$}
%%%%%%%%%%%%%%%%%%%%%%%%%%%%%%%%

\label{Symmetries_H0_0}

For $H^0 = 0$ the system of equations   (\ref{Lagrangian_Eqs})
becomes
\begin{subequations}
\label{Lagrangian_Eqs_H0_part_1}
\begin{gather}
\rho_t = -\rho^2 u_{{s}},
  \label{Lagrangian_Eq_1_H0}
\\
u_t = - p_{{s}} -        H^y    H^y   _s     -    H^z    H^z    _s ,
\qquad
x_t = u ,
% \qquad \mbox{maybe with $E$}
  \label{Lagrangian_Eq_2_H0}
\\
 p_t = -\gamma \rho  p  u_{{s}}
+ (\gamma -1) \sigma ( ( E^y )^2 + ( E^z )^2 )    ,
   \label{Lagrangian_Eq_7_H0}
\\
 H^y   _t  =   \rho (   -   H^y  u _s   +  E^z_s )  ,
  \label{Lagrangian_Eq_5_H0}
\\
 H^z   _t  =\rho (    -   H^z  u _s   - E^y _s  )  ,
  \label{Lagrangian_Eq_6_H0}
\\
\sigma E^y = -\rho H^z_{{s}},
\qquad
\sigma E^z = \rho H^y_{{s}} .
\end{gather}
\end{subequations}
% with  relations   (\ref{relation_Lagrange})
The remaining  four equations
\begin{subequations}
\label{Lagrangian_Eqs_H0_part_2}
\begin{gather}
v_t =    0  ,
\qquad
y_t = v ,
 \label{Lagrangian_Eq_3_H0}
\\
w_t =    0  ,
\qquad
z_t = w .
  \label{Lagrangian_Eq_4_H0}
\end{gather}
\end{subequations}
can be  analyzed independently.
In the rest of this section we discuss  the reduced system (\ref{Lagrangian_Eqs_H0_part_1}).
It implies that variables    $y$, $z$,  $v$ and $w$  are excluded from the consideration.

\begin{remark}
The equations   (\ref{Lagrangian_Eqs_H0_part_2})
can be easily solved as
\begin{equation*}
v = v _0 (s)  ,
\qquad
y = v _0 (s) t + y_0 (s, \eta, \zeta) ,
\end{equation*}
\begin{equation*}
w = w_0 (s)  ,
\qquad
z = w_0 (s) t + z_0 (s ,  \eta, \zeta) ,
\end{equation*}
with functions $v _0 (s) $,  $y _0 (s,  \eta, \zeta)  $,  $w _0 (s) $ and $z _0 (s ,  \eta, \zeta)  $
defined by the initial conditions.
\end{remark}

\begin{remark}    \label{remark_magnetic_2}
Similarly to Remark   \ref{remark_magnetic_1}
  equations (\ref{Lagrangian_Eq_5_H0}) and
(\ref{Lagrangian_Eq_6_H0}) can be rewritten  as the conservation laws
\begin{equation}
\label{correction_ Hy_H0}
\left(   { H^y  \over \rho } \right) _ t =   E^z  _s ,
\end{equation}
\begin{equation}
\label{correction_ Hz_H0}
\left(   { H^z  \over \rho } \right) _ t =    - E^y  _s  .
\end{equation}
\end{remark}

The {equivalence transformations}
of the reduced system~(\ref{Lagrangian_Eqs_H0_part_1})
are given in     (\ref{EqTrMLagrFlatH0=0}), Appendix B.
They can be used to scale $ \sigma  $.

The  Lie algebra admitted by system~(\ref{Lagrangian_Eqs_H0_part_1})
belongs to the extended algebra
whose {basis is defined}  by the generators~$ Y _1 $, $ Y _2$, ..., $ Y _8$ from~(\ref{LagrFlatClassOps}),
namely
\begin{multline} \label{LagrFlatClassOps8}
% \def\arraystretch{1.75}
% \begin{array}{c}
Y_1 = {\ddt},
\quad
Y_2 = {\dds},
\quad
Y_3 = {\ddx},
\quad
Y_4 = t {\ddx} + {\ddu},
\\
Y_5 = E^z {\ddEy} - E^y {\ddEz}
+ H^z {\ddHy} - H^y {\ddHz},
\\
Y_6 = t {\ddt} + 2 s {\dds}
    - u {\ddu}
    + 2\rho {\ddrho}
    - E^y {\ddEy} - E^z {\ddEz},
\\
Y_7 = -s  {\dds} + x {\ddx} + u {\ddu}
    - 2\rho {\ddrho}
    + E^y {\ddEy} + E^z {\ddEz},
\\
Y_8 = 2 s  {\dds}    + 2\rho {\ddrho} + 2 p {\ddp}
     + E^y {\ddEy} +  E^z {\ddEz}
    + H^y {\ddHy} +  H^z {\ddHz}.
% \end{array}
\end{multline}
Note that
these generators are truncated  generators (\ref{LagrFlatClassOps}):
the variables  $ y $,  $z$,  $v$ and  $ w $ are omitted.
In the subsequent discussion the corresponding group classification is denoted by~$\Theta_1$.

\subsubsection{Arbitrary  $ \sigma ( \rho , p)$}

The    kernel of the admitted Lie algebras   consists of the following five generators
\begin{multline} \label{kern01a}
% \def\arraystretch{1.75}
% \begin{array}{c}
 {X_1} = {\ddt},
\quad
 {X_2} = {\dds},
\quad
 {X_3} = {\ddx},
\quad
 {X_4} = t {\ddx} + {\ddu},
\\
 {X_5} = E^z {\ddEy} - E^y {\ddEz}
+ H^z {\ddHy} - H^y {\ddHz} ,
% \end{array}
\end{multline}
which are  admitted for all $  \sigma ( \rho , p) $.

\subsubsection{Special cases of $ \sigma ( \rho , p)$}

For $ H ^ 0 = 0$ the classifying system (\ref{LagrFlatClassSys}) reduces to
one equation
\begin{equation} \label{ClassifEqFlatHeq0}
{
2 (  a_6  - a_7   +  a_8) \rho \sigma_\rho
+  2 a_8 p \sigma_p
 =   (a_ 6 - 2 a_7  ) \sigma  .
}
% \boxed{
%    a_6 (2 \rho \sigma_\rho - \sigma)
%        - 2 a_7 (\rho \sigma_\rho - \sigma)
%        + 2 a_8 (\rho \sigma_\rho + p \sigma_p)
%         = 0.
% }
\end{equation}
% ~(\ref{Lagrangian_EqsH0=0}) (29)
% ~(\ref{LagrFlatClassOps8}) (31)
% ~(\ref{kern01a}) (32)
% inner automorphisms
% equivalence group of transformations
% group classification
% extended Lie algebra
% kernel of admitted Lie algebras
In this case  the  classification problem is more cumbersome  than that  for $ H^0 \neq 0 $.
To overcome these difficulties we use an approach,
based on the group properties of system~(\ref{Lagrangian_Eqs_H0_part_1}).
This approach is applicable when  transformations of the symmetry generators
under action of the equivalence transformations
coincide  with
transformations of the symmetry generators
under action of the inner automorphisms.
This method was applied in \cite{bk:Suriyawichitseranee2015}
(see also \cite{bk:Popovych_GC_2006}).
It is based on the following idea.
The group classification  of system~(\ref{Lagrangian_Eqs_H0_part_1})
supposes separation of the generators
into  dissimilar  classes
 with respect to the equivalence transformation group.
This separation leads to the  group classification $\Theta_1$
for  the extended Lie algebra (\ref{LagrFlatClassOps8}).
Another classification of the Lie algebra~(\ref{LagrFlatClassOps8}) can be obtained
for the inner automorphisms. We denote this classification by~$\Theta_2$.

If the action of the the equivalence transformation group
and the action of the inner automorphisms coincide,
then any {sub}algebra of  $ \Theta_1$ is a {sub}algebra of~$\Theta_2$.
Hence, for the group classification one can use {sub}algebras of $\Theta_2$.
This simplifies the group classification problem.
Instead of the  equivalence transformations,  which are  generally nonlinear,
one can consider the inner automorphisms,
which are presented by linear   transformations  of the generators.

Notice that any {sub}algebra of~$\Theta_1$ includes the kernel of admitted Lie algebras~(\ref{kern01a}).
Therefore, for obtaining~$\Theta_1$ from~$\Theta_2$ one can consider only {sub}algebras of~$\Theta_2$
containing the kernel~(\ref{kern01a}).
This allows avoiding analysis of all {sub}algebras of~$\Theta_2$
that  further   facilitates the group classification.

Thus, for the group classification
we can use the following algorithm.

\begin{enumerate}

\item

An optimal system of {sub}algebras $\Theta_2$ is constructed
(only {sub}algebras which contain  the kernel of the admitted Lie algebras.are needed).
This  optimal system of {sub}algebras  defines classes of
the non-equivalent   {sub}algebras
with respect to generator  transformations corresponding to the inner automorphisms.
As noted before, the inner automorphisms act similarly to the equivalence transformations.
Thus, it is possible to use this optimal system for the group classification.
From the optimal system of {sub}algebras $\Theta_2$ one chooses the {sub}algebras
which include the kernel of the admitted Lie algebras.
It significantly reduces the number of {sub}algebras to be considered.

\item

For each {sub}algebra of the optimal system   $\Theta_2$,
which contains the kernel of the admitted Lie algebras,
the coefficients of the basis elements are substituted into the classifying equation.
Here it is sufficient to consider the extension of the kernel.
Solving the    system of the equations obtained
for the function  $\sigma (\rho , p) $,
one obtains non-equivalent  cases $\sigma (\rho , p) $ for the group classification.
\end{enumerate}

The algorithm is applied  to the generators (\ref{LagrFlatClassOps8})
with the kernel of the admitted Lie algebras (\ref{kern01a})
 in Appendix   C.
The equivalence transformations are defined by  generators
given in     (\ref{EqTrMLagrFlatH0=0}), Appendix B.
Here we present only the results.
The kernel (\ref{kern01a})
can be extended by operators from the set   $ \{ Y_6 , Y_7, Y_8 \} $.
The possible extensions  are
one-dimensional {sub}algebras
\begin{equation} \label{extension_1}
\{ Y_7 \},
\qquad
\{ Y_6 + \alpha Y_7 \},
\qquad
\{ Y_8 + \alpha Y_6 + \beta Y_7 \} ,
\end{equation}
two-dimensional {sub}algebras
\begin{equation} \label{extension_2}
\{ Y_6, Y_7 \},
\qquad
\{ Y_8 + \alpha Y_6 ,  Y_7 \},
\qquad
\{ Y_8 + \alpha Y_7 , Y_6 + \beta Y_7 \}
\end{equation}
and
the three-dimensional {sub}algebra
\begin{equation} \label{extension_3}
\{ Y_6, Y_7, Y_8 \} .
\end{equation}
It remains to find out the corresponding functions  $ \sigma  ( \rho , p) {\not \equiv} 0  $.
Extensions  which lead to $ \sigma  ( \rho , p) {\equiv} 0  $
are to be discarded.

To solve the classifying equations (\ref{ClassifEqFlatHeq0})  for the
possible kernel extensions,
i.e. for {sub}algebras  given  in     (\ref{extension_1}),   (\ref{extension_2}) and   (\ref{extension_3}),
we consider the generators from the extensions.
For each basis generator of {sub}algebras,
the corresponding coefficients are substituted into equation  (\ref{ClassifEqFlatHeq0}).
It provides  the equations  which   the function $\sigma ( \rho , p )  $
must satisfy in order to admit the considered generator.

As an  illustrating {example} we consider the {sub}algebra $\{Y_8 + \alpha Y_7, Y_6 + \beta Y_7\}$.
The procedure described above leads to the system
\begin{equation}
\def\arraystretch{1.75}
\begin{array}{rl}
    a_6 = 0, \; a_7 = \alpha, \; a_8 = 1: &
        \qquad
        (1 - \alpha) \rho \sigma_\rho + p \sigma_p  = -  \alpha\sigma ,
    \\
    a_6 = 1, \; a_7 = \beta, \; a_8 = 0: &
        \qquad
        2(1 - \beta) \rho \sigma_\rho =  (1 - 2 \beta ) \sigma  .
\end{array}
\end{equation}
One can verify that $\beta = 1$ leads to the solution $\sigma \equiv  0$
that is excluded from the consideration.
Thus, the constraint~$\beta\neq 1$ is imposed on the {sub}algebra  generators.
The solution of the latter system is
\begin{equation}
\displaystyle
\sigma  ( \rho , p)  = C
\rho^{ \frac{2\beta - 1}{2(\beta - 1)}}
p^{ \frac{\alpha - 2 \beta + 1}{2(\beta - 1)}} ,
\end{equation}
where $C$ is constant.
By means of equivalence transformations (\ref{EqTrMLagrFlatH0=0}),
one can set
 $C=1$.

Similarly we consider the other  possible extensions.
The results of the calculations for all {sub}algebras
(\ref{extension_1}),   (\ref{extension_2}) and     (\ref{extension_3})
are presented in Table~2.
In the table $F$ is an arbitrary function
and $ C $  is an arbitrary constant,  which can be removed by scaling.
The cases corresponding to the solution $\sigma \equiv  0$ are excluded.

\begin{remark}
A particular case of the classification presented in Table~2
was carried out  for $H^z = 0$, $E^y = 0$   in~\cite{bk:DorMHDpreprint1976}.
The results obtained there provide  particular cases of the present classification.
For example, consider the exponential case from \cite{bk:DorMHDpreprint1976}
\begin{equation}
\sigma  ( \rho , p  )  = e^{ a p + b \rho} ,
\qquad
a, b = \text{const},
\qquad
a \geqslant 0,
\end{equation}
which  splits into three {sub}cases.

\begin{enumerate}

  \item

  If the coefficients $a$ and $b$ are arbitrary,
  the system admits only generators
  from the kernel~(\ref{kern01a}).
  This case corresponds to arbitrary $\sigma=\sigma(p,\rho)$.

  \item

  The case $a = 0$ and $b \neq 0$ corresponds to the extension
  $\{Y_8 + \alpha Y_6 + \beta Y_7\}$ with $\alpha = -2$ and  $\beta = -1$
(see Table~2).
  Thus, for $ F(\rho)=  e ^ { \rho } $  there exists the additional symmetry
  \begin{equation}
  -  ( Y_8 - 2 Y_6 - Y_7 ) =
    2 t   {\ddt}  +  s   {\dds}
    + x {\ddx}
    - u {\ddu}
    - 2 p {\ddp}
    - 2 E^z {\ddEz}
    - H^y {\ddHy}  .
  \end{equation}

  \item

  The case $a \neq 0$ and $b = 0$ corresponds to the extension
  $\{Y_6 + \alpha Y_7\}$ with $\alpha = 1/2$.
  It provides  $F(p) = e^{p} $  and the additional symmetry
  \begin{equation}
  2 Y_6 + Y_7 =
  2 t   {\ddt}
+  3 s  {\dds}
    + x {\ddx}
    - u {\ddu}
    + 2 \rho {\ddrho}
    - E^z {\ddEz}  .
  \end{equation}
\end{enumerate}
\end{remark}

\begin{table}[ht]
\def\arraystretch{1.5}
\centering
\begin{tabular}{|c|l|c|}
\hline
$\mbox{dim}  \   L $
& Extension of kernel~(\ref{kern01a})
& $\sigma(p, \rho)$
% & $\tilde{\sigma}(T)$
\\
\hline
${6}$
&  $ X_6 = Y_7 $
&  $\rho F(p)$
% & $T^{-1}$     for   $    F(p) \sim p^{-1}  $
\\
& $ X_6 = Y_6 + \alpha Y_7 $
&  $\rho^{ \frac{2\alpha - 1}{2(\alpha - 1)}} F(p)$
% & $T^{\frac{1}{2} \frac{2\alpha - 1}{1 - \alpha}}$   for   $    F(p) \sim p^{ \frac{2\alpha - 1}{2(\alpha - 1)}}  $
\\
& $ X_6 = Y_8 + \alpha Y_6 + \beta Y_7 $
&  $p^{\frac{\alpha}{2} - \beta} F\left( \rho p^{\beta - \alpha - 1} \right)$
% &  $T^{\frac{2\beta- \alpha}{2(\alpha- \beta)}}$       for   $    F(p) =  $
\\
\hline
${7}$&$ X_6 = Y_8 + \alpha Y_6, \quad X_7 = Y_7 $
&   $ C \rho p^{-\frac{\alpha + 2}{2}}$
% & $T^{-1} $ for $ \alpha = 0$
\\
&$  X_6 = Y_8 + \alpha Y_7, \quad X_7 = Y_6 + \beta Y_7 $
&   $ C \rho^{ \frac{2\beta - 1}{2(\beta - 1)}} p^{ \frac{\alpha - 2 \beta + 1}{ 2(\beta - 1)}} $
% & $T^{\frac{1 - 2\beta}{2(\beta-1)}}$   for $ \alpha = 0$
 \\
\hline
\end{tabular}
\label{tab:class_lagr_flat_H_eq_0}
\caption{Lie group extensions for  $H^0=0$.}
\end{table}

% { table checked 15.09.2021 }

\begin{remark}
For particular cases of  $ F(p)  $ in the cases   $   \mbox{dim}  \   L  = 6 $
and particular cases  of  $ \alpha$ in the cases   $   \mbox{dim}  \   L  = 7 $,
the conductivity can be  presented  as
a function of the temperature.
We get the following options

\begin{itemize}

\item

Case  $ X_6 = Y_7 $

If   $    F(p) = C  p^{-1}  $,
then
$ \tilde{\sigma}(T)    = \tilde{C}  T^{-1} $.

\item

Case  $ X_6 = Y_6 + \alpha Y_7 $

For  $    F(p) = C  p^{\frac{2\alpha - 1}{2(1 -  \alpha )}}  $
we get
$   \tilde{\sigma}(T)    = \tilde{C}   T^{ \frac{2\alpha - 1}{2(1 - \alpha)}}$.

\item

Case $ X_6 = Y_8 + \alpha Y_6 + \beta Y_7 $

Function    $    F(q ) =  C   q^{\frac{\alpha - 2 \beta}{2(\alpha- \beta)}}  $
provides
$    \tilde{\sigma}(T)   = \tilde{C}   T^{\frac{2\beta- \alpha}{2(\alpha- \beta)}}$.

\item

Case   $ X_6 = Y_8 + \alpha Y_6 $,   $  X_7 = Y_7  $

For $ \alpha = 0$
we obtain
$    \tilde{\sigma}(T)   =  \tilde{C}    T^{-1} $.

\item

Case    $  X_6 = Y_8 + \alpha Y_7 $,   $  X_7 = Y_6 + \beta Y_7 $

For $ \alpha = 0$
we get
$    \tilde{\sigma}(T)   = \tilde{C}   T^{\frac{2\beta -1}{2(1 - \beta)}}  $.

\end{itemize}

\end{remark}

% {ALL THESE are  checked, 15.09.2021}

\section{Conservation laws for the case of finite conductivity}

\label{CL_finite}

Conservation laws possessed by a system of PDEs with two independent  variables $ (t,s) $
have the form
\begin{equation} \label{CLgenform}
% \left[
D_t ^L ( {T}^t ) + D_s(  {T}^s )   % \right]|_{(\ref{Lagrangian_Eqs})}
= 0 .
\end{equation}
They hold on the solutions of the system.
The conservation law densities  $ {T}^t$ and $ {T}^s$ for the system  (\ref{Lagrangian_Eqs})
are functions of the independent and dependent variables
$ ( t, s, \mathbf{x}, \mathbf{u}, \rho, p,  E^y ,  E^z,  H^y ,  H^z  )  $.

There are several approaches to find conservation laws.
If equations have a variational structure,
i.e. have the form  of Euler-Lagrange equations for some Lagrangian function,
one can apply the Noether theorem~\cite{bk:Noether1918}.
It allows to use variational and divergence symmetries
of the Lagrangian function to obtain conservation laws.
% In a number of cases, conservation laws
% can be found using the known symmetries of the original equation.
% For example, for systems with a Lagrange's function,
% one can obtain conservation laws using Noether's
% theorem~\cite{bk:Noether1918, bk:Ibragimov1985}.
% This method requires  a Lagrangian function for the system.
For equations without variational structure  it is possible to introduce additional variables
and consider an extended system, which is variational.
This approach was called the adjoint equation method~\cite{bk:Bluman1997,Ibragimov_adj1[2011]}.

Conservation laws can also be found by direct computation.
First, the densities    $ {T}^t$ and $ {T}^s$
are differentiated  and some derivatives are eliminated
with the help of the considered equations and (if necessary)  their differential consequences.
In the case of evolutionary equations it is standard to eliminate time derivatives.
The resulting equation is split for the remaining derivatives {and later for the dependent variables}
in order to find the densities of the conservation law.
We will employ the direct method to find conservation laws in this section.

% {OLD} Consider a general equation~(\ref{CLgenform}),
% assuming the functions $\mathcal{T}^t$ and $\mathcal{T}^s$ to be unknown.
% These functions can be found by direct computation.
% Equation~(\ref{CLgenform}) is split with respect to
% the higher order derivatives and arbitrary functions.
% Then, the resulting overdetermined system of equations is solved by standard methods.

\begin{remark}
% Conservation laws in the Lagrangian and Eulerian coordinates}
Conservation laws
in the Lagrangian coordinates  (\ref{CLgenform})
% \begin{equation}
% D_{t} ^L (  T^{t} )   +  D_{s}  (  T^{s} ) = 0
% \end{equation}
can be rewritten
in the Eulerian  coordinates as
\begin{equation}
D_{t} ^E  (  { ^{e} T^{t} }  )  +  D_{x}  (  { ^{e} T^{x} }  ) = 0.
\end{equation}

The total differentiation  operators   $D_{t} ^L  $ and  $ D_{s} $
in  Lagrangian coordinates  $ (t,s)$
 and  the total differentiation  operators   $D_{t} ^E  $ and  $ D_{x} $
in Eulerian coordinates  $ (t,x)$
are related by
\begin{equation}
D_{t} ^L
% = D_{t}  ^E+ \varphi_{t}D_{x}
= D_{t}  ^E+ u D_{x}   ,
\qquad
D_{s}
% =\varphi_{s} D_{x}
=\frac{1}{\rho}D_{x} .
\end{equation}

The densities of the conservation laws in the Eulerian coordinates
are related to the  densities of the conservation laws in the Lagrangian coordinates as
\begin{equation}     \label{transformation_rule}
^{e}T^{t}=\rho T^{t}  ,
\ \
{}^{e}T^{x}=\rho u T^{t}+T^{s}  .
\end{equation}
This relation follows from the  identity
\begin{equation*}
D_{t} ^L ( T^{t} ) +D_{s}  ( T^{s} )
=  {  1 \over \rho }   \left(    D_{t}  ^E (\rho T^{t})    +    D_{x}(     \rho u T^{t}+T^{s} )    \right).
\end{equation*}
Here it is necessary to take into account that the mass Lagrangian coordinate $s$
is  a  {non}local dependent variable in the Eulerian coordinates,
i.e. it  is defined by the equations
$$
s_x = \rho ,
\qquad
s_t = - \rho u  ,
$$
which follow from the equations (\ref{introduction_s}).

% Note:
% Using~(\ref{potential}), we obtain
% \begin{equation}
% u_{x}= \varphi_{s}^{-1}    \varphi_{ts}
% \end{equation}
\end{remark}

\subsection{Case $H^0 \neq 0$}

\label{CL_finate_H_not_0}

\subsubsection{Arbitrary conductivity   $ \sigma (  \rho , p) $ }

Direct computation provides  the following 10  conservation laws
for the system~(\ref{Lagrangian_Eqs})
in the general case of $H^0$  and  $ \sigma (  \rho , p) $:

\begin{itemize}

\item

mass
\begin{equation}    \label{mass_LH}
%D_t\left[\psi_x \frac{1}{\rho} + \psi_s\right] - D_s\left[\psi_x u + \psi_t\right] = 0,
%\qquad
%\psi = \psi(t,s,x),
D_t  ^L \left(\frac{1}{\rho}\right) - D_s(u) = 0 ;
\end{equation}

\item

three momenta
\begin{equation}
D_t  ^L \left(u\right)       \label{momentum_LHx}
+ D_s\left(p + \frac{(H^y)^2 + (H^z)^2}{2}\right)
= 0,
\end{equation}
\begin{equation}       \label{momentum_LHy}
D_t  ^L (v) - D_s(H^0 H^y) = 0,
\end{equation}
\begin{equation}         \label{momentum_LHz}
D_t  ^L (w) - D_s(H^0 H^z) = 0 ;
\end{equation}

\item

the center of mass conservation laws
\begin{equation}           \label{center_LHx}
D_t  ^L \left(t u - x\right)
+ D_s\left\{t\!\left(p + \frac{(H^y)^2 + (H^z)^2}{2}\right)\right\}
= 0,
\end{equation}
\begin{equation}        \label{center_LHy}
D_t  ^L (t v - y) - D_s(t H^0 H^y) = 0,
\end{equation}
\begin{equation}         \label{center_LHz}
D_t ^L (t w - z) - D_s(t H^0 H^z) = 0 ;
\end{equation}

% \item

% angular momentum   in $YZ$  plane
% \begin{equation}       \label{angular_momentum_LH}
% D_t ^L (
%  {    z  v    -   y  w     }
% )
% +
% D_s (
%     {   H^0   (   y   H ^z      -     z    H^y     )         }
% )  = 0  ;
% \end{equation}

\item

magnetic fluxes
\begin{equation}        \label{Hy_LH}
D_t ^L \left(\frac{H^y}{\rho}\right)  -  D_s(  E^z  +  H^0 v) = 0,
\end{equation}
\begin{equation}         \label{Hz_LH}
D_t ^L \left(\frac{H^z}{\rho}\right) + D_s(E^y - H^0 w) = 0  ;
\end{equation}

\item

energy
\begin{multline}       \label{energy_LH}
D_t ^L \left\{
     \frac{1}{2}(u^2 + v^2 + w^2)
    + \frac{1}{\gamma - 1} \frac{p}{\rho}
    + \frac{(H^y)^2 + (H^z)^2}{2\rho}
\right\}
\\
+ D_s\left\{
    u \left(p + \frac{(H^y)^2 + (H^z)^2}{2}\right)
    + E^y H^z - E^z H^y
    - H^0 (v H^y   +  w H^z  )
\right\} = 0.
\end{multline}

\end{itemize}

\begin{remark}
The latter conservation law can be rewritten as
\begin{multline}    \label{energy _CL}
D_t ^L \left\{
        \frac{1}{2}|\mathbf{u}|^2
       + \frac{1}{\gamma - 1} \frac{p}{\rho}
       + \frac{1}{2\rho} |\mathbf{H}|^2
\right\}
\\
+ D_s\left\{
    u \left(p + \frac{1}{2}|\mathbf{H}|^2\right)
   +  [\mathbf{E} \times \mathbf{H}]_1
   %  \cdot \mathbf{e}_x
    %E^y H^z - E^z H^y
   - H^0 (\mathbf{u} \cdot \mathbf{H})
\right\} = 0,
\end{multline}
where    $  [\mathbf{E} \times \mathbf{H}]_1    $ stands for the first component of the vector.
\end{remark}

It is worth mentioning that some of these conservation laws are already present
in the system (\ref{Lagrangian_Eqs}).
For example, the conservation of momenta.
At the same time other conservation lows, e.g. conservation of energy,
hold due to several equations of the system (\ref{Lagrangian_Eqs}).

\subsubsection{Special cases of conductivity   $ \sigma ( \rho , p ) $ }

There are no particular cases  $ \sigma ( \rho , p ) $
which leads to  additional conservation laws.

\subsection{Case $H^0 = 0$}

\label{CL_finite_H_0}

For $H^0 =  0$ the system of MHD equations (\ref{Lagrangian_Eqs}) gets split
into the reduced system  (\ref{Lagrangian_Eqs_H0_part_1})
and the four equations  (\ref{Lagrangian_Eqs_H0_part_2}).
These two subsystems will be considered  separately.

\subsubsection{Arbitrary conductivity   $ \sigma ( \rho , p )$}

The conservation laws of the reduced system  (\ref{Lagrangian_Eqs_H0_part_1})
 are  obtained  by direct computation.
They represent

\begin{itemize}

\item

conservation of mass
\begin{equation}     \label{CL_H0_mass}
%D_t\left[\psi_x \frac{1}{\rho} + \psi_s\right] - D_s\left[\psi_x u + \psi_t\right] = 0,
%\qquad
%\psi = \psi(t,s,x),
D_t ^L  \left(\frac{1}{\rho}\right) - D_s\left(u\right) = 0 ;
\end{equation}

\item

conservation of momentum
\begin{equation}   \label{CL_H0_momenta}
D_t  ^L \left(u\right)
+ D_s\left(p + \frac{(H^y)^2 + (H^z)^2}{2}\right)
= 0 ;
\end{equation}

\item

motion of the center of mass
\begin{equation}     \label{CL_H0_center}
D_t  ^L  \left(t u - x\right)
+ D_s  \left\{ t\!\left(p + \frac{(H^y)^2 + (H^z)^2}{2}\right)\right\}
= 0 ;
\end{equation}

\item

conservation of magnetic fluxes
\begin{equation}    \label{CL_H0_flux_y}
D_t ^L \left(\frac{H^y}{\rho}\right) - D_s(E^z) = 0,
\end{equation}
\begin{equation}   \label{CL_H0_flux_z}
D_t ^L \left(\frac{H^z}{\rho}\right) + D_s(E^y) = 0 ;
\end{equation}

\item

conservation of energy
\begin{multline}        \label{CL_H0_energy}
D_t ^L \left\{
     \frac{1}{2} u^2
    + \frac{1}{\gamma - 1} \frac{p}{\rho}
    + \frac{(H^y)^2 + (H^z)^2}{2\rho}
\right\}
\\
+ D_s\left\{
    u \left(p + \frac{(H^y)^2 + (H^z)^2}{2}\right)
    + E^y H^z - E^z H^y
   \right\} = 0,
\end{multline}
which can be rewritten as  (\ref{energy _CL})  with $H^0 =0$.
\end{itemize}

The system (\ref{Lagrangian_Eqs_H0_part_2})   has conservation laws of the form
\begin{equation}    \label{many_CL}
D_t ^L  \left(
T ^t  ( v , w, y  - t v,  z - t w  )
 \right)  = 0 ,
\end{equation}
where $ T^t  $ is an arbitrary function.
These conservation laws include
conservation  of momenta  (\ref{momentum_LHy}) and  (\ref{momentum_LHz})
\begin{equation}
D_t ^L  ( v )  =0   ,
\end{equation}
\begin{equation}
D_t ^L (  w )  =0
\end{equation}
and conservation  laws for the motion of the center of mass  (\ref{center_LHy}) and  (\ref{center_LHz})
\begin{equation}
D_t ^L  ( y - t  v ) =0   ,
\end{equation}
\begin{equation}    \label{angular_momentum_LH_0}
D_t ^L ( z - t v )  =0
\end{equation}
as particular cases.
In contrast to the case $ H^0 \neq 0 $
there is  conservation of the angular momentum   %   (\ref{angular_momentum_LH})
\begin{equation}
D_t ^L ( z v  - y w )  = 0     .
\end{equation}
All these conservation laws have trivial coordinate density  $ T ^s \equiv 0 $
in the case $H^0 =  0 $.

The conservation laws
(\ref{CL_H0_mass})--(\ref{CL_H0_energy})
correspond to the  conservation laws
(\ref{mass_LH}), (\ref{momentum_LHx}), (\ref{center_LHx}),
(\ref{Hy_LH}), (\ref{Hz_LH}) and  (\ref{energy_LH})
of the case  $ H^0 \neq 0$.
Combing  these conservation laws
with the conservation laws   (\ref{many_CL}),
we conclude that there are more conservation laws for the case $ H^0 = 0$.
For example, conservation of the  angular momentum  (\ref{angular_momentum_LH_0})
has no analog for  $ H^0 \neq 0$.
In addition to this,
for $ H^0 = 0$   some of the conservation laws get simplified.

\subsubsection{Special case of conductivity   $ \sigma ( \rho , p )  = \rho $}

There are several cases with symmetry extensions,
which were described in point \ref{Symmetries_H0_0}.
For the conservation laws there is only one extension.

If    $\sigma (\rho, p)   = \rho$, there are two additional  conservation laws
\begin{equation}   \label{CL_extension_1}
D_t ^L \left(\frac{s H^z}{\rho}\right)  -  D_s (   s E^y   + H^z  ) = 0,
\end{equation}
\begin{equation}     \label{CL_extension_2}
D_t  ^L \left(\frac{s H^y}{\rho}\right)  +   D_s (  s E^z  -  H^y) = 0.
\end{equation}
Note that the condition $  \sigma    = \rho  $  has no analog
for the infinite  conductivity $  \sigma  =  \infty$.
Thus, conservation laws    (\ref{CL_extension_1}) and   (\ref{CL_extension_2})
hold only for the finite conductivity.

% {QUESTION: Shall we use $  \sigma \rightarrow \infty$ or  $  \sigma =  \infty$}

%%%%%%%%%%%%%%%%%%%%%%%%%%%%%%%%%%%%%%%%%%%%%%%%%%%%%%%%%%%%%%%%%%

\section{Symmetries for the case of infinite conductivity}

\label{Symmetries_infinite}

% In the case of  infinite conductivity  we consider the  system~(\ref{Lagrangian_Eqs_infinite})
% for $H^0 \neq 0$
% and the reduced system  (\ref{Lagrangian_Eqs_infinite_H0})
% for $H^0 = 0$.

MHD equations for the case of infinite conductivity ($ \sigma =  \infty$)
can be obtained  from the system (\ref{3D_system})
in the limiting case $ \sigma   \rightarrow   \infty$.
For the plane one-dimensional flows
in the mass Lagrangian coordinates  we derive from   (\ref{Lagrangian_Eqs}):
\begin{subequations}
\label{Lagrangian_Eqs_infinite}
\begin{gather}
\rho_t = -\rho^2 u_{{s}},
  \label{Lagrangian_Eq_1_infinite}
\\
u_t = - p_{{s}} -        H^y    H^y   _s     -    H^z    H^z    _s ,
\qquad
x_t = u ,
  \label{Lagrangian_Eq_2_infinite}
\\
v_t =     {  H^0   }     H^y   _s ,
\qquad
y_t = v ,
  \label{Lagrangian_Eq_3_infinite}
\\
w_t =     {  H^0    }     H^z  _s ,
\qquad
z_t = w ,
  \label{Lagrangian_Eq_4_infinite}
\\
p_t = -\gamma  \rho p u_{{s}}   ,
   \label{Lagrangian_Eq_7_infinite}
\\
 H^y   _t  =   \rho ( H^0   v _s   -   H^y  u _s )  ,
  \label{Lagrangian_Eq_5_infinite}
\\
 H^z   _t  =\rho ( H^0   w _s     -   H^z  u _s )  .
  \label{Lagrangian_Eq_6_infinite}
\end{gather}
\end{subequations}

\begin{remark}  Similarly to  Remark \ref{remark_magnetic_1}
it is possible to rewrite equations (\ref{Lagrangian_Eq_5_infinite}) and
(\ref{Lagrangian_Eq_6_infinite})  as the conservation laws
\begin{equation}
\label{correction_ Hy_infinite}
\left(   { H^y  \over \rho } \right) _ t = ( H^0   v     ) _s ,
\end{equation}
\begin{equation}
\label{correction_ Hz_infinite}
\left(   { H^z  \over \rho } \right) _ t = ( H^0   w   ) _s
\end{equation}
with the help of equation (\ref{Lagrangian_Eq_1_infinite}).

In the mass Lagrangian coordinates
we can rewrite the equations  (\ref{correction_ Hy_infinite}) and (\ref{correction_ Hz_infinite})
using functions  $ \varphi $, $\psi$ and $\chi$,
which describe the Eulerian coordinates  (\ref{introduction_xyz}),
as
\begin{equation*}
(\varphi_{s}H^{y})_{t}=(H^{0}\psi_{t})_{s},
\qquad
(\varphi_{s}H^{z})_{t}=(H^{0}\chi_{t})_{s}.
\end{equation*}
Integrating these equations   with respect to $t$, we get
\begin{equation*}
\varphi_{s}H^{y}=H^{0}\psi_{s}+g_{1}^{\prime} (s),
\qquad
\varphi_{s}H^{z}=H^{0}\chi_{s}+g_{2}^{\prime} (s),
\end{equation*}
where $g_{1}(s)$ and $g_{2}(s)$ are arbitrary functions of integration.

If $H^{0}\neq0$, then one can choose
the functions
% \footnote{\textbf{Not for the text: }the mass Lagrangian coordinates do not satisfy conditions (\ref{eq:sep14.2}). }
$\psi(t, s)$ and $\chi(t, s)$ such that $g_{1}(s)=0$ and $g_{2}(s)=0$.
% \[
% y_{s}H^{y}-z_{s}H^{z}=\psi_{s}H^{y}-\chi_{s}H^{z}
% =\varphi_{s}\left(\frac{H^{y}}{H^{0}}H^{y}-\frac{H^{z}}% {H^{0}}H^{z}\right)=
% \]
In this case
\begin{equation}  \label{for_angular_momentum}
H^{y}=H^{0} { \psi_{s} \over  \varphi_{s}  } ,
\qquad
H^{z}=H^{0} { \chi_{s} \over  \varphi_{s}  }
\end{equation}
 that leads to
 \begin{equation}     \label{yszs}
y_s =  { H^y  \over H^0 \rho } ,
 \qquad
 z_s =  { H^z  \over H^0 \rho }   .
 \end{equation}

% and
% \begin{equation*}
% y_{s}H^{z}-z_{s}H^{y}=\psi_{s}H^{z}-\chi_{s}H^{y}
% =\varphi_{s}\left(\frac{H^{y}}{H^{0}}H^{z}-\frac{H^{y}}  {H^{0}}H^{z}\right)=0
% \end{equation*}
\end{remark}

It is easy to see that equations   (\ref{Lagrangian_Eq_1_infinite})  and  (\ref{Lagrangian_Eq_7_infinite})    provide
\begin{equation*}
\left( {  p \over \rho ^{\gamma}  } \right)   _t   = 0 .
\end{equation*}
Therefore,  the entropy function  $ S $,     defined in (\ref{function_S}),
% \begin{equation*}
% S  =  {  p \over \rho ^{\gamma}  }
% \end{equation*}
satisfies the equation
\begin{equation}     \label{correction_ S}
S_t = 0
\end{equation}
that represents  the conservation of the  entropy along  trajectories.

% It will be useful to take    equation  (\ref{correction_ S}),
% which can be easily integrated,
% instead of  (\ref{Lagrangian_Eq_7_infinite}).

\subsection{Case $H^0 \neq 0$}

The equivalence transformations for system  (\ref{Lagrangian_Eqs_infinite})
are  given in      (\ref{infinite_equivalence_C_generators}),  Appendix B.
These transformations  can be used to scale  the constant $ H^0$.

The system~(\ref{Lagrangian_Eqs_infinite})   admits the symmetries
\begin{multline}
% \def\arraystretch{1.75}
% \begin{array}{c}
\displaystyle
X_1 = {\ddt},
\quad
X_2 = {\dds},
\quad
X_3 = {\ddx},
\quad
\displaystyle
X_4 = t  {\ddx} + {\ddu},
\\
\displaystyle
X_5 =     z {\ddy} - y {\ddz}
  + w {\ddv} - v {\ddw}
   + H^z {\ddHy} - H^y {\ddHz},
\\
% \displaystyle
% X_6 = t  {\ddt} +  s {\dds}
%     + x {\ddx} + y {\ddy} + z {\ddz},
% \\
\displaystyle
X_6 = t  {\ddt} + 2 s {\dds}
    - u {\ddu} - v {\ddv} - w {\ddw}
    + 2\rho {\ddrho},
\\
X_7 =
- s {\dds} +  x {\ddx} +  y {\ddy} +  z {\ddz}
    +  v {\ddv} +   u {\ddu} + w {\ddw}
    -  2\rho {\ddrho} ,
\\
X_{8} = q_1  \left(s, { p  \over \rho^{\gamma} }  \right)   {\ddy},
\quad
X_{9} = q_2 \left(s, { p  \over \rho^{\gamma} }  \right)    {\ddz},
\quad
X_{10} = t {\ddy} + {\ddv},
\quad
X_{11} = t {\ddz} + {\ddw}.
% \end{array}
\end{multline}
Note that $ { p /  \rho ^{\gamma} }$  is a function of the entropy (see    (\ref{function_S})).

\subsection{Case $H^0 =  0$}

For the  infinite conductivity  and  $ H^0 = 0  $
 the system (\ref{Lagrangian_Eqs_infinite})
is split into  the reduced system
\begin{subequations}
\label{Lagrangian_Eqs_infinite_H0}
\begin{gather}
\rho_t = -\rho^2 u_{{s}},
  \label{Lagrangian_Eq_1_infinite_H0}
\\
u_t = - p_{{s}} -        H^y    H^y   _s     -    H^z    H^z    _s ,
\qquad
x_t = u ,
  \label{Lagrangian_Eq_2_infinite_H0}
\\
p_t = -\gamma \rho    p  u_{{s}}   ,
   \label{Lagrangian_Eq_7_infinite_H0}
\\
 H^y   _t  =   -   \rho     H^y  u _s    ,
  \label{Lagrangian_Eq_5_infinite_H0}
\\
 H^z   _t  =  - \rho     H^z  u _s  .
  \label{Lagrangian_Eq_6_infinite_H0}
\end{gather}
\end{subequations}
and the remaining  four equations   (\ref{Lagrangian_Eqs_H0_part_2}),
namely
\begin{subequations}
\label{Lagrangian_Eqs_infinite_H0_part_2}
\begin{gather}
v_t =    0  ,
\qquad
y_t = v ,
 \label{Lagrangian_Eq_infinite_3_H0}
\\
w_t =    0  ,
\qquad
z_t = w .
  \label{Lagrangian_Eq_infinite_4_H0}
\end{gather}
\end{subequations}

% These two subsystems  of equations can be  analyzed independently.

\begin{remark}
Similarly to Remark   \ref{remark_magnetic_1}
we can  rewrite equations (\ref{Lagrangian_Eq_5_infinite_H0}) and
(\ref{Lagrangian_Eq_6_infinite_H0})  as
\begin{equation}
\label{correction_ Hy_infinite_H0}
\left(   { H^y  \over \rho } \right) _ t =  0 ,
\end{equation}
\begin{equation}
\label{correction_ Hz_infinite_H0}
\left(   { H^z  \over \rho } \right) _ t =   0  .
\end{equation}
\end{remark}

% We also note that the last equation
% can be replaced by (\ref{correction_ S}).

For symmetry properties we discuss only   the reduced system (\ref{Lagrangian_Eqs_infinite_H0}).
In the general case of the polytropic constant $ \gamma > 1  $
the system    admits the following Lie algebra
\begin{multline}
% \def\arraystretch{1.75}
% \begin{array}{c}
X_1 = {\ddt},
\quad
X_2 = {\dds},
\quad
X_3 = {\ddx},
\quad
X_4 = t {\ddx} + {\ddu},
\\
X_5 = h_1
% \left( s,  { p \over  \rho^{\gamma} } ,  { H^y  \over \rho }, { H^z \over \rho } \right)
    \left(
       {H^z}   {\ddHy} -   {H^y}  {\ddHz}
    \right),
\quad
% X_5 = t {\ddt} + s {\dds} + x {\ddx},
 X_6 = -s  {\dds} + x {\ddx} + u {\ddu}   -  2 \rho {\ddrho},
\\
X_7 = t {\ddt}  + 2 s {\dds} - u {\ddu} + 2 \rho {\ddrho},
\quad
X_8 = 2 s {\dds} + 2 \rho {\ddrho} + 2 p {\ddp}
    %+ H^z \left(1 + \left(\frac{H^y}{H^z}\right)^2 \right) \partial_{H^z}.
    + H^y {\ddHz} + H^z {\ddHz},
% \end{array}
\end{multline}
where
\begin{equation*}
h_1 = h_1 \left( s,  { p \over  \rho^{\gamma} } ,  { H^y  \over \rho }, { H^z \over \rho } \right)
\end{equation*}
is an arbitrary function.

% {Alternative form for $X_6$: }
% \begin{equation*}
% X_6 =  -   s {\dds}
%     + x {\ddx} + u {\ddu}
% -  2 \rho {\ddrho}
% \end{equation*}

For $\gamma = 2$ there is the additional generator
\begin{equation}
X_9 = \rho   h_2
% \left( s, { p \over  \rho^{\gamma} } ,  { H^y \over \rho }, { H^z \over \rho } \right)
  \left (    {\ddHy} - H^y {\ddp} \right) ,
\qquad
h_2 = h_2 \left( s,  { p \over  \rho^{\gamma} } ,  { H^y  \over \rho }, { H^z \over \rho } \right)  .
\end{equation}

% Here
% \begin{equation*}
% h_1 \left( s,  { p \over  \rho^{\gamma} } ,  { H^y  \over \rho }, { H^z \over \rho } \right)  ,
% \qquad
% h_2  \left( s, { p \over  \rho^{\gamma} } ,  { H^y \over \rho }, { H^z \over \rho } \right)
% \end{equation*}
% are arbitrary functions. Note that $ { p /  \rho ^{\gamma} }$  is a function of the entropy.

%%%%%%%%%%%%%%%%%%%%%%%%%%%%%%%%%%%%%%%%%%%%%%%%%%%%%%%%%%%%%%%%%%
%%%%%%%%%%%%%%%%%%%%%%%%%%%%%%%%%%%%%%%%%%%%%%%%%%%%%%%%%%%%%%%%%%

\section{Variational approach to conservation laws
in  the case of infinite conductivity}

\label{CL_infinite}

To the best of our knowledge  there is no Lagrangian formulation
of the MHD equations in the case  of the finite conductivity.
However,  for the  infinite conductivity  it is possible to bring
the plane one-dimensional  MHD flows  equations to a variational form.

In Section \ref{Symmetries_infinite} there was noticed
a crucial difference between the cases  of the finite  and   infinite conductivities:
 the conservation of  the entropy along the   trajectories      (\ref{correction_ S})
holds only for  the  infinite conductivity.
This can  be easily seen from
 equation (\ref{equation_S_Euler})
rewritten in the Lagrangian coordinates:
\begin{equation*}
  S  _t    =
      {    \gamma  -1    \over \rho   ^{\gamma} }    \   (  \textbf{i}      \textbf{E}  )    ,
\qquad
 \textbf{i}  = \sigma     \textbf{E}  =   \mbox{rot}  \ \textbf{H}   .
\end{equation*}

% This property allows to integrate the equation    (\ref{correction_ S})
% as
% \begin{equation*}
% S = S(s)
% \end{equation*}
% and use $S(s)$
% as a classifying function  for symmetries and conservation laws.

\subsection{Case  $H^0 \neq  0$}

% {Question:  Do we need to consider subcases    $ H^y = 0 $  and   $ H^z  = 0 $
% for additional CL}

\subsubsection{Variational formulation}

We start with the case  $H^0 \neq  0$.
The system (\ref{Lagrangian_Eqs_infinite})
with modified equations     (\ref{correction_ Hy_infinite}), (\ref{correction_ Hz_infinite})
and (\ref{correction_ S})
takes the form
\begin{subequations}
\begin{gather}
\left(   { 1  \over \rho  } \right) _t =   u _s     ,
  \label{Case_1B_Eq_1}
\\
u_t = -    \left(  p    +          {  ( H^y )^2    +   ( H^z  ) ^2   \over   2    }     \right)   _s ,
\qquad
x_t = u ,
  \label{Case_1B_Eq_2}
\\
v_t =     (   H^0    H^y )  _s ,
\qquad
y_t = v ,
  \label{Case_1B_Eq_3}
\\
w_t =     (    H^0    H^z )  _s ,
\qquad
z_t = w ,
  \label{Case_1B_Eq_4}
\\
S_t = 0   ,
   \label{Case_1B_Eq_7}
\\
\left(   { H^y   \over \rho  } \right) _t =  (  H^0 v ) _s     ,
  \label{Case_1B_Eq_5}
\\
\left(   { H^z   \over \rho  } \right) _t =  (  H^0  w ) _s     .
  \label{Case_1B_Eq_6}
\end{gather}
\end{subequations}
% where many equations already have the form of a conservation law.

% Potentials:

% Equations  (\ref{Case_1B_Eq_1}), (\ref{Case_1B_Eq_5})  and (\ref{Case_1B_Eq_6}),  which have the form of conservation laws,  can be used  to introduce  potentials   $ \varphi (t,s) $, $ \psi (t,s)$  and $ \chi (t,s)$ by
% \begin{equation*}
%  \varphi   _ t   = u  ,
% \qquad
% \varphi  _s =   {  1\over \rho  } ,
% \end{equation*}
% \begin{equation*}
% \psi  _t =  v  ,
% \qquad
% \psi _s =  { H^y   \over  H^0   \rho  } ,
% \end{equation*}
% and
% \begin{equation*}
% \chi   _t =  w  ,
% \qquad
% \chi  _s =   { H^z   \over  H^0  \rho  }   .
% \end{equation*}

Using functions  $ \varphi (t,s) $, $ \psi (t,s)$  and $ \chi (t,s)$,
which relate  the Eulerian and Lagrangian coordinates (\ref{introduction_xyz}),
we get
\begin{equation}
u =  \varphi   _ t   ,
\qquad
\rho   =  {  1\over \varphi  _s } ,
\end{equation}
\begin{equation}
v = { \psi _t  } ,
\qquad
H^y =  H^0     { \psi _s \over  \varphi _s  } ,
\end{equation}
and
\begin{equation}
w = { \chi _t  } ,
\qquad
H^z =  H^0     { \chi _s \over  \varphi _s  }    .
\end{equation}
This presentation of the physical variables
makes the equations  (\ref{Case_1B_Eq_1}), (\ref{Case_1B_Eq_5})  and (\ref{Case_1B_Eq_6})
satisfied.
% Note that $ \varphi  \equiv x $, $ \psi  \equiv y - \eta  $ and $ \chi   \equiv z - \zeta  $,
% i.e. they represent the Eulerian spatial coordinates,
% which are dependent variables in the mass Lagrangian coordinates $ ( t,s) $.
We also  solve the equation  (\ref{Case_1B_Eq_7})  as
\begin{equation}
S(s) ,
\end{equation}
where $S(s)$ is an arbitrary function.

% {Step 4.}

The remaining  three equations   (\ref{Case_1B_Eq_2}),   (\ref{Case_1B_Eq_3})  and   (\ref{Case_1B_Eq_4})
  can be presented as second-order PDEs
\begin{subequations}    \label{PDE_case_H_not_0}
\begin{gather}
\varphi  _{t t }
+
  \left(   {  S  \over \varphi  _{s}  ^{\gamma}  }
+
    ( H^0 ) ^2   {     \psi  _s  ^2  +  \chi   _s  ^2   \over   2    \varphi  _s  ^2 }  \right)   _s
= 0    ,
\\
\psi  _{t t }
-
{  (H^0)^2     }
\left(    { \psi  _s  \over   \varphi  _s  }      \right)   _s
= 0  ,
\\
\chi   _{t t }
-
{  (H^0)^2    }
\left(    { \chi   _s  \over   \varphi  _s  }      \right)   _s
= 0   ,
\end{gather}
\end{subequations}
which have  variational structure.
They  are Euler-Lagrange  equations (\ref{EL_equations}) for the Lagrangian
\begin{equation}     \label{LagrangianHnot0}
L
=    { 1\over 2 }
(  \varphi  _{t}   ^2      +      \psi _t  ^2    +      \chi  _t  ^2      )
 -   {  S   \over      \gamma  -1 }
   \varphi  _{s}  ^{1 -  \gamma}
-
  (H^0)^2  {   \psi  _s  ^2   +   \chi   _s   ^2
\over
2   \varphi  _s  }        .
\end{equation}

\begin{remark}
The Lagrangian function   (\ref{LagrangianHnot0})
has a  clear physical interpretation.
It equals to the kinetic energy  minus the potential  energy
\begin{equation*}
L
=   { 1\over 2 }    (     u  ^2  + v ^2 + w ^2   )
 -        {  S \over    \gamma  -1      }  \rho ^{\gamma -1}
-          {    ( H^y ) ^2 + ( H^z ) ^2 \over   2  \rho }  .
\end{equation*}
The potential energy consists of two terms:
the internal energy of the gas  and the magnetic field energy.
\end{remark}

The equivalence transformations  for the system  (\ref{PDE_case_H_not_0})
are given in (\ref{equivalence_S}), Appendix B.
They   can be used to scale the function $ S(s)$  and the constant  $ H ^0 $.

The symmetry generators of  the system  (\ref{PDE_case_H_not_0})   have the form
\begin{equation}     \label{symmetry_form_1}
X = \sum _{k = 1} ^{12}   k_i  Y_i   ,
\end{equation}
where
\begin{multline}
Y _1
=
{\ddt} ,
\quad
Y _2
=
{\dds} ,
\quad
Y  _3
=
{\ddphi} ,
\quad
Y  _4
=
 { \partial  \over \partial \psi} ,
\quad
Y _5
=
{ \partial  \over \partial \chi  } ,
\\
Y _6
=
t {\ddphi} ,
\quad
Y  _7
=
t  { \partial  \over \partial \psi} ,
\quad
Y  _8
=
t { \partial  \over \partial \chi  } ,
\quad
Y  _9
=
\chi   { \partial  \over \partial \psi}
-
\phi  { \partial  \over \partial \chi  } ,
\\
Y  _{10}
=
   t {\ddt}  ,
\quad
Y _{11}
=
s  {\dds}  ,
\quad
Y  _{12}
=
\varphi   {\ddphi}
+  \psi   { \partial  \over \partial \psi}
+  \chi  { \partial  \over \partial \chi  }    .
\end{multline}

Applying the prolonged generator   (\ref{symmetry_form_1})
to the equations  (\ref{PDE_case_H_not_0}),
we obtain the conditions on the coefficients $ k_i$
\begin{subequations}     \label{conditions_1}
\begin{gather}
( k_ {11}  s   +  k _{2} )  S_s = \gamma (  k_ {12} - k _{11} ) S   ,
\label{conditions_1b}
\\
k_{12} + k_ {11} - 2 k _{10} = 0  .
\label{conditions_1a}
\end{gather}
\end{subequations}

The condition  (\ref{conditions_1b})
is the classifying equation for $ S (s) $.
It can be rewritten as
\begin{equation}      \label{classifying_form}
(  \alpha _1 s  +  \alpha _0)   S_s =   \beta S   ,
\end{equation}
where $ \alpha _0 $, $ \alpha _1 $  and $ \beta $   are constant.
The same classifying equation was obtained for gas dynamics  equations
\cite{bk:AndrKapPukhRod[1998], DORODNITSYN2019201}.
It was shown that the equation specifies four cases of the entropy function:
the general case and three special cases. They are

\begin{itemize}

\item

arbitrary   $ S(s) $;

\item

$ S(s) =S_{0} $,  $ S_{0} = \mbox{const}$;

\item

$ S (s) =S_{0} s^{q}  $,  $ q \neq 0 $, $ S_0   = \mbox{const} $;

\item

$  S = S_{0} e^{q s}  $,  $  q\neq 0 $, $ S_0   = \mbox{const}    $.

\end{itemize}

\subsubsection{Symmetries}

Using the  equations (\ref{conditions_1}),
one obtains the following  symmetries   for arbitrary $ S(s) $:
\begin{multline}          \label{kernel_infinite_H0}
X_1 =  { \ddt}   ,
\quad
X_2 = {\ddphi}  ,
\quad
X_3 = { \partial  \over \partial \psi} ,
\quad
X_4 = { \partial  \over \partial \chi  } ,
\\
X_5 = t {\ddphi}  ,
\quad
X_6 = t { \partial  \over \partial \psi} ,
\quad
X_7 = t { \partial  \over \partial \chi  } ,
\quad
X_8 =    \chi  { \partial  \over \partial \psi}  -    \psi { \partial  \over \partial \chi  }  .
\end{multline}
They form the kernel of the admitted Lie algebras.
Note that these generators correspond to a subset of generators (\ref{kern01}).

Symmetry extensions for particular cases of $ S(s) $  are given in Table~3.

\begin{equation*}
\begin{array}{|c|c|l|l|}
\hline
\mbox{Case} &    S (s)
& \mbox{Symmetry Extension}
& \mbox{Conditions}
 \\
\hline
 & & & \\
1    &   S_0    &
{   \displaystyle
{\dds} ,
\qquad
 t {\ddt} + s {\dds}
+ \varphi  {\ddphi}
+   \psi  { \partial  \over \partial \psi}
+  \chi  { \partial  \over \partial \chi  }      }  &
\\
 & & & \\
\hline
 & & & \\
2       &   S_0  s^q    &
{   \displaystyle
 (2\gamma+q)  t {\ddt}
+ 2 \gamma s {\dds}
+ 2(\gamma+q)
\left(  \varphi  {\ddphi}
+   \psi  { \partial  \over \partial \psi}
+  \chi  { \partial  \over \partial \chi  }
\right)   }
 &
q \neq 0
\\
 & & & \\
\hline
 & & &  \\
3    &   S_0  e^{qx}   &
{   \displaystyle
 q   t {\ddt}
+ 2 \gamma  {\dds}
+ 2 q
\left(  \varphi  {\ddphi}
+   \psi  { \partial  \over \partial \psi}
+  \chi  { \partial  \over \partial \chi  }
\right)   }
 &
q \neq 0
\\
 & & & \\
\hline
\end{array}
\end{equation*}
\begin{center}
{Table 3:} Additional  symmetries  for $ H^0 \neq 0$.
$ S_0 $ is a nonzero constant. \\
% {Checked  14.06.2021  }
\end{center}

\subsubsection{Conservation laws}

\medskip
\noindent {\bf a)  Arbitrary    $ S (s) $}
\medskip

In the case of arbitrary $ S (s) $ the Lagrangian (\ref{LagrangianHnot0})
admits all eight symmetries  (\ref{kernel_infinite_H0}).
The symmetries $ X_5 $,   $ X_6 $ and   $ X_7 $ are divergence symmetries
with  $ (B_1, B_2 ) = (\varphi  , 0 )  $,
$ (B_1, B_2 ) =  (  \phi  , 0 )  $
and
$ (B_1, B_2 ) = (  \chi  , 0 )  $,
respectively.
The other  symmetries are variational.

Symmetries    $ X_1 $--$ X_7$  provide
conservation laws which also exist in the case of the finite conductivity:
energy   (\ref{energy_LH}),
momenta     (\ref{momentum_LHx}),   (\ref{momentum_LHy}),  (\ref{momentum_LHz}),
motion of center of mass    (\ref{center_LHx}),  (\ref{center_LHy}),  (\ref{center_LHz}).
% and angular momentum   (\ref{angular_momentum_LH}).

The symmetry  $X_8$ gives conservation of the angular momentum
\begin{equation}         \label{angular_momentum_LSa}
 D_t ^L
(       \chi    \psi _t     -   \psi  \chi_t      )
 -
 D_s \left(
         { ( H^0 ) ^2   \over \varphi_s   }   ( \chi    \psi _s     -   \psi  \chi_s        )
 \right)  = 0  .
\end{equation}
In physical variables this conservation law takes the form
\begin{equation}        \label{angular_momentum_LSb}
 D_t ^L (     {    z  v    -   y  w     }     )
 +
 D_s (
     {   H^0   (   y   H ^z      -     z    H^y     )         }
 )  = 0  .
\end{equation}
Note that the conservation of the angular momentum also
needs the equations  (\ref{yszs}) to hold.

% It holds provided
%  \begin{equation}     \label{angular_momentum_LSc}
% y_s =  { H^y  \over H^0 \rho } ,
%  \qquad
%  z_s =  { H^z  \over H^0 \rho }   .
%  \end{equation}

% that gives
% \begin{equation*}
% y_{s}H^{z}-z_{s}H^{y}=\psi_{s}H^{z}-\chi_{s}H^{y}
% =\varphi_{s}\left(\frac{H^{y}}{H^{0}}H^{z}-\frac{H^{y}} {H^{0}}H^{z}\right)=0
% \end{equation*}

There are also three other conservation laws,
namely conservation of mass
and magnetic fluxes
(in the case of finite conductivity they are
(\ref{mass_LH}),   (\ref{Hy_LH}) and    (\ref{Hz_LH}),
respectively),
which were used to introduce potentials $\varphi$, $\psi $ and $\chi$.
These three  conservation laws
as well as the conservation of entropy
were used to bring the equations to a variational form.
They can not be obtained from the Lagrangian.
We conclude that
for the arbitrary entropy  $S(s)$
we obtain all  conservation laws
as of the case of the finite conductivity
as well as
the conservation of  the entropy (\ref{correction_ S})
and  the angular momentum  (\ref{angular_momentum_LSb}).
% which holds under additional conditions (\ref{angular_momentum_LSc}).

\medskip
\noindent {\bf b) Special case of  $ S (s) $}
\medskip

Table~4 presents additional symmetries of the Lagrangian
which are admitted for particular cases of $S(s)$.
These symmetries are variational.
\begin{equation*}
\begin{array}{|c|c|l|l|}
\hline
\mbox{Case} &    S(s)
& \mbox{Symmetry Extension}
& \mbox{Conditions}
 \\
\hline
 & & & \\
1    &   S_0     &
{   \displaystyle
% Z_9  =
{\dds}   }  &
\\
 & & & \\
\hline
 & & & \\
2    &   S_0  s^q    &
{   \displaystyle
%  Z_9 =
 t  {\ddt} + 3s {\dds} -  \varphi  {\ddphi}
-  \psi  { \partial  \over \partial \psi}
-   \chi  { \partial  \over \partial \chi  }    }
 &
{   \displaystyle
q = - { 4 \over 3} \gamma  }
\\
 & & & \\
\hline
% & & & \\
% 3    &   S_0  e^{qs}     &
% \mbox{To check using the results from S.M. }
%  &
% \\
%  & & & \\
% \hline
\end{array}
\end{equation*}
\begin{center}
{Table 4:}    Additional variational symmetries   for  $ H^0 \neq 0 $.
$ S_0 $ is a nonzero constant. \\
\end{center}

% {Additional  symmetries for particular cases (the same as in table)}

% Case  $ S = S_0 $
% \begin{equation}
%  Z_9
% = {\dds}
% \end{equation}

% Case  $ S = S_0 s^q  $, $ q = - { 4 \over 3} \gamma$
% \begin{equation}
% Z_9 =
% t  {\ddt} + 3s {\dds} -  \varphi  {\ddphi}
% -  \psi  { \partial  \over \partial \psi}
% -    \chi  { \partial  \over \partial \chi  }
% \end{equation}

% \subsection{Conservation laws}

% Here we present the conservation laws
% for the variational and divergence symmetries of the Lagrangian (\ref{LagrangianHnot0}).

% It is convenient to split the symmetries  into groups according
% to the physical interpretation of the conservation laws.

The additional symmetries provide the following conservation laws:

\begin{itemize}

\item

Case  $ S (s) = S_0 $

The additional symmetry
\begin{equation*}
{\dds}
\end{equation*}
leads to the conservation law
\begin{equation}
D_t ^L  \left(
%  T _9 ^t =
  \varphi _s     \varphi _t   +     \psi _s   \psi _t    +   \chi  _s   \chi  _t
\right)
+
% \qquad
D_s  \left(
% T _ 9  ^s =
  -  { 1\over 2 }
(  \varphi  _{t}   ^2  +      \psi _t  ^2    +      \chi  _t  ^2  )
 +      {  \gamma S  \over      \gamma  -1 }
   \varphi  _{s}  ^{1 -  \gamma}
\right) = 0  .
\end{equation}

In the  physical    variables  it is given as
\begin{equation}
D_t ^L  \left(
%  T _9 ^t =
  { u \over \rho }   +    {  v H^y   + w H^z    \over H ^0 \rho    }
\right)
+
% \qquad
D_s  \left(
% T _ 9  ^s =
 - { 1\over 2 }
( u  ^2  +   v  ^2    +    w  ^2  )
 +      {  \gamma S  \over      \gamma  -1 }
   \rho   ^{\gamma - 1  }
\right)= 0    .
\end{equation}

% {checked 05.09.2021}

\item

Case $ S (s) = S_0 s^q  $.

For $ q = - { 4 \over 3} \gamma$ there is an additional scaling symmetry
\begin{equation*}
 t  {\ddt} + 3s {\dds} -  \varphi  {\ddphi}
-  \psi  { \partial  \over \partial \psi}
-   \chi  { \partial  \over \partial \chi  } ,
\end{equation*}
which provides
the conservation law
\begin{multline}
D_t ^L  \left\{
%  T  _ 9 ^t  =
 t \left(
       { 1\over 2 }
(   \varphi  _{t}   ^2  +     \psi _t  ^2    +      \chi  _t  ^2    )
 +    {  S    \over      \gamma  -1 }
   \varphi  _{s}  ^{1 -  \gamma}
 +
{  (H^0)^2    (\psi  _s  ^2   +   \chi   _s   ^2 )
\over
 2   \varphi  _s  }
\right)
\right.
\\
\left.
+
3s
(     \varphi _s     \varphi _t   +     \psi _s   \psi _t    +   \chi  _s   \chi  _t    )
 +
 \varphi    \varphi _t
+
 \psi       \psi _t
+
 \chi     \chi _t
\right\}
\\
+
% \end{multline}
% \begin{multline}
D_s  \left\{
% T _9 ^s  =
(  t     \varphi  _{t}   +   \varphi    )
\left(
 {  S  }     \varphi  _{s}  ^{ -  \gamma}
+
{  (H^0)^2    (  \psi  _s  ^2   +   \chi   _s   ^2  )
\over
 2   \varphi  _s   ^2 }
\right)
  -
(  t  \psi  _{t}  +    \psi )
  {  (H^0)^2       \psi   _s
\over
    \varphi  _s   }
\right.
\\
\left.
 -
 ( t \chi   _{t} +  \chi  )
  {  (H^0)^2     \chi  _s
\over
    \varphi  _s   }
+
3s \left(
 - { 1\over 2 }
(  \varphi  _{t}   ^2  +      \psi _t  ^2    +      \chi  _t  ^2  )
 +     {  \gamma S  \over      \gamma  -1 }
   \varphi  _{s}  ^{1 -  \gamma}
\right)
% \\
% +
% \varphi
% \left(
%   {  S  }     \varphi  _{s}  ^{ -  \gamma}
% +
% {  (H^0)^2     ( \psi  _s  ^2   +   \chi   _s   ^2 )
% \over
%   8 \pi  \varphi  _s   ^2 }
% \right)
% -
%   {  (H^0)^2   (  \psi   \psi   _s   +  \chi     \chi     _s )
% \over
%  4  \pi    \varphi  _s   }
\right\}= 0  .
\end{multline}

It takes the form
\begin{multline}    \label{CL_physical_1b}
D_t ^L  \left\{
%  T  _9   ^t  =
t \left(
    { 1\over 2 }    (     u  ^2  + v ^2 + w ^2   )
+         {  S \over    \gamma  -1      }  \rho ^{\gamma -1}
+           {    ( H^y ) ^2 + ( H^z ) ^2 \over   2   \rho }
\right)
\right.
\\
\left.
+
3s \left(
   { u \over \rho }   +    {  v H^y   + w H^z    \over H ^0 \rho    }
\right)
 +
 x u  +  y v +  z w
\right\}
\\
+
%  \end{multline}
%  \begin{multline}
D_s  \left\{
%  T  _9   ^s  =
(  t   u    +  x  )
\left(
 {  S  }     \rho   ^{   \gamma}
+
  { ( H^y ) ^2 + ( H^z ) ^2  \over     2}
\right)
  -
     (  t v + y )   {   H^0  H^y   }
\right.
\\
\left.
  -
     (  t w + z )   {   H^0    H^z   }
+
3s \left(
 - { 1\over 2 }
( u  ^2  +   v  ^2    +    w  ^2  )
 +     {  \gamma S  \over      \gamma  -1 }
   \rho   ^{\gamma - 1  }
\right)
% \\
% +
% x
% \left(
%    {  S  }     \rho   ^{\gamma}
% +
%   {( H^y ) ^2 + ( H^z ) ^2   \over     8 \pi   }
% \right)
% -
%    {  H^0   \over   4  \pi }   ( y H^y + z H_z )
\right\} = 0
\end{multline}
in  the physical    variables.
It should be noted that
for conservation law  (\ref{CL_physical_1b})
we also need the equations  (\ref{nonlocal_x}),
(\ref{nonlocals_y_z}) and (\ref{yszs}),
which define $x$, $y$ and $z$ as nonlocal variables.

\end{itemize}

\subsection{Case  $H^0 = 0$}

\subsubsection{Variational formulation}

For  $H^0 = 0$ we consider  the reduced system (\ref{Lagrangian_Eqs_infinite_H0}).
It can be presented  as
\begin{subequations}
  \label{Case_1_Eqs}
\begin{gather}
\left(   { 1 \over \rho  } \right) _t   =    u_{{s}},
  \label{Case_1_Eq_1}
\\
u_t = -  \left(     p   +     {   ( H^y )^2    +   ( H^z  ) ^2    \over 2    }     \right)    _s ,
\qquad
x_t = u ,
  \label{Case_1_Eq_2}
\\
S_t =  0  ,
%  p_t = -\gamma p \rho u_{{s}}   .
  \label{Case_1_Eq_5}
\\
\left(   { H^y   \over \rho  } \right) _t =  0   ,
  \label{Case_1_Eq_3}
\\
\left(   { H^z   \over \rho  } \right) _t =  0    ,
  \label{Case_1_Eq_4}
\end{gather}
\end{subequations}
where  modified equations
(\ref{correction_ S}),   (\ref{correction_ Hy_infinite_H0}) and  (\ref{correction_ Hz_infinite_H0})
are   taken into account.

\begin{remark}
The system (\ref{Lagrangian_Eqs_infinite_H0})
contains the closed subsystem
\begin{subequations}
\label{Lagrangian_Eqs_infinite_H0_part}
\begin{gather}
\rho_t = -\rho^2 u_{{s}},
 \\
u_t = - p_{{s}} -        H   _s     ,
\qquad
x_t = u ,
\\
p_t = -  \gamma  \rho p  u_{{s}},
\\
 H   _t  =   -   2 \rho     H u _s    .
\end{gather}
\end{subequations}
where
\begin{equation*}
H=\sqrt{\frac{ (H^y) ^{2}+(H^z) ^{2}}{2}}.
%  \qquad
%  \bar{p}=p+H^{2},
\end{equation*}
It is also possible to develop a variational formulation for this subsystem.

The subsystem  is not equivalent to the original system (\ref{Lagrangian_Eqs_infinite_H0}),
but will be equivalent if we add  one more equation,
namely  (\ref{Lagrangian_Eq_5_infinite_H0}) or
(\ref{Lagrangian_Eq_6_infinite_H0}).

% For variational approach we can use
% \begin{subequations}
%  \begin{gather}
% \left(   { 1 \over \rho  } \right) _t   =    u_{{s}},
%  \\
% u_t = -   (     p   +   H    )    _s ,
% \qquad
% x_t = u ,
%  \\
% S_t =  0  ,
%  p_t = -\gamma p \rho u_{{s}}   .
% \\
% \left(   { H    \over \rho ^2   } \right) _t =  0   ,
%  \end{gather}
% \end{subequations}
\end{remark}

\begin{remark}
It is easy to see that the equations  (\ref{Case_1_Eqs})
have conservation laws
\begin{equation*}
D _t ^L
\left\{
T^t \left( S, { H^y   \over \rho  } , { H^z   \over \rho  }
\right)
\right\} = 0 .
\end{equation*}
If we consider the compete system,
i.e. equations   (\ref{Case_1_Eqs})  and (\ref{Lagrangian_Eqs_H0_part_2}),
then there are more  conservation laws
with $ T^s \equiv 0 $, namely
\begin{equation*}
D _t ^L
\left\{
T^t \left( S, { H^y   \over \rho  } , { H^z   \over \rho  } ,
v, w, y - t v, z - t w
\right)
\right\}  = 0 .
\end{equation*}
\end{remark}

% {Potentials:}

The  system   (\ref{Case_1_Eqs}) can be reduced to one variational  PDE.
For the equation (\ref{Case_1_Eq_1})
we introduce the potential  $ \varphi (t,s)  \equiv x $:
\begin{equation*}
\varphi   _ t = u ,
\qquad
\varphi  _s  =  {  1\over \rho } .
\end{equation*}
% It is the Eulerian spatial coordinate $ \varphi  \equiv x $,
% which is a dependent variable in the mass Lagrangian coordinates $ ( t,s) $.
It is used to present the velocity and the density as
\begin{equation}     \label{potential_phi}
u = \varphi   ,
\qquad
\rho   =  {  1\over \varphi  _s  } .
\end{equation}
Equations  (\ref{Case_1_Eq_3})  and   (\ref{Case_1_Eq_4})
are solved as
\begin{equation}        \label{help_F_G}
H^y =   \rho F(s) =      { F(s) \over   \varphi  _s } ,
\qquad
H^z = \rho G(s) =       { G(s) \over   \varphi  _s } ,
\end{equation}
where $F(s) $ and $ G(s) $ are arbitrary functions.
Finally,  equation  (\ref{Case_1_Eq_5}) gives
\begin{equation}     \label{help_S}
S= S(s) ,
\end{equation}
where  $ S (s) $  is arbitrary.

Using       (\ref{potential_phi}),  (\ref{help_F_G}) and  (\ref{help_S}),
it is possible to rewrite the remaining  equations   (\ref{Case_1_Eq_2})
as the following second-order PDE
\begin{equation}     \label{PDE_case_H_0}
\varphi  _{t t }
+  \left(   {  S   \over \varphi  _{s}  ^{\gamma}  }
+            { A   \over      \varphi  _s  ^2 }    \right)   _s
= 0   ,
\qquad
A  (s)   =  {  F  ^2 (s) + G  ^2 (s)  \over 2    }   {  \not \equiv  } 0  .
\end{equation}

This PDE has variational structure (\ref{EL_equations}).
It is provided by the Lagrangian function
  \begin{equation}   \label{Lagrangian_case_H_0}
L
=    { 1\over 2 }     \varphi  _{t}   ^2
 -   {  S   \over      \gamma  -1 }     \varphi  _{s}  ^{1 -  \gamma}
-       { A \over    \varphi  _s   }         .
\end{equation}

Note that for $\gamma =2$ the last two terms in the PDE  (\ref{PDE_case_H_0})
and in  the Lagrangian  (\ref{Lagrangian_case_H_0}) merge:
we obtain  the PDE
\begin{equation}     \label{PDEB_case_H_0}
\varphi  _{t t }
+
  \left(        { B   \over    \varphi  _s  ^2 }    \right)   _s
= 0   ,
\qquad
B (s)  =  {  S  (s)    \over \gamma  - 1 }    +   { A (s)      }  ,
\end{equation}
which is given by the Lagrangian
\begin{equation}     \label{LagrangianB_case_H_0}
L
=    { 1\over 2 }     \varphi  _{t}   ^2
-       { B \over    \varphi  _s   }       .
\end{equation}
This particular case coincides with the gas dynamics
(MHD equations   (\ref{Case_1_Eqs})  in the absence of the magnetic field).
The complete analysis of this case is given in \cite{bk:AndrKapPukhRod[1998], DORODNITSYN2019201}.
This case needs to be presented separately
because of different symmetry and conservation properties.

% Notice that for $\gamma=2$ the latter equation coincide with the
% gas dynamic equation in absence of a magnetic field. As the complete
% analysis of this case is given in \cite{bk:AndrKapPukhRod[1998],  DORODNITSYN2019201},
% this case of $\gamma$ is {excluded} from our further consideration.

\begin{remark}
Searching for a Lagrangian of the form
${L}={L}(t, s,\varphi,\varphi_{t},\varphi_{s})$
for PDE (\ref{PDE_case_H_0}),
we obtain  the general form of the Lagrangian
\begin{equation*}
\bar{L}    = \alpha    L   + h ,
% \qquad
% {L}_0
% =   \frac{1}{2}\varphi_{t}^{2}
%  \frac{S_{0}}{\gamma-1}\varphi_{s}^{1-\gamma}
% - \frac{B_{0}}{\varphi_{s}}   ,
\qquad
\alpha \neq 0 ,
\qquad
\alpha  = \mbox{const} ,
\end{equation*}
\begin{equation*}
h=   D_t ^L ( C^t  ( t, s,  \varphi  ) )    +    D_s ( C^s  ( t, s,  \varphi  ) )     ,
\end{equation*}
where  $L$  is given  by   (\ref{Lagrangian_case_H_0})
and the functions $  C^t  ( t, s,  \varphi  )   $ and  $  C^s   ( t, s,  \varphi  )  $   are arbitrary.
Notice that
\begin{equation}
\frac{\delta h}{\delta\varphi}\equiv0 ,
\end{equation}
it follows  that  $h$  does not contribute to the Euler-Lagrange equations.

The Noether identity (\ref{Noether_identity}) gives
\begin{equation*}
X h + h   ( D_t ^L ( \xi  ^t  )  + D_s ( \xi ^s  )   )
 =  D_t ^L  ( N^t h )  + D_s ( N^s h )  ,
\end{equation*}
where
$  N^t  $ and $  N^s  $ are the Noether operators (\ref{Noether_operators}).
It follows that the general  Lagrangian $ \bar{L} $
provides the same  conservation laws as $L$.
Therefore the term $h$ (as well as the constant $\alpha$) can  be discarded.

% Noether identity (reference needed) gives
% \begin{equation}
% X h + h   ( D_t ^L ( \xi  ^t  )  + D_s ( \xi ^s  )   )
%  =  D_t ^L ( N^t h )  + D_s ( N^s h )  ,
% \end{equation}
% where
% $  N^t  $ and $  N^s  $ are Noether operators (reference needed).

% Using the same argument as in remark   \ref{general_L_1},
% we obtain
% that the general  Lagrangian $ \bar{L} $
% provides the same  conservation laws as $L$.
% Therefore we discard the term $h$  and the constant $\alpha$.
\end{remark}

\begin{remark}
In the physical variables the Lagrangian function
(\ref{Lagrangian_case_H_0}) takes the form
 \begin{equation*}
L
=    {    u  ^2  \over 2 }
 -        {  S \over   \gamma  -1   }  \rho ^{\gamma -1}
-          {  ( H^y ) ^2 + ( H^z ) ^2 \over   2  \rho }   .
\end{equation*}
\end{remark}

% \begin{remark}
% It is   possible  to add equations
% \begin{equation*}
%  \psi  _{tt}    = 0 ,
% \end{equation*}
% \begin{equation*}
%  \chi  _{tt}    = 0
% \end{equation*}
% and  consider Lagrangians
%  \begin{equation}
% L
% =    { 1\over 2 }  (    \varphi  _{t}   ^2  +   \psi  _{t}   ^2   +   \chi  _{t}   ^2  )
% -   {  S   \over      \gamma  -1 }     \varphi  _{s}  ^{1 -  \gamma}
% -       { B  \over   2 \varphi  _s   }
% \end{equation}
% and
% \begin{equation}
% L
% =    { 1\over 2 }     (    \varphi  _{t}   ^2  +   \psi  _{t}   ^2   +   \chi  _{t}   ^2  )
% -       { K  \over    \varphi  _s   }       .
% \end{equation}
% \end{remark}

In the following text  we will consider the cases  $ \gamma \neq 2 $ and  $ \gamma =  2 $  separately.

\subsubsection{Equivalence transformations}

The equivalence transformations for  PDEs (\ref{PDE_case_H_0}) and (\ref{PDEB_case_H_0})
are given in Appendix B.
The transformations   (\ref{equivalence_general}) for the PDE  (\ref{PDE_case_H_0})
can be used to scale the functions $ S(s) $ and $ A(s) $.
 The equivalence transformation of the PDE (\ref{PDEB_case_H_0})
are the same   as for the gas dynamics equation,  considered in  \cite{DORODNITSYN2019201}.
They are given in  (\ref{equivalence_gamma_2})
and can be used to scale the function $ B (s) $.

\subsubsection{Symmetries in the general case $\gamma \neq 2$ ($\gamma > 1$) }

\label{The_general_case_gamma}

Symmetries of  PDE  (\ref{PDE_case_H_0}) have the form
\begin{equation}    \label{symmetry_form_2}
X = \sum _{k = 1} ^{7}   k_i  Y_i    ,
\end{equation}
where
\begin{multline}
Y _1
=
{\ddt} ,
\quad
Y _2
=
{\dds} ,
\quad
Y  _3
=
{\ddphi} ,
\quad
Y _4
=
t {\ddphi} ,
\\
Y  _{5}
=
   t {\ddt}  ,
\quad
Y _{6}
=
s  {\dds}  ,
\quad
Y  _{7}
=
\varphi   {\ddphi}   .
\end{multline}
Applying generator  (\ref{symmetry_form_2}) to the PDE,
we get the following conditions for coefficients $ k_i$:
\begin{subequations}     \label{conditions_2}
\begin{gather}
( k_6 s + k_2 ) S_s
=   ( ( \gamma + 1) k_7 + (1 - \gamma) k_6 - 2 k_5 ) S  ;
\\
( k_6 s + k_2 ) A_s
=   ( 3 k_7  -   k_6 - 2 k_5 ) A  .
\end{gather}
\end{subequations}
If considered independently,
both conditions have the form  (\ref{classifying_form})
and lead to the same cases of  $ S(s) $  and  $ A(s) $ as discussed earlier.
Both       $ S(s) $  and  $ A(s) $
can be arbitrary,  constant, power or exponential functions.
However, not all pairs lead to additional symmetries.
We obtain the following  pairs for the consideration of symmetry  extensions:

\begin{itemize}

\item

arbitrary
$ ( S (s) ,   A (s)  ) $;

\item

constant
$ ( S (s) ,   A (s)  )  = ( S_0 ,  A_0 ) $,
$S_0 = \mbox{const}$,  $A_0 = \mbox{const}$;

\item

power
$ ( S (s) ,   A (s)  )  = ( S_0 s^{\alpha} ,  A_0 s^{\beta} ) $,
$\alpha  ^2 + \beta ^2 \neq 0  $,
$S_0 = \mbox{const}$,  $A_0 = \mbox{const}$;

\item

exponential
$ ( S (s) ,   A (s)  )  = ( S_0 e^{p s} ,  A_0 e^{q s} ) $,
$ p  ^2 + q ^2 \neq 0 $,
$S_0 = \mbox{const}$,  $A_0 = \mbox{const}$.

\end{itemize}

From equations   (\ref{conditions_2})
we obtain  that the kernel of the admitted  Lie algebras
is defined by the generators
\begin{equation}     \label{kernel0}
X  _1 =  {\ddt} ,
\quad
 X  _2 =  {\ddphi} ,
\quad
X  _3 =  t   {\ddphi}  .
\end{equation}
Particular cases  of  $  S (s) $  and  $ A (s) $
which lead to extensions of the admitted symmetry algebra
are presented in Table~5.

\begin{equation*}
\begin{array}{|c|c|c|l|l|}
\hline
\mbox{Case} &  S (s)    &  A (s)
& \mbox{Symmetry Extension}
& \mbox{Conditions}
 \\
\hline
 & & & & \\
1  & S_0  & A_0
&
{ \displaystyle
% X_4 =
{\dds}  ,
\qquad
% X_5 =
 t     { \ddt  }     + s  {\dds}+  \varphi  {\ddphi}    }
&
\\
 & & & & \\
\hline
%   & & & & \\
% 2  & 1 &   s^{\alpha}
% &
% { \displaystyle
%  X_4 = ( 2(\gamma -2 ) + 3 \alpha )  t   { \ddt  }
% +   2(\gamma -2 ) s    {\dds}
% + 2  (\gamma -2 + \alpha  )   \varphi   {\ddphi}   }
% \\
%  & & & &  \\
% \hline
%  & & & &  \\
% 3  & 1 &  e^{p s}
% &
% { \displaystyle
% X_4 = 3  p t    { \ddt  }
% +   2(\gamma -2 )      {\dds}
% + 2 p     \varphi  {\ddphi}   }
% \\
%  & & & & \\
% \hline
%  & & & & \\
% 4  &   s^{\beta}    &  1
% &
% { \displaystyle
% X_4
% =   (    2  (  \gamma - 2 )   -   \beta  (    \gamma  + 1  )     )     t   { \ddt  }
% + 2  (\gamma - 2 )  s     {\dds}
% +  2 (   \gamma -2  - \beta )  \varphi   {\ddphi}   }
% \\
%  & & & & \\
% \hline
 & & & & \\
2  &   S_0    s^{\alpha} &   A_0   s^{\beta}
&
{ \displaystyle
% X_4  =
(    2  (  \gamma - 2 )    + 3 \alpha  -   \beta   (    \gamma  + 1  )      )     t    { \ddt  }   }
% + 2  (\gamma -2 )  s    {\dds}
% + (  \gamma -2    -   \beta  + \alpha  )  \varphi   {\ddphi}   }
&    \alpha ^2   +  \beta ^2   \neq 0
\\
 & & & & \\
  &   &   &
{ \displaystyle
+ 2  (\gamma -2 )  s    {\dds}
+ (  \gamma -2   + \alpha   - \beta   )  \varphi   {\ddphi}   }
&
\\
 & & & & \\
\hline
% 6  &   s^{\beta}    &   e^{p  s}
% &    \mbox{none}
% \\
% \hline
%  & & & & \\
% 7  &   e^{q s}    &  1
% &
% { \displaystyle
% X_4 =   - q (\gamma +1 )     t    { \ddt  }
% + 2 ( \gamma - 2)      {\dds}
%  - 2 q  \varphi  {\ddphi}  }
% \\
%  & & & & \\
% \hline
% 8  &   e^{q s}    &  s^{\alpha}
% &    \mbox{none}
% \\
% \hline
 & & & & \\
3   &  S_0   e^{p s} &  A_0   e^{q s}
&
{ \displaystyle
%  X_4 =
(- q ( \gamma +1)  + 3 p  )     t    { \ddt  }
 + 2  (  \gamma - 2 )       {\dds}
+ 2 ( p  - q  )   \varphi   {\ddphi}   }
& p^2  +  q^2 \neq 0
\\
 & & & & \\
\hline
\end{array}
\end{equation*}
\begin{center}
{Table  5:} Additional symmetries for $ H^0 = 0 $,  $ \gamma \neq 2$. \\
$S_0$ and $A_0$ are nonzero  constants.
% {Note: there is no  need to separate $\gamma =3$}     \\
% {Note: to consider possibilities of more compact presentations \\
% by using only three cases: $ (1,1) $, $  ( s^{\beta}    ,    s^{\alpha} )$, $(e^{q s}   ,   e^{p s})$.}
\end{center}

% \begin{remark} {(to remove later)}
% The table is based on results form S.M.
% It gives the  additional symmetries of the form
% \begin{equation}
% P t {\ddt } +  ( Q_1 t + Q_2 )  {\dds} + R \varphi {  \ddphi} .
% \end{equation}
% \end{remark}

\subsubsection{Conservation laws  in the general case $\gamma \neq 2$ ($\gamma > 1$) }

\label{Conservation_laws_general}

\medskip
\noindent {\bf a)  Arbitrary  $ S(s)$ and $ A(s)$ }
\medskip

The symmetries (\ref{kernel0}) provide  conservation laws
of  energy, momentum and  motion of the center of mass.
These conservation laws exist for the finite conductivity.
They are    (\ref{CL_H0_energy}),   (\ref{CL_H0_momenta})
and     (\ref{CL_H0_center}), respectively.

The conservation of mass, magnetic fluxes and entropy
were used to rewrite the PDE in the variational form
and therefore they can not be  obtained from the Lagrangian.
For the finite conductivity  the conservation of mass and  magnetic fluxes
were given by    (\ref{CL_H0_mass}),  (\ref{CL_H0_flux_y}) and   (\ref{CL_H0_flux_z}).
The conservation law  of the entropy does not
hold for the finite conductivity.

We conclude that
we obtain the same conservation laws for the reduced system (\ref{Case_1_Eqs})
as for  the corresponding  system with the finite conductivity (\ref{Lagrangian_Eqs_H0_part_1}),
and  conservation law for the entropy.

\medskip
\noindent {\bf b) Special cases of $ S(s)$ and $ A(s)$  }
\medskip

There are particular  cases of
\begin{equation*}
S(s)
\qquad
\mbox{and}
\qquad
 A (s)     =  {  (H^y)^2 +   (H^z)^2  \over 2 \rho ^2  }
\end{equation*}
with additional  variational symmetries.
These cases  are specified in Table~6.

\begin{equation*}
\begin{array}{|c|c|c|l|l|}
\hline
\mbox{Case} &  S (s)   &  A (s)
& \mbox{Symmetry Extension}
& \mbox{Conditions} \\
\hline
  & & & &  \\
1  &   S_0   &    A_0
&
{\displaystyle
% Z _4  =
{\dds}    }
&
\\
  & & & &  \\
\hline
%   & & & &  \\
% 2  & 1 &   s^{\alpha}
% &
% { \displaystyle
% Z _4 = 5 t    { \ddt  }   - s   {\dds}    + 3 \varphi  {\ddphi}    }
% &
% \alpha =  - 4 ( \gamma - 2 )
% \\
%   & & & &  \\
% \hline
%  & & & &  \\
% 3  &   s^{\beta}    &  1
% &
% {\displaystyle
% Z _4  = ( 3 \gamma -1 )   t     { \ddt  }
% + (\gamma -3 )  s     {\dds}
% + (\gamma +1 )  \varphi  {\ddphi}   }
% &
% {\displaystyle  \beta =   -4   {   \gamma - 2   \over  \gamma - 3   }   } ,
% \quad
% \gamma  \neq 3
% \\
%   & & & &  \\
% \hline
  & & & &   \alpha ^2    +    \beta^2     \neq 0    \\
2   &   S_0    s^{\alpha}   &   A_0    s^{\beta}
&
 {\displaystyle
% Z _4  =
( 2  \beta  + 5 )   t    { \ddt  }
-  s   {\dds}
+ ( \beta  +  3  )  \varphi  {\ddphi}   }
&
\alpha +  \beta  ( \gamma -3 )    =  - 4 ( \gamma -2 )
\\
  & & & &  (\mbox{note:}  \ \  \gamma \neq 3   \ \  \mbox{if} \ \  \alpha = 0)  \\
\hline
%   & & & &  \\
% 5  &   e^{q s}    &  1
% &
% {\displaystyle
% Z _4  = 2 q    t     { \ddt  }   -   {\dds}  + q  \varphi {\ddphi}   }
% &
% \gamma  =  3
% \\
%   & & & &  \\
% \hline
  & & & &    p^2   +  q^2 \neq 0 \\
3     &   S_0   e^{p s}    &  A_0    e^{q s}
&
{\displaystyle
% Z _4  =
2 q    t     { \ddt  }   -     {\dds}    + q  \varphi  {\ddphi}   }
&
p   +   q (\gamma -3)  =  0
\\
  & & & &   (\mbox{note:}  \ \  \gamma = 3   \ \  \mbox{if} \ \  p = 0)  \\
\hline
\end{array}
\end{equation*}
\begin{center}
{Table 6:} Additional variational symmetries for $ H^0 = 0 $, $ \gamma \neq 2$. \\
$S_0$ and $A_0$ are nonzero  constants.  \\
\end{center}

The  additional symmetries provide the following  conservation laws:

\begin{itemize}

\item

 Case  $ (  S(s)  , A(s) )   =  (S_0  ,  A_0  ) $

There exist the additional conservation law
\begin{equation}
D_t ^L \left(
%  T _4 ^t   =
 \varphi_{s} \varphi_{t}
\right)
+
D_s \left(
%  T _4 ^s =
  -    { \varphi_{t}^{2}   \over 2 }
+   { \gamma S \over  \gamma - 1 }         \varphi_{s}^{1-\gamma}
+  {  2 A \over  \varphi_{s}  }
\right) = 0   .
\end{equation}

In the physical variables it is rewritten as
\begin{equation}
D_t ^L \left(
% T _4 ^t  =
   { u \over  \rho }
\right)
+
D_s \left(
%  T _4 ^s =
  -  { u ^{2}   \over 2 }
+   { \gamma  S \over  \gamma - 1 }     \rho ^{\gamma -1}
+   {   (H^y) ^2 +   (H^z) ^2  \over  \rho  }
\right) = 0   .
\end{equation}

% {checked}

\item

Case $ (  S(s)  , A(s) )   =   (   S_0  s^{\alpha}  ,   A_0     s^{\beta}   ) $

For $ \alpha +  \beta  ( \gamma -3 )    =  - 4 ( \gamma -2 )   $
there  is the additional  conservation law
\begin{multline}
D_t ^L \left\{
% T _4 ^t =
  ( 2 \beta + 5 )   t
\left(
   { 1\over 2 }     \varphi  _{t}   ^2
 +    {  S   \over      \gamma  -1 }     \varphi  _{s}  ^{1 -  \gamma}
 +       { A \over   \varphi  _s   }
 \right)
  -   s
   \varphi_{s} \varphi_{t}
  -    ( \beta  +  3  )  \varphi
      \varphi_{t}
\right\}
\\
+
% \end{equation}
% \begin{equation}
D_s \left\{
% T _4 ^s  =
(   ( 2 \beta + 5 )   t  \varphi_{t} -  ( \beta  +  3  )  \varphi )
  \left(   S      \varphi_{s}^{-\gamma}
+   { A  \over   \varphi  _s  ^2  }
\right)
  +  s
\left(
    { \varphi_{t}^{2}   \over 2 }
 -    { \gamma S \over  \gamma - 1 }       \varphi_{s}^{1-\gamma}
 -    { 2 A  \over   \varphi  _s   }
 \right)
% \\
% -  ( \beta  +  3  )  \varphi
% \left(
%   S      \varphi_{s}^{-\gamma}
% +   { B  \over   2 \varphi  _s  ^2  }
% \right)
\right\} = 0     .
\end{multline}

It is presented  in  the physical    variables as
\begin{multline}    \label{physical_A_2}
D_t ^L \left\{
%  T _4 ^t =
( 2 \beta + 5 )   t
\left(
  {  u ^{2}  \over 2 }
+ {S  \over  \gamma - 1 }     \rho  ^{\gamma - 1 }
+       {  ( H^y )  ^2 +   ( H^z )  ^2 \over   2 \rho   }
 \right)
 -  s
   { u \over  \rho }
-    ( \beta  +  3  )  x
     u
\right\}
\\
+
% \end{equation}
% \begin{multline}
D_s \left\{
%  T _4 ^s =
(    ( 2 \beta + 5 )   t     u   -   ( \beta  +  3  )  x )
\left(
S  \rho ^{\gamma}
+    {  ( H^y )  ^2 +   ( H^z )  ^2 \over   2   }
\right)
\right.
\\
\left.
  +  s
\left(
   { u ^{2}   \over 2 }
-  { \gamma  S \over  \gamma - 1 }     \rho ^{\gamma -1}
-   {  ( H^y )  ^2 +   ( H^z )  ^2 \over   \rho    }
 \right)
% -  ( \beta  +  3  )  x
% \left(
% S    \rho  ^{\gamma}
% +    {  ( H^y )  ^2 +   ( H^z )  ^2 \over   2   }
% \right)
\right\} = 0     .
\end{multline}

\item

 Case   $  (  S(s)  , A (s) )   =  (   S_0   e^{p s}    ,   A _0    e^{q s}    ) $

If  $ p   +   q (\gamma -3)  =  0  $, there is the  conservation law
\begin{multline}
D_t ^L \left\{
%  T _4 ^t =
 2 q    t
\left(
  { 1\over 2 }     \varphi  _{t}   ^2
 +   {  S(s) \over      \gamma  -1 }     \varphi  _{s}  ^{1 -  \gamma}
 +     { A  \over    \varphi  _s   }
 \right)
  -
    \varphi_{s} \varphi_{t}
-  q  \varphi
    \varphi_{t}
\right\}
\\
+
% \end{equation}
% \begin{equation}
D_s \left\{
%  T _4 ^s =
q  (   2   t     \varphi_{t}   -    \varphi   )
\left(   S      \varphi_{s}^{-\gamma}
+   { A \over    \varphi  _s  ^2  }
\right)
  +
   { \varphi_{t}^{2}   \over 2 }
-  { \gamma S \over  \gamma - 1 }       \varphi_{s}^{1-\gamma}
-  { 2 A  \over    \varphi  _s   }
% -  q  \varphi
% \left(
%   S      \varphi_{s}^{-\gamma}
% +   { B  \over   2 \varphi  _s  ^2  }
% \right)
\right\} = 0   .
\end{multline}

In the physical  variables it takes the form
\begin{multline}     \label{physical_A_3}
D_t ^L \left\{
%  T _4 ^t =
 2 q    t
\left(
  {  u ^{2}  \over 2 }
 +   {S  \over  \gamma - 1 }     \rho  ^{\gamma - 1 }
 +     {  ( H^y )  ^2 +   ( H^z )  ^2 \over   2 \rho   }
 \right)
  -
  { u \over  \rho }
-  q x
   u
\right\}
\\
+
% \end{equation}
% \begin{equation}
D_s \left\{
%  T _4 ^s  =
q    (   2     t     u      -    x    )
\left(
S  \rho ^{\gamma}
+    {  ( H^y )  ^2 +   ( H^z )  ^2 \over   2   }
\right)
   +
   { u ^{2}   \over 2 }
-  { \gamma  S \over  \gamma - 1 }     \rho ^{\gamma -1}
 -   {  ( H^y )  ^2 +   ( H^z )  ^2 \over   \rho     }
% \\
% -  q  x
% \left(
% S    \rho  ^{\gamma}
% +    {  ( H^y )  ^2 +   ( H^z )  ^2 \over   2   }
% \right)
\right\}  = 0    .
\end{multline}

% {checked}

\end{itemize}

We remark that conservation laws    (\ref{physical_A_2})  and   (\ref{physical_A_3})
in addition to the reduced MHD system (\ref{Lagrangian_Eqs_infinite_H0})
(equivalently   (\ref{Case_1_Eqs}))
need equations    (\ref{nonlocal_x}),
which  define $x$ as a {non}local variable.

\subsubsection{Symmetries in the special case  $\gamma =2 $}

For $\gamma =2  $  we consider the variational  PDE   (\ref{PDEB_case_H_0})
and the corresponding Lagrangian (\ref{LagrangianB_case_H_0}).
This case is equivalent to that  of gas dynamic,
i.e. MHD equation without the magnetic field.
The gas dynamic equations were analyzed  in  \cite{DORODNITSYN2019201}.
Therefore we can rely on the results obtained there.

The kernel of the Lie algebras admitted by equation (\ref{PDEB_case_H_0})
consists of the generators
\begin{equation}     \label{kernel1}
X  _1 =  {\ddt} ,
\quad
 X  _2 =  {\ddphi} ,
\quad
X  _3 =  t   {\ddphi}  ,
\quad
X_4 = 3 t    {\ddt}  + 2 \varphi  {\ddphi}    .
\end{equation}
The first three symmetries are the same as for  $\gamma \neq 2$.
For particular cases  $B(s)$ there are additional symmetries,
given in Table~7.

\begin{equation*}
\begin{array}{|c|c|l|l|}
\hline
\mbox{Case} &  B  (s)
& \mbox{Symmetry Extension}
& \mbox{Conditions} \\
\hline
 & & &  \\
1  & B_0
&
{\displaystyle
% X_5 =
{\dds}  ,
\qquad
% X_6 =
 t     {\ddt}   +  s    {\dds}   +  \varphi  {\ddphi} }
&
\\
 & & &   \\
\hline
%  & & &  &  \\
% 2  & 1 & s^{\beta}
% &
% {\displaystyle
% X_5 =   t     {\ddt}   -     ( s+1)      {\dds}   }
% &
% \alpha = 1
% \\
%   & & & &   \\
% \hline
% 3  & 1 & e^{ps}
% &
% \mbox{none}
% &
% \\
% \hline
%  & & & &   \\
% 4  &  s^{\beta}  &   1
% &
% {\displaystyle
% X_5 =   t     {\ddt}  -     ( s+1)     {\dds}    }
% &
% \beta  = 1
% \\
%  & & & &   \\
% \hline
 & & &   \\
2  &   B_0    s^{\beta}
&
{\displaystyle
% X_5 =
 ( \beta  + 1 ) t     {\ddt}   -  2 s   {\dds}  }
&
 \beta  \neq 0
\\
 & & &  \\
\hline
%  6  & s^{\alpha } &   e^{p  s}
% &
% \mbox{none}
% &
% \\
% \hline
% 7   & e^{q s}   &   1
% &
% \mbox{none}
% &
% \\
% \hline
% 8  & e^{q s}   &   s^{\alpha }
% &
% \mbox{none}
% &
% \\
% \hline
 & & &   \\
3    &  B_0   e^{q s}
&
{\displaystyle
% X_5 =
q    t     {\ddt}   -   2    {\dds}    }
&  q\neq 0
\\
 & & &    \\
\hline
\end{array}
\end{equation*}
\begin{center}
{Table 7:} Additional symmetries for $ H^0 = 0 $, $ \gamma = 2$.
$B_0 $ is a nonzero constant.  \\
\end{center}

\subsubsection{Conservation laws in the special case  $\gamma =2 $}

\medskip
\noindent {\bf a)  Arbitrary    $ B(s) $    }
\medskip

In the general case of   $B(s)$  we obtain the same variational  symmetries as
in the case   $\gamma \neq 2$ with arbitrary  $S(s)$ and  $A(s)$,
namely        (\ref{kernel0}).
The corresponding conservation laws were discussed in point \ref{Conservation_laws_general}.

\medskip
\noindent {\bf b)  Special cases of $ B(s) $     }
\medskip

For particular cases of
\begin{equation*}
B(s)
=  { S(s) \over \gamma -1}
+   {  (H^y)^2 +   (H^z)^2  \over 2 \rho ^2  }
\end{equation*}
the Lagrangian   (\ref{LagrangianB_case_H_0})
has additional variational symmetries given in Table~8.

\begin{equation*}
\begin{array}{|c|c|l|l|}
\hline
\mbox{Case} &  B  (s)
& \mbox{Symmetry Extension}
& \mbox{Conditions} \\
\hline
 & & &   \\
1  & B_0
&
{\displaystyle
%  Z _4  =
 {\partial  _s } ,
\qquad
%   Z _5 =
  5 t     {\ddt}   - s    {\dds}  + 3 \varphi    {\ddphi}   }
&
\\
 & & &  \\
\hline
%  & & & &   \\
% 2  & 1 & s^{\alpha}
% &
% {\displaystyle
%  Z _4 =   7 t     {\ddt}    - ( s +1)    {\dds}  + 4 \varphi  {\ddphi}  }
% &     \alpha =1
% \\
%  & & & &   \\
% \hline
%  & & & &   \\
% 3  & s^{\alpha}  &  1
% &
% {\displaystyle
%  Z _4 =   7 t     {\ddt}   - ( s +1)    {\dds}  + 4 \varphi   {\ddphi}   }
% &     \alpha  =1
% \\
%  & & & &   \\
% \hline
 & & &  \\
2  & B_0  s^{\beta}
&
{\displaystyle
%  Z _4 =
 ( 2   \beta  + 5 ) t    {\ddt}  - s    {\dds}  + (\beta  +3)  \varphi    {\ddphi}  }
&
 \beta  \neq 0
\\
 & & &   \\
\hline
 & & &    \\
3   &   B_0  e^{q s}
&
{\displaystyle
%   Z _4 =
 2 q    t     {\ddt}   -      {\dds} + q  \varphi   {\ddphi}   }
&
q  \neq 0
\\
 & & &  \\
\hline
\end{array}
\end{equation*}
\begin{center}
{Table 8:} Additional variational symmetries for $ H^0 = 0$, $ \gamma = 2$. \\
$ B_0 $ is a nonzero constant. \\
% {Note: the symmetries   $Z_4$ of case $ (1,1) $, $  ( s^{q}    ,    s^{q} )$, $(e^{q s}   ,   e^{q s})$ \\
% are included into the case of general $\gamma$ \\
% New  symmetries :    \\
% $Z_5$ of case $ (1,1) $;   \\
\end{center}

The comparison of these symmetries   with those for $\gamma \neq 2 $
shows  that for   $\gamma = 2 $
we obtain the symmetries of the generic case $\gamma \neq 2 $ and one extension.
This extension is the scaling  symmetry admitted for    $B(s) = B _0$.
It provides the  conservation law
\begin{equation}
D_t ^L \left\{
% T _5 ^t  =
   5 t
\left(
  { 1\over 2 }     \varphi  _{t}   ^2
 +        { B  \over    \varphi  _s   }
 \right)
   -  s
  \varphi_{s} \varphi_{t}
 -  3 \varphi
  \varphi_{t}
\right\}
+
% \qquad
% \end{equation}
% \begin{equation}
D_s \left\{
%  T _5 ^s =
(  5 t    \varphi_{t}   -  3 \varphi   )
   { B \over    \varphi  _s  ^2  }
 +  s
\left(
   { \varphi_{t}^{2}   \over 2 }
- { 2 B  \over    \varphi  _s   }
 \right)
% -  3 \varphi
%      { K   \over    \varphi  _s  ^2  }
\right\}= 0   .
\end{equation}
In the physical  variables it takes the form
\begin{multline}
D_t ^L \left\{
% T _5 ^t =
    5 t
\left(
   {  u ^{2}  \over 2 }
 +  {S  \over  \gamma - 1 }     \rho  ^{\gamma - 1 }
  +       {  ( H^y )  ^2 +   ( H^z )  ^2 \over   2 \rho   }
 \right)
   - s
 { u \over  \rho }
 -  3 x
    u
\right\}
\\
+
% \end{equation}
% \begin{equation}
D_s \left\{
%  T _5 ^s =
(    5 t   u     -  3 x  )
\left(
S  \rho ^{\gamma}
+    {  ( H^y )  ^2 +   ( H^z )  ^2 \over   2   }
\right)
   +  s
\left(
   { u ^{2}   \over 2 }
-  { \gamma  S \over  \gamma - 1 }     \rho ^{\gamma -1}
-   {  ( H^y )  ^2 +   ( H^z )  ^2 \over   \rho    }
 \right)
% \\
% -  3 \varphi
% \left(
% S    \rho  ^{\gamma}
% +    {  ( H^y )  ^2 +   ( H^z )  ^2 \over   2   }
% \right)
\right\} = 0    .
\end{multline}
% { checked }
Note that for verification of this conservation law,
presented in  the physical variables,
we  also need  equations   (\ref{nonlocal_x}).

\section{Concluding remarks}

\label{Concluding}

The paper is devoted to Lie point symmetries and conservation laws
of the plane one-dimensional MHD flows, described in the mass Lagrangian
coordinates by equations (\ref{Lagrangian_Eqs}).

The analysis leads to four cases for the electric conductivity $ \sigma ( \rho , p)$
and $ H^x = H^0 =  \mbox{const} $:
\begin{enumerate}

\item

Finite electric conductivity  and $ H^0 \neq 0$;

\item

Finite electric conductivity and $ H^0 = 0$;

\item

Infinite electric conductivity and $ H^0 \neq 0$;

\item

Infinite electric conductivity and $ H^0 = 0$.
\end{enumerate}
The latter case splits for the value of the polytropic constant.
We get the generic {sub}case $   \gamma  \neq 2 $
and the special  {sub}case  $   \gamma = 2 $.
For $   \gamma  =  2 $ the equations are
equivalent to the equations describing
the plane one-dimensional flows of the gas dynamics,
i.e. the equations without the magnetic field.

For all cases given above we found the admitted Lie point
symmetries. For the cases with the finite conductivity it results in
the Lie group classifications for $ \sigma ( \rho , p)$. For  the
infinite electric conductivity the classifications have
the entropy $S$  as an arbitrary element.

The conservation laws for the finite electric conductivity are found
by direct computation. For the cases with the infinite  conductivity
the equations can be brought into the variational forms. Further, the
conservation laws were found using the Noether theorem, which was
applied to the variational equations. Finally, the conservation laws
were converted into the original physical variables.

\section*{Acknowledgements}

The research was supported by Russian Science Foundation Grant no. 18-11-00238
"Hydrodynamics-type equations: symmetries, conservation laws, invariant difference
schemes".
The authors thank G. Webb for discussions.
E.I.K. sincerely appreciates the hospitality of the Suranaree University of Technology.

\bigskip

%\bibliographystyle{unsrt}
%\bibliography{references_kaptsov_kozlov}
% \bibliography{references_MSV}

\begin{thebibliography}{99}

\bibitem{bk:Ovsyannikov[1962]}
L.~V. Ovsiannikov.
\newblock {\em Group Analysis of Differential Equations}.
\newblock Academic, New York, 1982.

\bibitem{bk:Ibragimov1985}
N.~H. Ibragimov.
\newblock {\em Transformation Groups Applied to Mathematical Physics}.
\newblock Reidel, Boston, 1985.

\bibitem{bk:Olver}
P.~J. Olver.
\newblock {\em Applications of Lie Groups to Differential Equations}.
\newblock Springer, New York, 1986.

\bibitem{bk:Bluman1989}
G.W. Bluman and S.~Kumei.
\newblock {\em Symmetries and Differential Equations}.
\newblock Applied Mathematical Sciences. Springer New York, 2013.

\bibitem{bk:Gridnev1968}
N.P. Gridnev.
\newblock Study of the magnetohydrodynamics equations' group properties and
  invariant solutions.
\newblock {\em Journal of Applied Mechanics and Technical Physics},
  (6):103--107, 1968.
\newblock in Russian.

\bibitem{bk:DorMHDpreprint1976}
V.~A. Dorodnitsyn.
\newblock On invariant solutions of one-dimensional nonstationary
  magnetohydrodynamics with finite conductivity.
\newblock {\em Keldysh Institute preprints}, 143, 1976.
\newblock in Russian.

\bibitem{art:Rogers1969}
C.~Rogers.
\newblock Invariant transformations in non-steady gasdynamics and
  magneto-gasdynamics.
\newblock {\em Zeit angew. Math. Phys.}, 20:370--382, 1969.

\bibitem{bk:HandbookLie_v2}
N.~H. Ibragimov, editor.
\newblock {\em {CRC} Handbook of {L}ie Group Analysis of Differential
  Equations}, volume~2.
\newblock CRC Press, Boca Raton, 1995.

\bibitem{art:MeleshkoMoyoWebb2021}
S.~V. Meleshko, S.~Moyo, and G.~M. Webb.
\newblock Solutions of generalized simple wave type of magnetic fluid.
\newblock {\em Commun. Nonlinear. Sci. Numer. Simulat.}, 103:105991, 2021.

\bibitem{bk:PaliathanasisMHD2021}
A.~Paliathanasis.
\newblock Group properties and solutions for the 1{D} {H}all {MHD} system in
  the cold plasma approximation.
\newblock {\em Eur. Phys. J. Plus}, 136(538), 2021.

\bibitem{bk:Oliveri_F_2005}
F.~Oliveri and Speciale M.P.
\newblock Exact solutions to the ideal magneto-gas-dynamics equations through
  lie group analysis and substitution principles.
\newblock {\em J. Phys. A: Math. Gen.}, 38(40):8803--8820, 2005.

\bibitem{bk:Picard_P_Y_2008}
P.Y. Picard.
\newblock Some exact solutions of the ideal mhd equations through symmetry
  reduction.
\newblock {\em J. Math. Anal. Appl.}, 337(1):360--385, 2008.

\bibitem{bk:Golovin_2009}
S.V. Golovin.
\newblock Regular partially invariant solutions of defect 1 of the equations of
  ideal magnetohydrodynamics.
\newblock {\em Journal of Applied Mechanics and Technical Physics},
  50(2):171--180, 2019.

\bibitem{bk:Golovin_2011}
S.V. Golovin.
\newblock Natural curvilinear coordinates for ideal mhd equations.
  non-stationary flows with constant total pressure.
\newblock {\em Physics Letters, Section A: General, Atomic and Solid State
  Physics}, 375(3):283--290, 2011.

\bibitem{bk:Golovin_2019}
S.V. Golovin and L.T. Sesma.
\newblock Exact solutions of stationary equations of ideal magnetohydrodynamics
  in the natural coordinate system.
\newblock {\em Journal of Applied Mechanics and Technical Physics},
  60(2):234--247, 2019.

\bibitem{bk:Noether1918}
E.~Noether.
\newblock Invariante variations problem.
\newblock {\em Konigliche Gesellschaft der Wissenschaften zu Gottingen,
  Nachrichten, Mathematisch-Physikalische Klasse Heft 2}, pages 235--257, 1918.
\newblock English translation:~Transport Theory and Statist. Phys., 1(3), 1971,
  183-207.

\bibitem{bk:Webb2018}
G.~Webb.
\newblock {\em Magnetohydrodynamics and Fluid Dynamics: Action Principles and
  Conservation Laws}.
\newblock Springer, Heidelberg, 2018.
\newblock Lecture Notes in Physics, v. 946.

\bibitem{bk:DorodnitsynKozlov[2011]}
V.~A. Dorodnitsyn and R.~V Kozlov.
\newblock {L}agrangian and {H}amiltonian formalism for discrete equations:
  Symmetries and first integrals.
\newblock In D.~Levi, P.~Olver, Z.~Thomova, and P.~Winternitz, editors, {\em
  Symmetries and Integrability of Difference Equations}, London Mathematical
  Society Lecture Note Series, page 7–49. Cambridge University Press, 2011.

\bibitem{bk:KulikovskiiLubimov1965}
A.~G. Kulikovskii and G.~A. Lyubimov.
\newblock {\em Magnetohydrodynamics}.
\newblock Addison-Wesley Educational Publishers, Berlin/Boston, 1965.

\bibitem{bk:SamarskyPopov_book[1992]}
A.~A. Samarskii and Y.~P. Popov.
\newblock {\em Difference methods for solving problems of gas dynamics}.
\newblock Nauka, Moscow, 1980.
\newblock in Russian.

\bibitem{bk:Landau_Electrodynamics}
Landau L.D., Pitaevskii~L. P., and E.M.~Lifshitz E.M.
\newblock {\em Electrodynamics of Continuous Media}.
\newblock Elsevier Science \& Technology, Oxford, United Kingdom, 1984.

\bibitem{Davidson2001}
P.~A. Davidson.
\newblock {\em An Introduction to Magnetohydrodynamics}.
\newblock Cambridge Texts in Applied Mathematics. Cambridge University Press,
  2001.

\bibitem{Galtier2016}
S.~Galtier.
\newblock {\em Introduction to Modern Magnetohydrodynamics}.
\newblock Cambridge University Press, 2016.

\bibitem{Brushlinskii}
K.~V. Brushlinskii.
\newblock {\em Mathematical and Computational Problems of
  Magnetohydrodynamics}.
\newblock Laboratorija Znanij, Moscow, 2014.
\newblock in Russian.

\bibitem{bk:Chernyi_gas}
G.~G. Chernyi.
\newblock {\em Gas dynamics}.
\newblock Nauka, Moscow, 1988.
\newblock in Russian.

\bibitem{bk:Ovsyannikov[2003]}
L.~V. Ovsiannikov.
\newblock {\em Lectures on the gas dynamics equations}.
\newblock Institute of Computer Studies, Moscow--Izhevsk, 2003.
\newblock in Russian.

\bibitem{Landau}
L.~D. Landau and E.~M. Lifshitz.
\newblock {\em Fluid Mechanics}.
\newblock Pergamon Press, 2nd. ed., 1987.

\bibitem{Chorin}
A.~J. Chorin and J.~E. Marsden.
\newblock {\em A Mathematical Introduction to Fluid Mechanics}.
\newblock Springer-Verlag, 1990.

\bibitem{Toro}
E.~F. Toro.
\newblock {\em Riemann Solvers and Numerical Methods for Fluid Dynamics}.
\newblock Springer-Verlag, Berlin-Heidelberg, 1997.

\bibitem{bk:Suriyawichitseranee2015}
A.~Suriyawichitseranee, Yu.~N. Grigoriev, and S.~V. Meleshko.
\newblock Group analysis of the {F}ourier transform of the spatially
  homogeneous and isotropic {B}oltzmann equation with a source term.
\newblock {\em Commun. Nonlinear. Sci. Numer. Simulat.}, 20(3):719--730, 2015.

\bibitem{bk:Popovych_GC_2006}
R.O. Popovych.
\newblock Classification of admissible transformations of differential
  equations.
\newblock {\em Collection of Works of Institute of Mathematics, Kyiv},
  3(2):239--254, 2006.

\bibitem{bk:Bluman1997}
S.~C. Anco and G.~W. Bluman.
\newblock Direct construction of conservation laws from field equations.
\newblock {\em Physical Review Letters}, 78:2869--2873, 04 1997.

\bibitem{Ibragimov_adj1[2011]}
N.~H. Ibragimov.
\newblock Nonlinear self-adjointness and conservation laws.
\newblock {\em Journal of Physics A: Mathematical and Theoretical}, 44:432002,
  10 2011.

\bibitem{bk:AndrKapPukhRod[1998]}
V.~K. Andreev, O.~V. Kaptsov, V.~V. Pukhnachov, and A.~A. Rodionov.
\newblock {\em Applications of Group-Theoretic Methods in Hydrodynamics}.
\newblock Kluwer, Dordrecht, 1998.

\bibitem{DORODNITSYN2019201}
V.~A. Dorodnitsyn, R.~Kozlov, and S.~V. Meleshko.
\newblock One-dimensional gas dynamics equations of a polytropic gas in
  {L}agrangian coordinates: Symmetry classification, conservation laws,
  difference schemes.
\newblock {\em Commun. Nonlinear. Sci. Numer. Simulat.}, 74:201--218, 2019.

\bibitem{bk:Ovsiannikov[1993opt]}
L.~V. Ovsiannikov.
\newblock On optimal system of subalgebras.
\newblock {\em Dokl. RAS}, 333(6):702--704, 1993.

\bibitem{bk:PateraWinternitz1977}
J.~Patera and P.~Winternitz.
\newblock Subalgebras of real three‐ and four‐dimensional {L}ie algebras.
\newblock {\em Journal of Mathematical Physics}, 18(7):1449--1455, 1977.

\end{thebibliography}
%%%%%%%%%%%%%%%%%%%%%%%%%%%%%%%%%%%%%%%%%%%%%%%%%%%%%%

%%%%%%%%%%%%%%%%%%%%%%%%%%%%%%%%%%%%%%%%%%%%%%%%%%%%%%%%%%%%%%%%%%%%%%%%%%%%

\bigskip
\bigskip

\noindent{\bf \Large Appendices}

\appendix

\section{Lagrangian variables}

\label{Appendix_Lagrangian_variables}

Let $(\xi,\eta,\zeta)$ be Lagrangian spatial variables.
The Eulerian coordinates $(x,y,z)$ and Lagrangian coordinates are related by
the equations
\begin{subequations}     \label{eq:sep14.0}
\begin{gather}
x=\tilde{\varphi}(t, \xi,\eta,\zeta),
\\
y=\tilde{\psi}(t, \xi,\eta,\zeta),
\\
z=\tilde{\chi}(t, \xi,\eta,\zeta),
\end{gather}
\end{subequations}
where the functions $\tilde{\varphi}$, $\tilde{\psi}$ and $\tilde{\chi}$
are smooth functions, satisfying the Cauchy problem
\begin{subequations}    \label{eq:sep14.0b}
\begin{gather}
\tilde{\varphi}_{t}=u(t, \tilde{\varphi},\tilde{\psi},\tilde{\chi}),
\qquad
\tilde{\varphi}(0, \xi,\eta,\zeta)=\xi,
\label{eq:sep14.1}
\\
\tilde{\psi}_{t}=v(t, \tilde{\varphi},\tilde{\psi},\tilde{\chi}),
\qquad
\tilde{\psi}(0, \xi,\eta,\zeta)=\eta,
\label{eq:sep14.2}
\\
\tilde{\chi}_{t}=w(t, \tilde{\varphi},\tilde{\psi},\tilde{\chi}),
\qquad
\tilde{\chi}(0, \xi,\eta,\zeta)=\zeta.
\label{eq:sep14.3}
\end{gather}
\end{subequations}

Notice also that due to Euler's theorem
\begin{equation*}
\frac{\partial J}{\partial t}=J  \   \mbox{div} _{e}    \textbf{u}  ,
\end{equation*}
where  $ \mbox{div} _{e}  \textbf{u} $ is the divergence of the velocity $  \textbf{u} $
in the Eulerian coordinates with substituted (\ref{eq:sep14.0}),
the general solution of the conservation of mass equation has the
form
\begin{equation}
\rho( t, \tilde{\varphi}( t, \xi,\eta,\zeta ), \tilde{\psi}(t, \xi,\eta,\zeta ),\tilde{\chi}(t, \xi,\eta,\zeta))
=
\frac{\rho_{0}(\xi,\eta,\zeta)}{J(t, \xi,\eta,\zeta)},
\label{eq:sep14.5}
\end{equation}
where $\rho_{0}$ is an arbitrary function and
\begin{equation*}
J(t, \xi,\eta,\zeta)=\frac{\partial(\tilde{\varphi},\tilde{\psi},\tilde{\chi})}{\partial(\xi,\eta,\zeta)}
\end{equation*}
is the Jacobian.

One can show that for the plane one-dimensional flow
\begin{equation}
u=u(t, x),
\qquad
v=v(t, x),
\qquad
w=w(t, x),
\label{eq:sep14.3b}
\end{equation}
where the functions $u$, $v$ and $w$ are continuously differentiable
functions, it is necessary and sufficient that the functions $\tilde{\varphi}$,
$\tilde{\psi}$ and $\tilde{\chi}$ have the form
\begin{subequations}   \label{eq:sep14.4}
\begin{gather}
\tilde{\varphi}(t, \xi,\eta,\zeta)=\hat{\varphi}(t, \xi),
\\
\tilde{\psi}(t, \xi,\eta,\zeta)=\eta+\hat{\psi}(t. \xi),
\\
\tilde{\chi}(t, \xi,\eta,\zeta)=\zeta+\hat{\chi}(t, \xi).
\end{gather}
\end{subequations}
Indeed, assume that $u$, $v$ and $w$ satisfy (\ref{eq:sep14.0b}).
Consider the function $ \tilde{\varphi} $.
Differentiating (\ref{eq:sep14.1})  with respect
to $\eta$, we obtain that the function $g(t , \xi,\eta,\zeta)=\tilde{\varphi}_{\eta}(t, \xi,\eta,\zeta)$
% (or $g(\xi,\eta,\zeta,0)=\tilde{\varphi}_{\zeta}(\xi,\eta,\zeta,t)$)
satisfies the Cauchy problem
\begin{equation*}
g_{t}=u_{y} g,
\qquad
g(0, \xi,\eta,\zeta)=0.
\end{equation*}
As $  g(t, \xi,\eta,\zeta) \equiv  0$ is a solution of the latter Cauchy problem, and due to uniqueness
of the solution of this problem, we conclude  that $\tilde{\varphi}_{\eta}=0$.
Similarly, we obtain  $\tilde{\varphi}_{\zeta}=0$.
Therefore,   $\tilde{\varphi}(t, \xi,\eta,\zeta)=\hat{\varphi}(t, \xi)$.

Consider the function $\tilde{\psi}(t, \xi,\eta,\zeta)$.
Noting that the function $h =\tilde{\psi}_{\eta}$ satisfies the Cauchy problem
\begin{equation*}
h _{t}=0,
\qquad
h (0, \xi,\eta,\zeta )=1,
\end{equation*}
one obtains that $\tilde{\psi}_{\eta}=1$.
Similarly,  we derive
that $\tilde{\psi}_{\zeta}=0$,  $\tilde{\chi}_{\eta}=0$ and $\tilde{\chi}_{\zeta}=1$.
This gives that
\begin{equation*}
\tilde{\psi}(t, \xi,\eta,\zeta)=\eta+\hat{\psi}(t, \xi),
\qquad
\tilde{\chi}(t, \xi,\eta,\zeta)=\zeta+\hat{\chi}(t, \xi).
\end{equation*}

Converse, assume that the relations between the Lagrangian and Eulerian
coordinates have the form (\ref{eq:sep14.4}). Differentiating (\ref{eq:sep14.4})
with respect to $t$, and noting that due to the inverse function
theorem the equation $x-\hat{\varphi}(t, \xi)=0$ can be solved with
respect to $\xi$, one obtains (\ref{eq:sep14.0b}).

Considering (\ref{eq:sep14.5}) at $t=0$, we  obtain  that $\rho_{0}=\rho_{0}(\xi)$.
Hence, (\ref{eq:sep14.5}) becomes
\begin{equation*}
\rho(t, \hat{\varphi}(t, \xi))=\frac{\rho_{0}(\xi)}{\hat{\varphi}_{\xi}(t,  \xi)}.
\end{equation*}

\medskip

For the {\it mass Lagrangian coordinates}  one applies the change $s=\alpha(\xi)$,
where $\alpha^{\prime}(\xi)=\rho_{0}(\xi)$. Hence, the function $\varphi(t, s)$
such that $\varphi(t, \alpha(\xi))=\hat{\varphi}(t, \xi)$ satisfies
the conditions
\begin{equation*}
\varphi_{t}(t, s)=u(t, \varphi(t, s)),
\qquad
\varphi_{s}(t, s)=\frac{1}{\rho(t, \varphi(t, s) )}.
\end{equation*}
Using similar relations
\begin{equation*}
\psi(t, \alpha(\xi))=\hat{\psi}(t, \xi),
\qquad
\chi(t, \alpha(\xi))=\hat{\chi}(t, \xi),
\end{equation*}
one obtains
\begin{equation*}
\psi_{t}(t, s)=v(t, \varphi(s,t)),
\qquad
\chi_{t}(t, s)=w(t, \varphi(s,t)).
\end{equation*}

\section{Equivalence transformations}

Here we  provide  the equivalence transformations for the different  MHD systems
considered in the paper.
Equivalence transformations allow to change  arbitrary elements while
preserving the structure of the  equations.
The algorithm for finding equivalence transformations is given in \cite{bk:Ovsyannikov[1962]}.

\subsection{The case of finite conductivity $\sigma (\rho , p) $ and  $ H^0 \neq 0 $}

The generators of the equivalence transformations for  system~(\ref{Lagrangian_Eqs})
have  the form
\begin{multline}
X ^e  = \zeta^t \frac{\partial}{\partial{t}}
    + \zeta^s \frac{\partial}{\partial{s}}
    + \zeta^x \frac{\partial}{\partial{x}}
    + \zeta^y \frac{\partial}{\partial{y}}
    + \zeta^z \frac{\partial}{\partial{z}}
    + \zeta^u \frac{\partial}{\partial{u}}
    + \zeta^v \frac{\partial}{\partial{v}}
    + \zeta^w \frac{\partial}{\partial{w}}
      \\
    + \zeta^\rho \frac{\partial}{\partial{\rho}}
    + \zeta^p \frac{\partial}{\partial{p}}
    + \zeta^{E^y} \frac{\partial}{\partial{E^y}}
    + \zeta^{E^z} \frac{\partial}{\partial{E^z}}
    + \zeta^{H^y} \frac{\partial}{\partial{H^y}}
    + \zeta^{H^z} \frac{\partial}{\partial{H^z}}
    + \zeta^{\sigma} \frac{\partial}{\partial{\sigma}}
    + \zeta^{H^0} \frac{\partial}{\partial{H^0}},
\end{multline}
where $\zeta^t$, $\zeta^s$,  ... , $\zeta^{\sigma}$,   $\zeta^{H^0}$    are functions
of $t$, $s$, $\mathbf{x}$, $\mathbf{u}$, $\rho$, $p$,  $E^y$, $E^z$,  $H^y$, $H^z$, $\sigma$ and $H^0$.
Computation provides  the generators
\begin{multline}     \label{equivalence_2}
% \def\arraystretch{1.75}
% \begin{array}{c}
X ^e _1 =  {\ddt},
\quad
X ^e _2 =  {\dds},
\quad
X ^e _3 =   {\ddx},
\quad
X ^e  _4 =  t {\ddx} + {\ddu} ,
\\
X ^e  _5 =   z {\ddy} - y {\ddz}
    + w {\ddv} - v {\ddw}
   + E^z {\ddEy} - E^y {\ddEz}
   + H^z {\ddHy} - H^y {\ddHz}  ,
\\
% Y_8 = t {\ddt}  + s {\dds}  + x {\ddx}  + y {\ddy}  + z {\ddz}  - \sigma {\partial  \over  \partial \sigma } ,
% \\
X ^e _{6} =
    t {\ddt}
    + 2 s {\dds}
    - v {\ddv}  - u {\ddu} - w {\ddw}
    + 2 \rho {\ddrho}
    - E^y {\ddEy}
    - E^z {\ddEz}
    + \sigma { \partial \over \partial \sigma } ,
\\
X ^e _{7} =
    -  s {\dds}
    + x {\ddx}  + y {\ddy}  + z {\ddz}
    + v {\ddv}  + u {\ddu} +  w {\ddw}
    -  2 \rho {\ddrho}
    + E^y {\ddEy}
    +  E^z {\ddEz}
    -  2 \sigma { \partial \over \partial \sigma }  ,
\\
X ^e _8 =   2s {\dds}
    + 2\rho {\ddrho}
    + 2p {\ddp}
    + E^y {\ddEy}
    + E^z {\ddEz}
    + H^y {\ddHy}
    + H^z {\ddHz}
    + H^0 { \partial \over \partial {H^0}  } ,
\\
X ^e _{9} = {\phi}^{1} (s) {\ddy},
\quad
X ^e  _{10} = {\phi}^{2} (s) {\ddz},
\quad
X ^e _{11} =   t {\ddy}  + {\ddv} ,
\quad
X ^e _{12} =   t {\ddz} + {\ddw} ,
% \end{array}
\end{multline}
where ${\phi}^1 (s) $ and ${\phi}^{2} (s) $    are arbitrary functions.

% { to operators were CHANGED, they should be CHECKED !}

\subsection{The case of finite conductivity $\sigma (\rho , p) $ and $ H^0  =  0 $}

The {equivalence transformations}
of the reduced  system~(\ref{Lagrangian_Eqs_H0_part_1})
are provided by the generators of the form
\begin{multline} \label{equivalence_02}
X ^e  = \zeta^t \frac{\partial}{\partial{t}}
    + \zeta^s \frac{\partial}{\partial{s}}
    + \zeta^x \frac{\partial}{\partial{x}}
    + \zeta^u \frac{\partial}{\partial{u}}
   + \zeta^\rho \frac{\partial}{\partial{\rho}}
    + \zeta^p \frac{\partial}{\partial{p}}
\\
    + \zeta^{E^y} \frac{\partial}{\partial{E^y}}
    + \zeta^{E^z} \frac{\partial}{\partial{E^z}}
    + \zeta^{H^y} \frac{\partial}{\partial{H^y}}
    + \zeta^{H^z} \frac{\partial}{\partial{H^z}}
    + \zeta^{\sigma} \frac{\partial}{\partial{\sigma}},
\end{multline}
where $\zeta^t$, $\zeta^s$,  ... , $\zeta^{\sigma}$  are functions
of $t$, $s$, $x$, $u$, $\rho$, $p$,  $E^y$, $E^z$,  $H^y$, $H^z$ and $\sigma$.
We obtain
\begin{multline} \label{EqTrMLagrFlatH0=0}
% \def\arraystretch{1.75}
% \begin{array}{c}
X^e  _1 = {\ddt},
\quad
X^e  _2 = {\dds},
\quad
X^e  _3 = {\ddx},
\quad
X^e  _4 = t {\ddx} + {\ddu},
\\
X^e  _5 = % v \partial_w - w \partial_v +
    E^z {\ddEy}  -   E^y {\ddEz}
   + H^z {\ddHy}  -  H^y {\ddHz} ,
\\
X^e  _6 = t  {\ddt} + 2 s  {\dds}
    - u {\ddu} % - v \partial_{v} - w \partial_{w}
    + 2\rho {\ddrho}
    - E^y {\ddEy} - E^z {\ddEz}
     + \sigma { \partial \over \partial \sigma },
\\
X^e  _7 = -s {\dds} + x {\ddx} + u {\ddu}
    - 2\rho {\ddrho}
    + E^y {\ddEy} + E^z {\ddEz}
- 2  \sigma { \partial \over \partial \sigma },
\\
X^e  _8 = 2 s  {\dds} + 2\rho {\ddrho}  + 2 p {\ddp}
    + E^y {\ddEy} +  E^z {\ddEz}
    + H^y {\ddHy} +  H^z {\ddHz}.
% \end{array}
\end{multline}

% {checked, operator $X^e  _7$ was corrected: sigma term}

\subsection{The case of infinite conductivity and   $ H^0  \neq  0 $}

The equivalence transformations for system  (\ref{Lagrangian_Eqs_infinite})
have the generators
\begin{multline} \label{infinite_equivalence_A}
X ^e  = \zeta^t \frac{\partial}{\partial{t}}
    + \zeta^s \frac{\partial}{\partial{s}}
    + \zeta^x \frac{\partial}{\partial{x}}
    + \zeta^y \frac{\partial}{\partial{y}}
    + \zeta^z \frac{\partial}{\partial{z}}
    + \zeta^u \frac{\partial}{\partial{u}}
    + \zeta^v \frac{\partial}{\partial{v}}
    + \zeta^w \frac{\partial}{\partial{w}}
      \\
    + \zeta^\rho \frac{\partial}{\partial{\rho}}
    + \zeta^p \frac{\partial}{\partial{p}}
    + \zeta^{E^y} \frac{\partial}{\partial{E^y}}
    + \zeta^{E^z} \frac{\partial}{\partial{E^z}}
    + \zeta^{H^y} \frac{\partial}{\partial{H^y}}
    + \zeta^{H^z} \frac{\partial}{\partial{H^z}}
    + \zeta^{H^0} \frac{\partial}{\partial{H^0}} ,
 \end{multline}
where the coefficients  $\zeta^t$, $\zeta^s$,  ... ,   $\zeta^{H^0}$    are functions
of $t$, $s$, $\mathbf{x}$, $\mathbf{u}$, $\rho$, $p$,  $E^y$, $E^z$,  $H^y$, $H^z$ and $H^0$.
Computations lead to  the following generators
\begin{multline}   \label{infinite_equivalence_C_generators}
X_1 ^e  = \frac{\partial}{\partial t},
\quad
X_2 ^e  = \frac{\partial}{\partial s},
\quad
X_3 ^e  = \frac{\partial}{\partial x},
\quad
X_4 ^e  = t \frac{\partial}{\partial x} + \frac{\partial}{\partial u},
\\
X_5 ^e  =  z \frac{\partial}{\partial y} - y \frac{\partial}{\partial z}
+ w \frac{\partial}{\partial v} - v \frac{\partial}{\partial w}
+ H^z \frac{\partial}{\partial H^y} - H^y \frac{\partial}{\partial H^z},
\\
X_6 ^e  = t \frac{\partial}{\partial t}
+ 2 s \frac{\partial}{\partial s}
- u \frac{\partial}{\partial u}
- v \frac{\partial}{\partial v}
- w \frac{\partial}{\partial w}
+ 2\rho \frac{\partial}{\partial \rho},
\\
X_7 ^e  = -s \frac{\partial}{\partial s}
+ x \frac{\partial}{\partial x}
+ y \frac{\partial}{\partial y}
+ z \frac{\partial}{\partial z}
+ u \frac{\partial}{\partial u}
+ v \frac{\partial}{\partial v}
+ w \frac{\partial}{\partial w}
- 2\rho \frac{\partial}{\partial \rho},
\\
X_{8} ^e  = 2 s \frac{\partial}{\partial s}
+ 2 p \frac{\partial}{\partial p}
+ 2 \rho \frac{\partial}{\partial \rho}
+ H^y \frac{\partial}{\partial H^y}
+ H^z \frac{\partial}{\partial H^z}
+ H^0 \frac{\partial}{\partial H^0},
\\
X_{9} ^e  = f_1 \left( s,  { p \over \rho^{\gamma}  } \right) \frac{\partial}{\partial y} ,
\quad
X_{10} ^e  = f_2 \left( s,  { p \over \rho^{\gamma} } \right) \frac{\partial}{\partial z} ,
\quad
X_{11} ^e  = t \frac{\partial}{\partial y} + \frac{\partial}{\partial v},
\quad
X_{12}  ^e  = t \frac{\partial}{\partial z} + \frac{\partial}{\partial w},
\end{multline}
where $f_1$ and $f_2$ are arbitrary functions of their arguments.

% \emph{(The generator $X_{12}$ is new)}.

% \subsection{The case of infinite conductivity and   $ H^0 =   0 $}

% {DO we need these equivalence transformations}

% {Probably NOT because there is nothing to scale  }

\subsection{Variational approach
for infinite conductivity  and  $ H^0  \neq  0 $}

The general form of the equivalence transformation generators
for system  (\ref{PDE_case_H_not_0})  is
\begin{equation}
X ^e    = \zeta^t \frac{\partial}{\partial{t}}
    + \zeta^s \frac{\partial}{\partial{s}}
    + \zeta^{\varphi} { \partial  \over \partial \varphi}
    + \zeta^{\psi} { \partial  \over \partial \psi}
    + \zeta^{\chi } { \partial  \over \partial \chi }
    + \zeta^{H^0} \frac{\partial}{\partial{H^0}}
    + \zeta^{S} \frac{\partial}{\partial{S}},
\end{equation}
where the coefficients depend on $( t, s ,  \varphi ,    \psi,   \chi ,  H^0 , S )  $.
We obtain the following generators
\begin{multline}     \label{equivalence_S}
X ^e  _1
=
{\ddt} ,
\quad
X ^e  _2
=
{\dds} ,
\quad
X ^e  _3
=
{\ddphi} ,
\quad
X ^e  _4
=
 { \partial  \over \partial \psi} ,
\quad
X ^e  _5
=
{ \partial  \over \partial \chi  } ,
\\
X ^e  _6
=
t {\ddphi} ,
\quad
X ^e  _7
=
t  { \partial  \over \partial \psi} ,
\quad
X ^e  _8
=
t { \partial  \over \partial \chi  } ,
\quad
X ^e  _9
=
\chi   { \partial  \over \partial \psi}
-
\phi  { \partial  \over \partial \chi  } ,
\\
X ^e  _{10}
=
   t {\ddt}
+ s  {\dds}
+ \varphi   {\ddphi}
+  \psi   { \partial  \over \partial \psi}
+  \chi  { \partial  \over \partial \chi  }    ,
\\
X ^e  _{11}
=    t {\ddt}
+ 2 s  {\dds}
 - 2  \gamma    S  {\partial \over  \partial  S  }    ,
\quad
X ^e  _{12}
=
( 1 - \gamma)     t {\ddt}
+ 2 s  {\dds}
+  \gamma H^0  {\partial \over  \partial  H^0  } .
\end{multline}

\subsection{Variational approach
for infinite conductivity and     $ H^0  =  0 $}

PDEs (\ref{PDE_case_H_0}) and (\ref{PDEB_case_H_0}),
which correspond to cases  $\gamma \neq 2 $ and $\gamma = 2$,
have arbitrary functions.
The equivalence transformations
for these PDEs
have generators of the forms
\begin{equation}
X^{e}
= \xi^{t} \frac{\partial}{\partial t}
+\xi^{s} \frac{\partial}{\partial s}
+\eta^{\varphi} \frac{\partial}{\partial\varphi}
+\eta^{S} \frac{\partial}{\partial S}
+\eta^{A} \frac{\partial}{\partial A}
\end{equation}
% for for  equation (\ref{PDE_case_H_0})
and
\begin{equation}
X^{e}
= \xi^{t} \frac{\partial}{\partial t}
+\xi^{s} \frac{\partial}{\partial s}
+\eta^{\varphi} \frac{\partial}{\partial\varphi}
+\eta^{B} \frac{\partial}{\partial B}  .
\end{equation}
% for equation   (\ref{PDEB_case_H_0}).
% respectively.
The coefficients of these generators depend on  $  ( t , s.  \varphi, S, A ) $  and $  ( t , s.  \varphi, B ) $,
respectively.

Computations provide  the generators
\begin{multline}       \label{equivalence_general}
X ^e  _1 =  {\ddt} ,
\quad
X ^e  _2 =  {\dds},
\quad
 X ^e  _3 =  {\ddphi} ,
\quad
X ^e  _4 =  t   {\ddphi} ,
\quad
X ^e  _5 =   t {\ddt}   + s  {\dds}  + \varphi   {\ddphi} ,
\\
X ^e  _6
= t  {\ddt}  - 2s  {\dds} + 2 (\gamma-2) S  {\partial \over  \partial  S  }  ,
\quad
X ^e  _7
= (1-\gamma)  t  {\ddt}  + 2s  {\dds} + 2 (\gamma-2) A {\partial \over \partial  A  }
\end{multline}
for equation (\ref{PDE_case_H_0}).
For PDE  (\ref{PDEB_case_H_0}),
we obtain  the generators
\begin{multline}      \label{equivalence_gamma_2}
X_{1}^{e}=  \frac{\partial}{\partial t} ,
\quad
X_{2}^{e}=\frac{\partial}{\partial s } ,
\quad
X_{3}^{e}=   \frac{\partial}{\partial\varphi}   ,
\quad
X_{4}^{e}=t  \frac{\partial}{\partial\varphi} ,
\\
X_{5}^{e}
=   ( 1 - \gamma) t  \frac{\partial}{\partial t }
 + 2 s   \frac{\partial}{\partial  s  }  ,
\quad
% X_{6}^{e}
% =   ( \gamma  + 1 ) t  \frac{\partial}{\partial t }
%  + 2 \varphi    \frac{\partial}{\partial\varphi}  ,
X_{6}^{e}
 =   t  \frac{\partial}{\partial t }
 +  s   \frac{\partial}{\partial  s  }
 +   \varphi    \frac{\partial}{\partial\varphi}  ,
\quad
X_{7}^{e}
= t  \frac{\partial}{\partial t }
- 2 B \frac{\partial}{\partial  B } .
\end{multline}

% {To compare with what  S.M. sent   }

\section{Lie algebra  extensions
for finite conductivity $\sigma  ( \rho , p)  $  and     $H^0 = 0 $ }

% {

% Details:

% 1. Operators for equivalence transformations (\ref{EqTrMLagrFlatH0=0})

% 2. The extended algebra  generators  (\ref{LagrFlatClassOps8})

% 3. kernel    (\ref{kern01a})

% }

Here we find extensions of the kernel of the admitted Lie algebras   (\ref{kern01a}),
which belong to the extended Lie algebra   (\ref{LagrFlatClassOps8}),
by the other generators of the  extended Lie algebra,
namely  by the generators from the set $ \{ Y_6, Y_7, Y_8 \} $.

For this purpose  we show that the action of the equivalence transformations
defined by the generators  (\ref{EqTrMLagrFlatH0=0})
and the action of the  inner automorphisms of  the  extended Lie algebra   (\ref{LagrFlatClassOps8})
coincide.
Since the kernel is an ideal of the  extended algebra
it can be extended by {sub}algebras formed by the remaining generators.
Therefore we take the optimal  system of {sub}algebras
for the  {sub}algebra  of the  remaining  operators  $ \{ Y_6, Y_7, Y_8 \} $.
It provides possible extensions  of the kernel.

\subsection{Action of the equivalence transformations}

Consider the change of the coefficients of the generators  (\ref{LagrFlatClassOps8})
under the variables  changes
given by the equivalence transformations  with generators (\ref{EqTrMLagrFlatH0=0}).
The equivalence transformation groups  corresponding to the generators $X ^e _1$, ..., $  X ^e  _{8}$
act as follows
(the unchanged variables are omitted)
\begin{equation}
\def\arraystretch{1.5}
\begin{array}{rl}
X ^e _1:&   \bar{t} = t + a ;
\\
X ^e _2: &   \bar{s} = s + a ;
\\
X ^e _3: &  \bar{x} = x + a ;
\\
X ^e _4:&   \bar{x} = x + a t,
    \quad
    \bar{u} = u + a  ;
\\
X ^e _5:&
    %\bar{v} = v \cos{a} - w \sin{a}, \quad
    %\bar{w} = w \cos{a} + v \sin{a},
    %\\
    %&
    \bar{E}^y = E^y \cos{a} + E^z \sin{a}, \quad
    \bar{E}^z = E^z \cos{a} - E^y \sin{a},
    \\
    &
    \bar{H}^y = H^y \cos{a} + H^z \sin{a}, \quad
    \bar{H}^z = H^z \cos{a} - H^y \sin{a} ;
\\
X ^e _6:&
    \bar{t} = e^a t, \quad
    \bar{s} = e^{2a} s, \quad
  %\bar{v} = e^{-a} v, \quad
    \bar{u} = e^{-a} u, \quad
    %\bar{w} = e^{-a} w, \quad
    \bar{\rho} = e^{2a} \rho, \quad
 \\&
    \bar{E}^y = e^{-a} E^y, \quad
    \bar{E}^z = e^{-a} E^z,  \quad
    \bar{\sigma} = e^{a} \sigma ;
\\
X ^e _7:&
    \bar{s} = e^{-a} s, \quad
    \bar{x} = e^{a} x, \quad
    \bar{u} = e^{a} u, \quad
    \bar{\rho} = e^{-2a} \rho,
    \\&
    \bar{E}^y = e^{a} E^y, \quad
    \bar{E}^z = e^{a} E^z, \quad
     \bar{\sigma} = e^{-2a} \sigma ;
\\
X ^e _8:&
    \bar{s} = e^{2a} s, \quad
    \bar{\rho} = e^{2a} \rho, \quad
    \bar{p} = e^{2a} p,
    \\&
    \bar{E}^y = e^{a} E^y, \quad
    \bar{E}^z = e^{a} E^z, \quad
    \bar{H}^y = e^{a} H^y, \quad
    \bar{H}^z = e^{a} H^z.
\end{array}
\end{equation}
Here $a$  is a group parameter.

Consider {transformations}  defined by a generator of the form
\begin{equation}\label{XbyBasis}
\displaystyle X = \sum_{i=1}^8 \xbasis^i  Y_i
\end{equation}
under the action of the equivalence transformations.
An equivalence transformation changes this  generator
into the  generator
\begin{equation}\label{XbyHatBasis}
X = \sum_{i=1}^8 \hat{\xbasis}^i \hat{Y}_i  ,
\end{equation}
where the basis generators in the new  variables are
\begin{multline}
% \def\arraystretch{1.75}
% \begin{array}{c}
\hat{Y}_1 = {\partial \over \partial {\bar{t}} },
\quad
\hat{Y}_2 =  {\partial \over \partial {\bar{s}} },
\quad
\hat{Y}_3 =  {\partial \over \partial {\bar{x}} },
\quad
\hat{Y}_4 = \bar{t} {\partial \over   \partial {\bar{x}} }
+ {\partial \over  \partial {\bar{u}} },
\\
\hat{Y}_5 = \bar{E}^z   {\partial \over   \partial {\bar{E}^y} }
   - \bar{E}^y {\partial \over   \partial {\bar{E}^z}  }
     + \bar{H}^z {\partial \over  \partial {\bar{H}^y}  }
   - \bar{H}^y {\partial \over   \partial {\bar{H}^z} } ,
\\
\hat{Y}_6 = \bar{t}   {\partial \over  \partial_{\bar{t}}  }
+ 2 \bar{s} {\partial \over  \partial {\bar{s}} }
    - \bar{u} {\partial \over   \partial {\bar{u}}  } %- \bar{v} \partial_{\bar{v}} - \bar{w} \partial_{\bar{w}}
    + 2 \bar{\rho} {\partial \over  \partial {\bar{\rho}} }
    - \bar{E}^y {\partial \over   \partial {\bar{E}^y} }
- \bar{E}^z {\partial \over  \partial {\bar{E}^z} } ,
\\
\hat{Y}_7 = -\bar{s}  {\partial \over  \partial {\bar{s}} }
  + \bar{x} {\partial \over  \partial {\bar{x}} }
   + \bar{u} {\partial \over  \partial {\bar{u}} }
    - 2\bar{\rho} {\partial \over   \partial {\bar{\rho}} }
    + \bar{E}^y {\partial \over   \partial {\bar{E}^y} }
  + \bar{E}^z {\partial \over   \partial {\bar{E}^z} } ,
\\
\hat{Y}_8 = 2 \bar{s} {\partial \over   \partial {\bar{s}} }
 + 2\bar{\rho} {\partial \over   \partial {\bar{\rho}} }
   + 2 \bar{p} {\partial \over   \partial   {\bar{p}} }
    + \bar{E}^y {\partial \over   \partial {\bar{E}^y}  }
+  \bar{E}^z {\partial \over   \partial {\bar{E}^z} }
    + \bar{H}^y {\partial \over   \partial {\bar{H}^y} }
   +  \bar{H}^z {\partial \over   \partial {\bar{H}^z} } .
%\\
%\hat{X}_9 = \bar{f}_1 (\bar{t} \partial_{\bar{y}} + \partial_{\bar{v}}),
%\quad
%\hat{X}_{10} = \bar{f}_2 (t \partial_{\bar{z}} + \partial_{\bar{w}}),
%\quad
%\hat{X}_{11} = \bar{f}_3 \partial_{\bar{y}},
%\quad
%\hat{X}_{12} = \bar{f}_4 \partial_{\bar{z}}.
% \end{array}
\end{multline}

For example, consider the change of the generators $Y_1$, $Y_2$, ..., $Y_8$
under the transformation corresponding to $ X ^e _1$:
\begin{equation*}
\bar{t} = t + a.
\end{equation*}
The other variables stay unchanged.
According to the variables change in the differential operator formula  \cite{bk:Ovsyannikov[1962]},
the generators~$Y_4$ and~$Y_6$ become
\begin{multline*}
Y_4 = Y_4(\bar{x}) {\partial \over  \partial {\bar{x}} }
+ Y_4(\bar{u}) {\partial \over  \partial {\bar{u}} } + \cdots
%= X_4(x)  \partial_{\bar{x}}    + X_4(u) \partial_{\bar{u}}
\\
= t   {\partial \over  \partial {\bar{x}} }
  + {\partial \over  \partial {\bar{u}} }
= (\bar{t} - a)   {\partial \over  \partial {\bar{x}} }
 + {\partial \over \partial {\bar{u}} }
= \bar{t}  {\partial \over   \partial {\bar{x}} }
  +  {\partial \over   \partial {\bar{u}}  }
- a  {\partial \over   \partial {\bar{x}} }
= \hat{Y}_4 - a \hat{Y}_3 ,
\end{multline*}
\begin{multline*}
Y_6 = Y_6(\bar{t}) {\partial \over  \partial {\bar{t}} }
+ Y_6(\bar{s}) {\partial \over   \partial {\bar{s}} } + \cdots
%= X_6(t + a) \partial_{\bar{t}} + X_6(s) \partial_{\bar{s}} + \cdots
\\
= t  {\partial \over   \partial {\bar{t}}  }
+ s   {\partial \over   \partial {\bar{s}} } + \cdots
= (\bar{t} - a)  {\partial \over   \partial {\bar{t}}  }
+ \bar{s}  {\partial \over   \partial {\bar{s}} }   + \cdots
= \hat{Y}_6 - a \hat{Y}_1.
\end{multline*}
The remaining generators stay unchanged
\begin{equation*}      % \label{XeqXhat}
Y_i = \hat{Y}_i , \qquad i\neq 4, 6.
\end{equation*}
From~(\ref{XbyBasis}) and~(\ref{XbyHatBasis}) we find
\begin{equation*}
 \xbasis^4 (\hat{Y}_4 - a \hat{Y}_3)
%  + \xbasis^5 \hat{X}_5
+ \xbasis^6 (\hat{Y}_6 - a \hat{Y}_1)
% + \xbasis^7 \hat{X}_7
% + \xbasis^8 \hat{X}_8
+ \sum_{i  \neq  \{  4, 6   \} }     \xbasis^i \hat{Y}_i
=
\sum_{i=1}^{8} \hat{\xbasis}^i \hat{Y}_i.
\end{equation*}
Hence,
\begin{equation} \label{etrkappas}
\hat{\xbasis}^1 = \xbasis^1 - a \xbasis^6,
\qquad
\hat{\xbasis}^3 = \xbasis^3 - a \xbasis^4,
\qquad
\hat{\xbasis}^i = \xbasis^i
\quad
\text{for}
\quad
i \neq 1, 3.
\end{equation}

Similarly we derive the transformations of the the coefficients
related to the generators $X^e _2$, ..., $X ^e _8$.  They are
%\begin{equation}
%\def\arraystretch{1.5}
%\begin{array}{rl}
%%Y_1 :&
%%    X_4 = \Hat{X}_4 - A \Hat{X}_3,
%%    \Quad
%%    X_6 = \Hat{X}_6 - A \Hat{X}_1,
%%    \\
%Y_2 :&
%    X_6 = \hat{X}_6 - 2 a \hat{X}_2,
%    \quad
%    X_7 = \hat{X}_7 + a \hat{X}_2,
%    \quad
%    X_8 = \hat{X}_8 - 2 a \hat{X}_2,
%    \\
%Y_3 :&
%    X_7 = \hat{X}_7 - a \hat{X}_3,
%    \\
%Y_4 :&
%    X_1 = \hat{X}_1 + a \hat{X}_3,
%    \quad
%    X_6 = \hat{X}_6 + a \hat{X}_4,
%    \quad
%    X_7 = \hat{X}_7 - a \hat{X}_4,
%    \\
%Y_6 :&
%    X_1 = e^a \hat{X}_1,
%    \quad
%    X_2 = e^{2a} \hat{X}_2,
%    \quad
%    X_4 = e^{-a}\hat{X}_4,
%\\
%Y_7 :&
%    X_2 = e^{-a} \hat{X}_2,
%    \quad
%    X_3 = e^{a} \hat{X}_3,
%    \quad
%    X_4 = e^{a} \hat{X}_4,
%\\
%Y_8 :&
%    X_2 = e^{a} \hat{X}_2,
%\end{array}
%\end{equation}
%where only changeable generators are presented.
%Coordinates of the generator $X$ are changed as follows
\begin{equation} \label{eqtrgroup01}
\def\arraystretch{1.5}
\begin{array}{rl}
%Y_1 :&
%    \hat{\xbasis}^1 = \xbasis^1 - a \xbasis^6,
%    \quad
%    \hat{\xbasis}^3 = \xbasis^3 - a \xbasis^4,
%    \\
X^e_2 :&
    \hat{\xbasis}^2 = \xbasis^2 - a (2 \xbasis^6  - \xbasis^7 + 2 \xbasis^8 )  ;
    \\
X^e_3 :&
    \hat{\xbasis}^3 = \xbasis^3 - a \xbasis^7 ;
    \\
X^e_4 :&
    \hat{\xbasis}^3 = \xbasis^3 + a \xbasis^1,
    \quad
    \hat{\xbasis}^4 = \xbasis^4 + a (\xbasis ^6 - \xbasis^7);
    \\
X^e_6 :&
    \hat{\xbasis}^1 = e^a \xbasis^1
    \quad
    \hat{\xbasis}^2 = e^{2a} \xbasis^2
    \quad
    \hat{\xbasis}^4 = e^{-a} \xbasis^4;
\\
X^e_7 :&
    \hat{\xbasis}^2 = e^{-a} \xbasis^2
    \quad
    \hat{\xbasis}^3 = e^{a} \xbasis^3
    \quad
    \hat{\xbasis}^4 = e^{a} \xbasis^4;
\\
X^e_8 :&
    \hat{\xbasis}^2 = e^{2a} \xbasis^2;
\end{array}
\end{equation}
where the unchanged coefficients are omitted.

% {checked 03.10.2021}

\subsection{Action of the inner automorphisms}

%Further we consider actions of inner automorphisms on the coefficients of the generators~(\ref{LagrFlatClassOps8}).
The  inner automorphisms are constructed
with the help of the commutator table   \cite{bk:Ovsyannikov[1962], bk:Olver}.
We obtain the following commutator table  for the generators    (\ref{LagrFlatClassOps8})
\begin{equation} \label{comtab2}
\begin{array}{c|cccccccc}
    & Y_1    & Y_2    & Y_3    & Y_4   & Y_5   & Y_6    & Y_7    & Y_8  \\ \hline
Y_1 &  0     &  0     &   0    &  Y_3  &   0   &  Y_1   &   0    &   0  \\ %\hline
Y_2 &  0     &  0     &   0    &   0   &   0   & 2Y_2   & -Y_2   & 2Y_2 \\ %\hline
Y_3 &  0     &  0     &   0    &   0   &   0   &   0    &  Y_3   &   0  \\ %\hline
Y_4 & -Y_3   &  0     &   0    &   0   &   0   & -Y_4   &  Y_4   &   0  \\ %\hline
Y_5 &   0    &  0     &   0    &   0   &   0   &   0    &   0    &   0  \\ %\hline
Y_6 & -Y_1   & -2Y_2  &   0    &   Y_4 &   0   &   0    &   0    &   0  \\ %\hline
Y_7 &   0    &  Y_2   & -Y_3   &  -Y_4 &   0   &   0    &   0    &   0  \\ %\hline
Y_8 &   0    &-2Y_2   &   0    &   0   &   0   &   0    &   0    &   0
\end{array}
\end{equation}
% {commutator table checked 15.09.2021}
The inner automorphism $A_j$ corresponds to the Lie group
of transformations with the generator~\cite{bk:Ovsyannikov[1962]}
(the minus sign is chosen for convenience)
\begin{equation*}
- \xbasis^\alpha C^\gamma_{\alpha j} \frac{\partial}{\partial{\xbasis^\gamma}} ,
\end{equation*}
where the structure constants $C^\gamma_{\alpha j}$ are found from the commutator table.

As a particular example, consider the inner automorphisms corresponding to the generator~$Y_1$:
\begin{equation*}
E_1 = - \xbasis^\alpha C^\gamma_{\alpha 1} \frac{\partial}{\partial{\xbasis^\gamma}}
= - \xbasis^4 \frac{\partial}{\partial{\xbasis^3}} -  \xbasis^6 \frac{\partial}{\partial{\xbasis^1}}.
\end{equation*}
We obtain the one-parameter group of the  inner automorphisms for~$E_1$ integrating
the Lie equations
\begin{equation*}
\frac{d\tilde{\xbasis}^1}{da}= - \tilde{\xbasis}^6,
\qquad
\frac{d\tilde{\xbasis}^3}{da}= - \tilde{\xbasis}^4,
\qquad
\frac{d\tilde{\xbasis}^i}{da}= 0, \quad i \ne 1, 3
\end{equation*}
with the initial conditions
\begin{equation*}
{\tilde{\xbasis}^j}|_{a = 0} = \xbasis^j, \quad j = 1, ..., 8.
\end{equation*}
The solution of this Cauchy problem is %(see~(\ref{etrkappas}))
\begin{equation} \label{iakappas}
\tilde{\xbasis}^1 = \xbasis^1 -  a \xbasis^6,
\qquad
\tilde{\xbasis}^3 = \xbasis^3 - a \xbasis^4,
\qquad
\tilde{\xbasis}^i = \xbasis^i
\quad
\text{for}
\quad
i \neq 1, 3.
\end{equation}

Similarly  we  obtain the  inner automorphisms
for the transformations corresponding to the other generators
\begin{equation} \label{eqtr01}
\def\arraystretch{1.25}
\begin{array}{ll}
%X_1:
%    %& \xbasis^6 \partial_{\xbasis^1} + \xbasis^4 \partial_{\xbasis^3} \Rightarrow
%    & \tilde{\xbasis}^1 = \xbasis^1 + a \xbasis^6, \quad \tilde{\xbasis}^3 = \xbasis^3 + a \xbasis^4,\\
Y_2:
    %& (2 \xbasis^6 - \xbasis^7 + 2 \xbasis^8 )\partial_{\xbasis^2} \Rightarrow
    & \tilde{\xbasis}^2 = \xbasis^2 - a (2 \xbasis^6 - \xbasis^7 + 2 \xbasis^8 ) ;  \\
Y_3:
    %& \xbasis^7 \partial_{\xbasis^3} \Rightarrow
    & \tilde{\xbasis}^3 = \xbasis^3 - a \xbasis^7 ;  \\
Y_4:
    %& \xbasis^1 \partial_{\xbasis^3} + (\xbasis^6 - \xbasis^7) \partial_{\xbasis^4} \Rightarrow
    & \tilde{\xbasis}^3 = \xbasis^3 + a \xbasis^1,
\quad \tilde{\xbasis}^4 = \xbasis^4 + a (\xbasis^6 - \xbasis^7) ;  \\
Y_6:
    %& \xbasis^1 \partial_{\xbasis^1} + 2 \xbasis^2 \partial_{\xbasis^2} - \xbasis^4 \partial_{\xbasis^4} \Rightarrow
    & \tilde{\xbasis}^1 = e^a \xbasis^1 ,
\quad \tilde{\xbasis}^2 = e^{2a} \xbasis^2 ,
\quad \tilde{\xbasis}^4 = e^{-a} \xbasis^4  ;  \\
Y_7:
    %& \xbasis^2 \partial_{\xbasis^2} - \xbasis^3 \partial_{\xbasis^3} - \xbasis^4 \partial_{\xbasis^4} \Rightarrow
    & \tilde{\xbasis}^2 =  e^{-a} \xbasis^2 ,
\quad \tilde{\xbasis}^3 = e^a \xbasis^3 ,
\quad \tilde{\xbasis}^4 = e^a \xbasis^4 ;  \\
Y_8:
    %& \xbasis^2 \partial_{\xbasis^2} \Rightarrow
    & \tilde{\xbasis}^2 =  e^{2a} \xbasis^2 ;
\end{array}
\end{equation}
where the unchanged coefficients are skipped.

\subsection{Extensions of the kernel of the admitted Lie algebras}

We observe that the coefficients changes~(\ref{etrkappas}), (\ref{eqtrgroup01})
corresponding to the equivalence transformations
coincide with the coefficients changes~(\ref{iakappas}), (\ref{eqtr01}) for the inner automorphisms.
%This property allows one can use an optimal system of {sub}algebras for classifying
% equations~\cite{bk:Suriyawichitseranee2015}.
It means that
the  equivalence transformations act
on the generators of the extended  Lie algebra
{the same way} as the inner automorphisms.
The partition of the admitted Lie algebras into classes with respect to the inner automorphisms
{coincides} with the
dissimilar {sub}algebras with respect to the  equivalence transformations.
This allows to use the optimal system of {sub}algebras for the group classification.
Moreover, for the group classification %there is no need to study all {sub}algebras from the optimal system.
it is necessary to study only {sub}algebras which include the kernel~(\ref{kern01a}).
This realizes a significant advantage of the chosen approach:
one needs to consider the minimal number of {sub}algebras.
%  to classify system~(\ref{Lagrangian_Eqs_H0_part_1}).

       %\todo{describe the two-step algorithm in more detail}
% As it was discussed  in point  \ref{two_step_method},
       %for low-dimensional Lie algebras calculation of an optimal system of subalgebras is easy enough.
% for high-dimensional Lie algebras the problem becomes complicated
% and the two-step algorithm~\cite{bk:Ovsiannikov[1993opt]} can be used.
%This algorithm reduces the problem of constructing an optimal system of high-dimensional subalgebras to a problem of lower %dimension.

% \section{Two-step algorithm for classification of high dimensional Lie algebras}

% \label{two_step_method}

For low-dimensional Lie algebras calculation of the optimal system of {sub}algebras
(also called the representative list of {sub}algebras)
is relatively easy.
For high-dimensional Lie algebras the problem becomes complicated
because it requires extensive computations.
The difficulties can be facilitated by a two-step algorithm
proposed  in~\cite{bk:Ovsiannikov[1993opt]}.
This algorithm replaces the problem of constructing
the optimal system of high-dimensional {sub}algebras
by a similar problem for lower dimensional {sub}algebras.
{Shortly, it can be described as follows. }

Let $L$ be a Lie algebra $L$ with the basis $\left\{ X_{1},X_{2},\ldots,X_{r}\right\} $.
Assume that the Lie algebra $L$ is decomposed as  $L=I\oplus F$,
where $I$ is a  proper ideal of the algebra $L$ and $F$ is a {sub}algebra.
Then the  set of the inner automorphisms $A=\mbox{Int} \ L$ of the Lie algebra $L$
is decomposed $A=A_{I}A_{F}$,
where
\begin{equation*}
AI\subset I,
\quad
A_{F}F\subset F,
\quad
(A_{I}X)_{F}=X,
\quad
\forall X \in F.
\end{equation*}
This means the following \cite{bk:Ovsiannikov[1993opt]}.
Let $x\in L$ be decomposed as $x=x_{I}+x_{F}$, where $x_{I}\in I$,
and $x_{F}\in F$. Any automorphism $B\in A$ can be written as $B=B_{I}B_{F}$,
where $B_{I}\in A_{I},\;B_{F}\in A_{F}$. The automorphisms $B_{I}$
and $B_{F}$ have the properties:
\begin{equation*}
\begin{array}{c}
B_{I}x_{F}=x_{F},
\quad
\forall x_{F}\in F,
\quad
\forall B_{I}\in A_{I}, \\
B_{F}x_{I}\in I,
\quad
B_{F}x_{F}\in F,
\quad
\forall x_{I}\in I,
\quad
\forall x_{F}\in F,
\quad
\forall B_{F}\in A_{F}.
\end{array}
\end{equation*}

At the first step, an optimal system of {sub}algebras $\Theta_{A_{F}}(F)=\{ F_{0}, F_{1},F_{2},...,F_{p},F_{p+1}\}$
of the algebra $F$ is formed. Here  $F_{0}=\{0\}$,   $F_{p+1}=\{ F \}$ and the optimal
system of the algebra $F$ is constructed with respect to the automorphisms
$A_{F}$. For each {sub}algebra $F_{j}$, $ j=0, 1,2,...,p+1 $ one has to
find its stabilizer $ \mbox{St} (F_{j})\subset A$:
\begin{equation*}
 \mbox{St}   (F_{j})=\{B\in A\ |\ B(F_{j})=F_{j}\}.
\end{equation*}
Note that $  \mbox{St}  (F_{p+1})=A$.

The second step consists of forming
% optimal systems $\Theta_{St(F_{j})}(I\oplus F_{j})$.
the optimal system of {sub}algebras $\Theta_{A}(L)$ of the algebra
$L$ as a collection of $\Theta_{  \mbox{St}  (F_{j})} (I\oplus F_{j})$, $ j=0, 1,2,...,p+1$.

If the {sub}algebra $F$ can be decomposed, then the two-step algorithm
can be used for construction of $\Theta_{A_{F}}(F)$.

% \textbf{Definition.}  Let $F$ be a {sub}algebra of a Lie algebra $L$. A maximum {sub}algebra of the Lie algebra $L$ among all {sub}algebras such that $F$ is an ideal of these {sub}algebras is called a {normalizer}  $NF$ of the {sub}algebra $F$.

% It is useful to require that an optimal system of {sub}algebras be normalized \cite{bk:Ovsiannikov[1993opt]}, that is that along any {sub}algebra from the optimal system its {normalizer } is also included in the optimal system.

\medskip  % OK

Following the two-step algorithm,
we split the extended algebra   (\ref{LagrFlatClassOps8})
 into the ideal $\{Y_1, Y_2, Y_3, Y_4, Y_5\}$
and the {sub}algebra $\{ Y_6, Y_7, Y_8\}$.
In the considered  case, the ideal coincides with the kernel and the {sub}algebra is Abelian.

The first step consists of classifying the {sub}algebra.
The optimal system of {sub}algebras for the
Abelian 3-dimensional algebra $\{X_6, X_7, X_8\}$ was obtained in \cite{bk:PateraWinternitz1977}.
It consists of three one-dimensional {sub}algebras
\begin{equation*}
\{Y_7\},
\qquad
\{Y_6 + \alpha Y_7 \},
\qquad
\{Y_8 + \alpha Y_6 + \beta Y_7\},
\end{equation*}
three two-dimensional {sub}algebras
\begin{equation*}
\{Y_6,Y_7\},
\qquad
\{ Y_8 + \alpha Y_6, Y_7\},
\qquad
\{ Y_8 + \alpha Y_7, Y_6 + \beta Y_7\}
\end{equation*}
and the whole three-dimensional {sub}algebra
\begin{equation*}
\{Y_6,Y_7,Y_8\}  .
\end{equation*}

Finally,
the cases of the optimal {sub}algebras are added to the ideal.
It turns out to be trivial because the ideal
coincides with the kernel $\{ Y_1, Y_2, Y_3, Y_4, Y_5\}$
of the admitted algebras.

\end{document}